\PassOptionsToPackage{unicode}{hyperref}
\PassOptionsToPackage{hyphens}{url}
\documentclass[
]{article}
\usepackage{xcolor}
\usepackage{amsmath,amssymb}
\usepackage{iftex}
\ifPDFTeX
  \usepackage[T1]{fontenc}
  \usepackage[utf8]{inputenc}
  \usepackage{textcomp} % provide euro and other symbols
\else % if luatex or xetex
  \usepackage{unicode-math} % this also loads fontspec
  \defaultfontfeatures{Scale=MatchLowercase}
  \defaultfontfeatures[\rmfamily]{Ligatures=TeX,Scale=1}
\fi
\usepackage{lmodern}
\ifPDFTeX\else
\fi
\IfFileExists{upquote.sty}{\usepackage{upquote}}{}
\IfFileExists{microtype.sty}{% use microtype if available
  \usepackage[]{microtype}
  \UseMicrotypeSet[protrusion]{basicmath} % disable protrusion for tt fonts
}{}
\makeatletter
\@ifundefined{KOMAClassName}{% if non-KOMA class
  \IfFileExists{parskip.sty}{%
    \usepackage{parskip}
  }{% else
    \setlength{\parindent}{0pt}
    \setlength{\parskip}{6pt plus 2pt minus 1pt}}
}{% if KOMA class
  \KOMAoptions{parskip=half}}
\makeatother
\usepackage{longtable,booktabs,array}
\usepackage{calc} % for calculating minipage widths
\usepackage{etoolbox}
\makeatletter
\patchcmd\longtable{\par}{\if@noskipsec\mbox{}\fi\par}{}{}
\makeatother
\IfFileExists{footnotehyper.sty}{\usepackage{footnotehyper}}{\usepackage{footnote}}
\makesavenoteenv{longtable}
\usepackage{graphicx}
\makeatletter
\newsavebox\pandoc@box
\newcommand*\pandocbounded[1]{% scales image to fit in text height/width
  \sbox\pandoc@box{#1}%
  \Gscale@div\@tempa{\textheight}{\dimexpr\ht\pandoc@box+\dp\pandoc@box\relax}%
  \Gscale@div\@tempb{\linewidth}{\wd\pandoc@box}%
  \ifdim\@tempb\p@<\@tempa\p@\let\@tempa\@tempb\fi% select the smaller of both
  \ifdim\@tempa\p@<\p@\scalebox{\@tempa}{\usebox\pandoc@box}%
  \else\usebox{\pandoc@box}%
  \fi%
}
\def\fps@figure{htbp}
\makeatother
\providecommand{\tightlist}{%
  \setlength{\itemsep}{0pt}\setlength{\parskip}{0pt}}
\usepackage{bookmark}
\IfFileExists{xurl.sty}{\usepackage{xurl}}{} % add URL line breaks if available
\hypersetup{
  hidelinks,
  pdfcreator={LaTeX via pandoc}}
\usepackage[a4paper,margin=2.3cm]{geometry}
\usepackage{float}
\usepackage{caption}
\AtBeginEnvironment{longtable}{\small}
\usepackage{newunicodechar}
\newunicodechar{−}{\textminus}
\newunicodechar{→}{\ensuremath{\rightarrow}}
\newunicodechar{≈}{\ensuremath{\approx}}
\newunicodechar{×}{\ensuremath{\times}}
\newunicodechar{¹}{\textsuperscript{1}}
\title{\textbf{The Main Barrier to AI Adoption in the Public Sector Is Lack of Training:}\\[0.45em]\large How a Structured Method Accompanied Productivity Gains in Two Brazilian Government Cases}
\author{Vinicius Santana Gomes\thanks{Special Advisor, Internal Control Unit, Department of Economic Development, Labor and Income of the Government of the Federal District; Official Instructor in Artificial Intelligence, Federal District School of Government. Contact: vinicius.gomes@sedet.df.gov.br; gomes.vs@gmail.com .}}
\date{}
\begin{document}
{\centering
{\LARGE\bfseries The Main Barrier to AI Adoption in the Public Sector Is Lack of Training:\par}
\vspace{0.45em}
{\large How a Structured Method Accompanied Productivity Gains in Two Brazilian Government Cases\footnotemark[2]\par}
\vspace{1.0em}
{\large Vinicius Santana Gomes\footnotemark[1]\par}
\vspace{1.2em}
\par}
\footnotetext[1]{Special Advisor, Internal Control Unit, Department of Economic Development, Labor and Income of the Government of the Federal District; Official Instructor in Artificial Intelligence, Federal District School of Government. Contact: vinicius.gomes@sedet.df.gov.br; gomes.vs@gmail.com .}
\footnotetext[2]{The Portuguese version of this document begins on page~31 of this document.}

\begin{abstract}
The adoption of generative artificial intelligence in the public sector has been treated predominantly as a technological problem, with the expectation that productivity gains would follow from the availability of increasingly capable models. This paper argues, drawing on two auditable cases in the Brazilian Public Service, that the determining barrier to adoption observed in these units was not technological but training-related, and describes the four-layer structured pedagogical methodology developed by the author. The method was applied in two units with distinct institutional profiles: the Sectoral Internal Control Office of the Federal District Department of Health (SES/CONT) throughout 2024, and the Internal Control Unit of the Federal District Department of Economic Development, Labor and Income (UCI/SEDET) throughout 2025. In both cases, the official indicators from the Electronic Information System of the Federal District Government (SEI-GDF), verifiable by third parties, recorded gains that accompanied the method's rollout: average processing time fell by 18.2\% at SES/CONT and by 50\% at UCI/SEDET, with UCI also recording an 85\% increase in technical-report production, the issuance of 286 formal recommendations to public managers, and the analysis of cases and matters whose total value, per the unit's own signed statistics, was US\$ 94.8 million, the value of the matters submitted to technical analysis. For SES/CONT, the institutional report stated that no AI-related problems were reported during the period reviewed. For UCI/SEDET, the institutional report documents that every AI-assisted draft was subject to a mandatory human-review procedure before issuance, under a Human-in-the-Loop policy operationalized through a three-step Triple Review. The analysis is consistent with the hypothesis that the method is portable across agencies with distinct mandates, operates within protocols designed to comply with international and national data-protection law and with the principles of public administration, and is accessible to public entities under budget constraints, since it used free AI models.
\end{abstract}

\noindent\textbf{Keywords:}Generative artificial intelligence; public sector; AI governance; training methodology; institutional replicability; data protection.

\section{I. Introduction}\label{i.-introduction}

There is a recurring statement in presentations on artificial intelligence in the public sector: AI gains will come once the technology matures. The statement postpones the discussion. The models already exist, are in domestic and institutional use, and continue to be improved on a monthly cycle. The delay lies elsewhere.

When the author of this paper assumed leadership of the Advisory Office at the Sectoral Internal Control Office of the Federal District Department of Health, the team was multidisciplinary, with no homogeneous legal background, manually processing a caseload whose average processing time the SEI-GDF recorded at 17 days, 22 hours, and 11 minutes. These were public data, available to anyone.

The question put to the team at that moment was not whether AI could help. It was a different one: if the tool exists, is free, and any public servant can open a browser tab and use it, why was it not being used in the unit to increase productivity, improve quality of work, and reduce case processing time and the accumulated backlog?

The answer to that question, and the methodology built over the following two years to answer it in practice, is the object of this paper.

The first piece of evidence revealed by both cases is that the main barrier between AI availability and value delivery is not technological. It is training-related. Increasingly capable models reach the desk of an operator who has no repertoire to use them safely or methodically. The second is that generic AI training does not resolve this bottleneck in public service. What resolved it was a structured method, with legal governance embedded throughout, separating what the tool does, what the operator needs to do, and what the public office requires before any document leaves the unit.

The method described here has auditable empirical evidence in two agencies of the Federal District Government: the Sectoral Internal Control Office of the Department of Health (SES/CONT, 2024) and the Internal Control Unit of SEDET-DF (UCI/SEDET, 2025). Both datasets come from the SEI-GDF (the official information system of the Brazilian Federal District Government), are verifiable by third parties, and are consolidated in formal institutional reports signed by the respective unit heads. Equally important, the method was implemented using the free-tier versions of commercial AI platforms, as the agencies lacked the budget to procure commercial AI solutions at the time.

The main text follows the order in which the method was actually built and validated. Sections II through VI cover the Sectoral Internal Control Office of the Department of Health: the operational diagnosis found in 2023, the construction of the method in partnership with the Federal District School of Government (EGOV-DF), the training of the unit's public servants who enrolled as students in the same official course open to the entire Federal District Government, and the measurement of 2024 against the previous year. Sections VII through X cover the transfer of the method to the Internal Control Unit of SEDET-DF throughout 2025, under the question that the second block must answer: does the method survive a change of agency, of subject matter, and of team? Section XI discusses what the two cases teach together, with attention also to what did not work and to the limits identified during implementation. Section XII presents the study's conclusions.

For the sake of transparency, the author declares that during the analyzed period he accumulated the roles of proponent of the method, instructor of the course at the School of Government, manager of the two units studied, and signatory of the institutional report that is the primary source in the case of the Sectoral Internal Control Office of the Department of Health. The institutional report for the second case (UCI/SEDET) was signed by the author's hierarchical superior in the unit's structure.

The technical appendix at the end of the paper gathers the documentary references and tables of official data that supported the analysis.

\subsection{I.1. Related work}\label{i.1.-related-work}

This paper enters three bodies of literature that rarely speak to one another: regulatory frameworks for artificial intelligence governance, studies on technology adoption in the public sector, and research on training in the use of generative models. The method is an attempt to integrate these three fronts into a single pedagogical architecture, and the purpose of this section is to situate the contribution in relation to each.

At the governance level, public-sector frameworks such as the NIST AI Risk Management Framework (NIST, 2023) and the U.S. Government Accountability Office's framework on artificial intelligence accountability (GAO, 2021) articulate governance structures for managing AI-related risks through iterative cycles of risk identification, measurement, and treatment. The EU Artificial Intelligence Act, adopted in 2024, translates high-level principles into legally binding risk-based categories of AI systems. The OECD Principles on Artificial Intelligence, adopted in 2019, provide a widely endorsed international normative baseline for trustworthy AI. While these instruments indicate what should be observed and how compliance or adherence may be assessed, the method presented in Section III operates on a complementary plane by translating such principles into the operational conduct of public servants, subject to procedural confidentiality and data protection laws.

At the organizational level, research on technology adoption in the public sector has emphasized since the 1990s that technology availability alone is not sufficient for value delivery. Effective performance depends on absorptive capacity (Cohen \& Levinthal, 1990), understood as the organization's ability to recognize the value of new external information, assimilate it, and apply it to its operations. The literature on digital-era governance (Dunleavy et al., 2006) argues that public-sector transformation increasingly depends on reintegration, needs-based holism, and digitization. Studies on e-government implementation (Heeks, 2003) show that adoption failures often stem from design--reality gaps between system design and organizational realities. The method detailed here operates at this frontier in artificial intelligence: it works on the team's repertoire before introducing the tool and measures results through the unit's official indicators.

On the pedagogical front, recent literature on AI literacy (Long \& Magerko, 2020) identifies the competencies required for users to understand and critically engage with AI systems, while work on large generative language models and prompting (Brown et al., 2020; Wei et al., 2022) characterizes technical patterns such as prompt design, few-shot prompting, and chain-of-thought prompting. These contributions, however, largely treat users in the abstract and do not address the specific legal regimes within which users operate. The method described here integrates these technical elements into a layer of legal governance embedded from the first class onward and evaluates its effects in a public sector context.

The method's distinct contribution is the simultaneous integration of the three planes into a single applied architecture. Regulatory frameworks, without operationalization, remain normative instruments; absorptive capacity, without pedagogical method, remains diagnostic; AI literacy research, without legal and institutional anchoring, remains a technical recommendation. The method described below unites the three and offers auditable empirical evidence of the effect of this integration in two units of the public service.

\section{II. Diagnosis: SES/CONT (2023)}\label{ii.-diagnosis-sescont-2023}

The Sectoral Internal Control Office of the Federal District Department of Health (SES/CONT) was established by Decree 39.546 of December 19, 2018 (Internal Bylaws of the State Department of Health, Art. 38). Decree 45.128 of October 31, 2023, established within this structure the Advisory Office for the Adjudication of Administrative Proceedings (ASJULG, Art. 39-B), regulated by Ordinance 1.290/2023 - SES/DF, subsequently amended by Ordinance 1.166/2024. The unit therefore operated within a formal normative framework, with regulated competencies to produce decisions, orders, adjudications, and other technical documents that sustain the correctional activity of the Department.

The first feature of the picture found in 2023, and the one that defines the entire problem the method would address, is the profile of the team. Because the Federal District Department of Health has no specific position for the correctional analysis function with legal training, it is common for the public servants who staff the advisory office to have training in Health, not Law. The team nevertheless produces documents whose requirements are legal and administrative: normative grounding, admissibility analysis, interpretation of disciplinary cases, and management of legal deadlines and statutes of limitation. The mismatch between the predominant training of the team and the nature of the expected deliverable is what makes the unit a paradigmatic case of the training bottleneck described in the previous section.

The second element of the picture is volume. In 2023, the unit processed 1,752 cases. The average processing time consolidated by the SEI-GDF was 17 days, 22 hours, and 11 minutes (17.92 days). Documentary output totaled 4,846 documents among Decisions, Memoranda, and Adjudications, with 434 Official Letters and 28 Advisory Notices issued.

In August 2023, the unit produced 69 Adjudications in a single month. That number alone represented 17.1\% of the entire annual Adjudication output. It was not the result of regular operation: it was a batch of cases with imminent risk of statute-of-limitations expiration, forwarded simultaneously to the Internal Control Office after accumulated delay in another sector. To absorb the demand within the legal deadline, an emergency task force had to be mobilized. Several public servants worked beyond normal capacity that week, and part of the output was produced on non-working days to avoid losing cases due to the expiration of the statute of limitations. No case in that batch was lost. The episode, however, made plain what the aggregate average processing time did not say: the unit operated at the limit, and any external demand peak pushed regular work into holidays and the public servants' days off.

A specific structural restriction to the subject matter handled by the Internal Control Office added to the previous elements. Disciplinary cases involve confidential data, identification of individuals, and information whose exposure to external systems is prohibited under the law. Any artificial-intelligence tool to be introduced into the routine would need to assist the team without exposing protected data. This restriction, at first sight a limitation, would prove to be a decisive element of the methodological design and is described in detail in Sections III and IV.

The 2023 diagnosis, therefore, was not that of a malfunctioning unit. It was that of a unit delivering above its own sustainable capacity, with a willing team but pedagogically unassisted regarding the available tools, and within a legal perimeter that made generic market solutions unfeasible. The method had to be born of this intersection.

\section{III. The Method}\label{iii.-the-method}

The diagnosis of the previous section left three parallel fronts to resolve: a team without homogeneous legal repertoire needing to produce documents that require technical grounding; an operation whose effective capacity was constantly exceeded by demand; and a confidentiality and data-protection perimeter that prohibited direct use of external tools. There was no program in the Brazilian AI-training market at that time that addressed these three vectors at the same time. Prompt-engineering courses taught technique without legal context; AI-governance courses discussed regulatory frameworks without grounding in the public servant's routine; generic ``AI for managers'' courses were too superficial to sustain real documentary production.

The choice was to build the repertoire by hand, integrating tracks that normally do not speak to one another: technical fundamentals of generative AI, prompt engineering, ethics and risk, regulatory governance, and applied AI. The directed study path, over a little more than a year, articulated courses and certifications from international institutions covering four axes: (i) conceptual and managerial base in AI applied to business and public administration; (ii) operational technique of using generative models and prompt engineering; (iii) regulatory governance and ethics of AI; and (iv) people management during technology adoption. No single track sufficed; the useful repertoire was the intersection.

This repertoire, organized by the demands of the problem and not by the ease of the content, had to be transferable to a team that would have neither the time nor the mandate to retrace the same path. It was from this transferability requirement that the methodological structure was born: the structure that would come to sustain both the official course of the Federal District School of Government and the internal operation of the two units. This is the AI House, a four-layer pedagogical method for the adoption of Artificial Intelligence in the public sector:

\begin{figure}[H]
\centering
\includegraphics[width=0.85\linewidth]{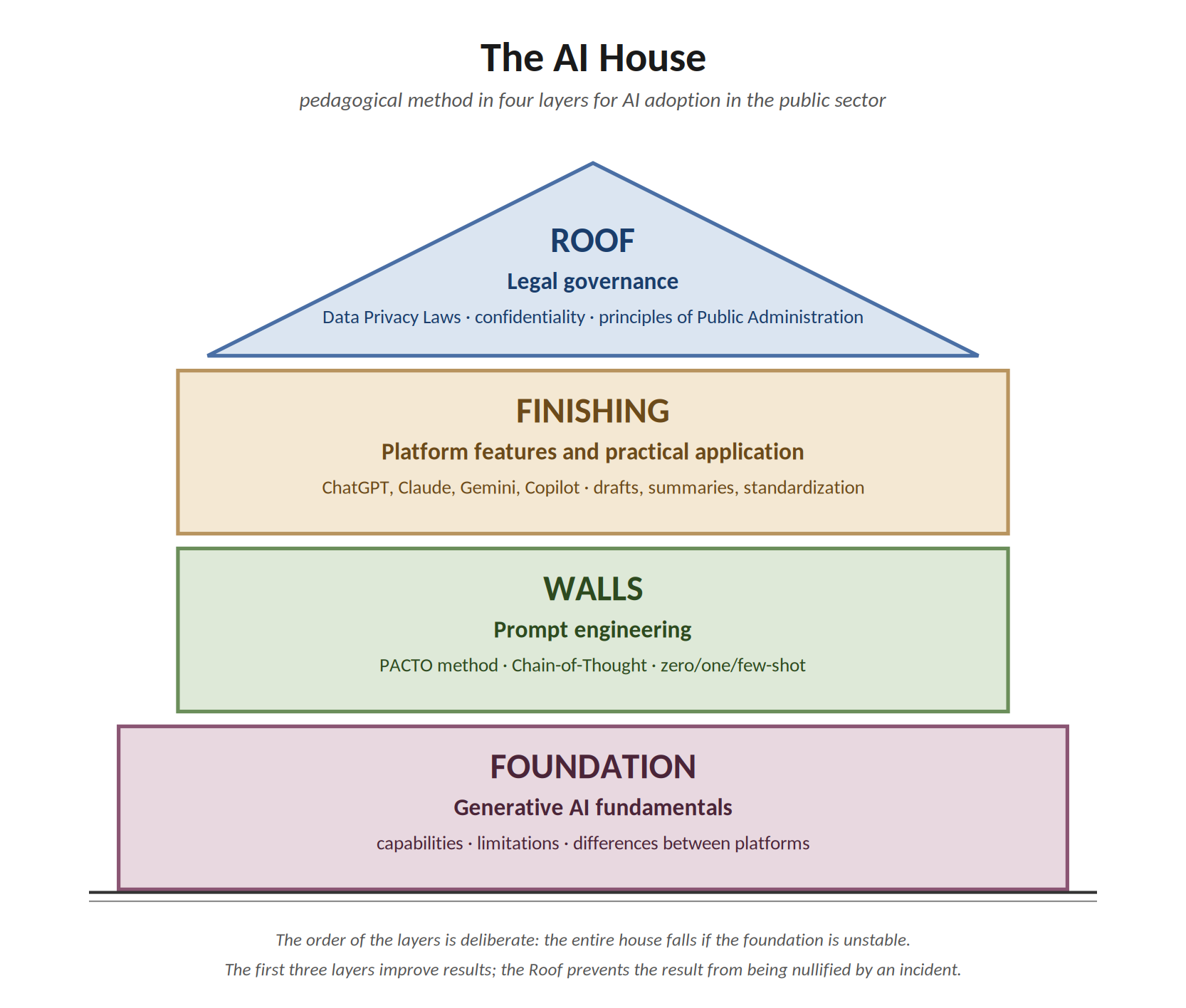}
\par\smallskip
\textit{Figure 1: The AI House: four-layer structure, with examples of applied content in each.}
\end{figure}

Each layer fulfills a function in the public servant's learning, and skipping any of them generates a specific, predictable mode of failure. The Foundation teaches what generative artificial intelligence is and how the model produces text. It is the foundation of the mental model the public servant needs to have about the tool before operating it. The Walls teach the correct technique of prompt construction: the PACTO method (Persona, Action, Context, Tone, Observations, with an optional Example), Chain of Thought, and the proper use of zero-shot, one-shot, and few-shot modes. Without a well-structured prompt, even the most advanced tool functionality fails to produce usable results. The Finishing brings AI functionalities applied to the concrete activities of the public servant's job (drafting of motions, synthesis of long documents, standardization of technical writing, summarization of case-law) and the functional differences between the main platforms available to the public servant. It is in this layer that each functionality is tied to the type of document the public servant produces, with clear identification of which part the AI can prepare, which part the public servant must write, and the point at which human review is mandatory. The Roof integrates from the outset the Data Protection Law, procedural confidentiality, the principles of Public Administration, external oversight, and the international AI governance frameworks applicable to the public sector (NIST AI Risk Management Framework, GAO AI Accountability Framework, EU AI Act, OECD recommendations).

The order of the layers is deliberate, and the house collapses entirely if the foundation is unstable. Without the Foundation, the public servant uses the tool as a search engine and replicates any hallucinations in an official document. Without the Walls, the public servant writes vague prompts and receives equally vague answers, wasting time in trial-and-error without reaching a usable result. Without the Finishing, knowledge remains abstract and does not descend to the concrete task of the public servant: it is known what AI can do in theory, but not where it fits into the document that must be produced by the end of the day. Without the Roof, the previous three layers produce data-protection incidents, poorly grounded administrative decisions, and liability for the Administration and the public servant, precisely the kind of AI adoption that paralyzes public projects. The first three layers raise output; the Roof prevents the output from being voided by an incident.

There is one design choice that merits explicit mention. The method was built to operate on the free-tier versions of commercial AI platforms (ChatGPT, Claude, Gemini, Copilot, NotebookLM, Perplexity, POE). At no point did the method presuppose corporate licensing, customized platforms, or proprietary AI infrastructure. The reason has three layers: technical, strategic, and the motivational effect on managers. Technical because the empirical evidence presented in Sections VI and X shows that free-tier tools were sufficient for most analytical and writing tasks of the average public servant, provided they know how to operate them within the governance Roof. Strategic because tying the methodology to a specific corporate license would create an initial adoption barrier that would exclude most public agencies, especially those whose budgets cannot accommodate that kind of acquisition. Motivational because concrete results obtained with free-tier versions provide institutional managers with the empirical basis to justify, when and if necessary, the subsequent acquisition of paid corporate versions of the same platforms. The method had to be replicable in any public agency, regardless of scale or jurisdiction. The accessibility of the method was a design parameter.

One operational element of this design deserves conceptual emphasis. The custom AI functionality, available in the main commercial platforms (Projects in ChatGPT and Claude, Gems in Gemini, for example), even in the free-tier versions, makes possible the effective use of generative AI in public agencies without the need for corporate licensing or proprietary platforms. By loading into the custom AI the normative base and the correct documents pertinent to the unit's subject matter, the public servant obtains answers anchored in that base, with persona and tone configured for the institutional context. In both units studied in this paper, SES/CONT and UCI/SEDET, the construction of custom AIs was the operational key tool of the method.

With the repertoire consolidated and the pedagogical structure designed, the next step was to materialize the method as an official course, open to any public servant of the Federal District Government. This is the object of Section IV.

\section{IV. Official course at the School of Government}\label{iv.-official-course-at-the-school-of-government}

The Federal District School of Government (EGOV-DF) is the official training institution of public servants of the Federal District Government. Building the method as an EGOV-DF course, rather than as in-house training, had three direct effects. The first, institutional: the course becomes part of the official training catalog of the GDF, with a formal pedagogical mandate, and public servants who complete it receive certification from a recognized institution. The second, scalar: the cohort is no longer limited to the staff of one Unit, and any public servant of the GDF, from any Department, can enroll, opening the training to a much broader audience than that of a single unit. The third, evidential: by applying the course in the Unit after it already exists externally, the team becomes a testbed for something that exists independently of it, and the productivity results acquire internal validity that ad-hoc training built within the unit could never have.

The course was structured into five in-person classes, totaling 20 hours of instruction (four hours per class), under the official title Artificial Intelligence in the public sector: techniques, risks, and applications. The 20-hour course load was a deliberate decision: longer courses compete with the public servant's operational routine, disperse focus, and raise dropout before completion. Experience with cohorts in formats of 12, 16, and 20 hours showed that this range is the most effective, precisely because it does not compete with the public servant's routine and preserves the pedagogical depth of the classes. The proposal was filed by the author of this paper with the School of Government (Memorandum No.~4/2024 from SES/CONT/ASJULG) and approved by the School of Government (Memorandum No.~166/2024 from SEEC/SEGEA/EGOV, SEI case 04044-00021535/2024-26). The modality was in-person and intended for public agents of agencies and entities of the Direct and Indirect Administration of the Federal District, in civil or military careers.

\begin{longtable}[]{@{}
  >{\raggedright\arraybackslash}p{(\linewidth - 4\tabcolsep) * \real{0.0975}}
  >{\raggedright\arraybackslash}p{(\linewidth - 4\tabcolsep) * \real{0.8058}}
  >{\raggedright\arraybackslash}p{(\linewidth - 4\tabcolsep) * \real{0.0967}}@{}}
\toprule\noalign{}
\begin{minipage}[b]{\linewidth}\raggedright
Class
\end{minipage} & \begin{minipage}[b]{\linewidth}\raggedright
Topic
\end{minipage} & \begin{minipage}[b]{\linewidth}\raggedright
Hours
\end{minipage} \\
\midrule\noalign{}
\endhead
\bottomrule\noalign{}
\endlastfoot
1 & Introduction and Applications of Artificial Intelligence & 4h \\
2 & Prompt Engineering for Artificial Intelligence: Techniques and Approaches & 4h \\
3 & Artificial Intelligence Platforms & 4h \\
4 & Limitations, Risks, and Ethical Issues of Artificial Intelligence & 4h \\
5 & Artificial Intelligence in Public Service & 4h \\
\multicolumn{2}{@{}>{\raggedright\arraybackslash}p{(\linewidth - 4\tabcolsep) * \real{0.9033} + 2\tabcolsep}}{%
\textbf{Total}} & \textbf{20h} \\
\end{longtable}

The correspondence between the classes and the structure of the method is direct: Class 1 supports the Foundation; Class 2 forms the Walls; Classes 3 and 5 compose the Finishing; and Class 4 concentrates the Roof, as detailed in Section III.

The modality combines lecture-based classes in a conventional classroom with practical classes in a computer lab, allowing the public servant to leave the training having already operated the tools in exercises and projects directly applicable to their routine.

For the specific case of the Sectoral Internal Control Office of the Department of Health, a practical extension of the method was built that deserves separate note: a custom AI, with a knowledge base composed of ten legal instruments directly applicable to the unit: Decree 39.701/2019 (disciplinary procedures), Normative Instructions 1/2021 (Conduct Adjustment Term), 2/2021 (Admissibility Assessment and Preliminary Investigation Procedure) and 2/2016 (Mediation), Law 4.938/2012 (SICOR, Federal District Correctional System), Complementary Law 840/2011 (Legal Regime of Federal District Public Servants), the Federal District Organic Law, Decree 37.297/2016 (Code of Ethics), and the Theoretical and Practical Manuals of Disciplinary Procedures of the Federal District Comptroller General. The custom AI operated as an internal legal assistant to the team: it answered questions about legislation, suggested structures for legal grounding, and pointed to relevant administrative precedents, always from the loaded documentary base, without the need to query external systems with data from the concrete case.

This architecture resolves the structural restriction identified in Section II. Queries to the custom AI were made in a de-identified manner: the public servant described the normative doubt without informing names, case numbers, or any identifiable data of the actual situation. The AI responded from the loaded legal content. Three technical-legal considerations sustain the design of the method. First: the data treatment by the Data Anonymization Protocol (PAD), adopted before any submission to the platform, ensures that the content sent ceases to be personal data under Article 12 of the Brazilian Data Protection Law (Law No.~13.709/2018), which removes anonymized data from the Law's scope. Second: to mitigate residual risk in the free-tier versions of the platforms, the protocol guides the deactivation of conversation-training authorization functionalities and the use of momentary or temporary conversation modes offered by the providers, which prevent the use of submitted data for model training and establish a limited log-retention window. Third: administrative liability for the issued document remains entirely with the public servant who signs it, in accordance with the Human-in-the-Loop principle adopted in both units and formalized in the UCI's AI Governance Framework (SEI Doc. 194251158).

The final element of the course design, present from the first class, is the principle of integral human supervision. All content generated by AI passes through technical review by the document's author and validation by the unit head before any formalization. The AI is an assistant; the public servant remains responsible. Under this principle, throughout the entire cycle of the Internal Control Office no administrative decision, adjudication, or technical report was issued without integral human review. The institutional report stated that no AI-related problems were reported during the observed period.

With the course designed, the EGOV-DF catalog receptive to the offering, and the custom AI built for the legal environment of the Internal Control Office, what remained was the step that would make all of this measurable: applying the method to the public servants who actually produce the unit's documents. This is the object of Section V.

\section{V. Application at SES/CONT}\label{v.-application-at-sescont}

The course Artificial Intelligence in the public sector: techniques, risks, and applications was open to any public servant of the Federal District. The public servants of the Sectoral Internal Control Office of the Department of Health enrolled as students in the course, the same course offered to colleagues from other Departments, under the official instruction of EGOV-DF, and with certification at the end. The public servants who would operate AI in the unit learned the method alongside public servants from other agencies, in an egalitarian and auditable pedagogical environment.

The internal operation was assembled with five elements, all derived from the EGOV-DF course classes and translated into the unit's routine following the training. The first, fundamentals of generative AI and its limits, ensured that each public servant understood what the tool could and could not deliver. The second, prompt-engineering techniques applied to the drafting of official documents, transformed generic prompts into commands with persona, action, context, tone, observations, and an example when necessary, according to the PACTO method already described. The third, use of the custom AI with the legal base of the Internal Control Office. The fourth, security and data-protection protocols, was trained as a reflex: no identifiable data, no case number, no confidential matter enters the tool without prior de-identification. The fifth, critical review and human validation, transformed AI into a formal assistant: every generated document went through technical review by the responsible public servant and validation by the unit head before any formalization.

In daily practice, the method materialized in clear usage patterns. To draft a Decision, the public servant would formulate the normative question in a de-identified manner in the custom AI, obtain the outline of the legal grounding from the unit's legal base, and then take up the text incorporating the specific data of the concrete case outside the tool. At no point was AI the author of the document. It was an operational assistant that expanded the speed and consistency of human work.

The unit operated under this regime throughout 2024. What happened to the operational indicators during that year, compared to the previous year, is the object of Section VI.

\section{VI. Results: SES/CONT (2023→2024)}\label{vi.-results-sescont-20232024}

The year 2024 was the unit's first year operating within the method. The numbers below, extracted entirely from the official SEI-GDF statistics and consolidated in the SES/CONT Executive Management Report (SEI Doc. 197403428, case 00060-00582291/2024-11), compare 2024 performance with that of 2023.

The unit's central indicator, average processing time, fell from 17 days, 22 hours, and 11 minutes (17.92 days) in 2023 to 14 days, 15 hours, and 44 minutes (14.66 days) in 2024. The reduction was 3.26 days per case, or 18.2\% in the aggregate.

\begin{figure}[H]
\centering
\includegraphics[width=0.80\linewidth,height=7cm,keepaspectratio]{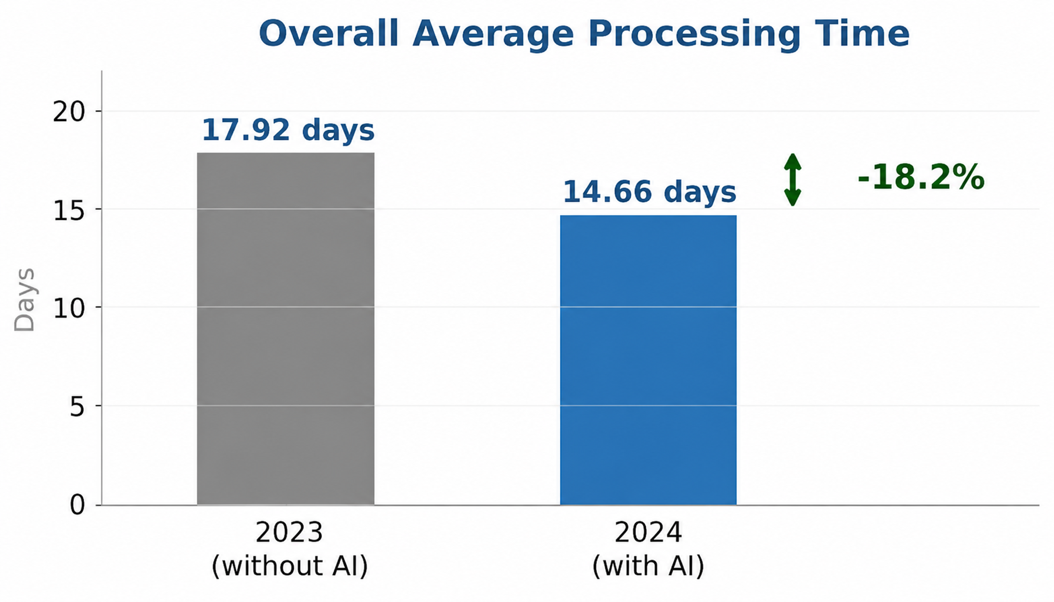}
\par\smallskip
\textit{Figure 2: Overall average processing time SES/CONT: 2023 vs.~2024.}
\end{figure}

\begin{longtable}[]{@{}
  >{\centering\arraybackslash}p{(\linewidth - 6\tabcolsep) * \real{0.2499}}
  >{\centering\arraybackslash}p{(\linewidth - 6\tabcolsep) * \real{0.2499}}
  >{\centering\arraybackslash}p{(\linewidth - 6\tabcolsep) * \real{0.2499}}
  >{\centering\arraybackslash}p{(\linewidth - 6\tabcolsep) * \real{0.2502}}@{}}
\toprule\noalign{}
\begin{minipage}[b]{\linewidth}\centering
Indicator
\end{minipage} & \begin{minipage}[b]{\linewidth}\centering
2023
\end{minipage} & \begin{minipage}[b]{\linewidth}\centering
2024 (with AI)
\end{minipage} & \begin{minipage}[b]{\linewidth}\centering
Variation
\end{minipage} \\
\midrule\noalign{}
\endhead
\bottomrule\noalign{}
\endlastfoot
Average processing time of the Unit's cases & 17.92 days & 14.66 days & \textbf{−18.2\%} \\
D+M+A (Decisions + Memoranda + Adjudications) produced per case & 2.77 & 2.78 & +0.4\% \\
\end{longtable}

Documentary output per case remained stable (2.78 versus 2.77 in 2023), meaning that the unit did not cut deliveries to deliver faster. Each case continued to generate approximately the same number of Decisions, Memoranda, and Adjudications, in 18.2\% less time.

The reduction in time was not distributed uniformly. It concentrated precisely in the types of cases that demand more technical analysis, normative grounding, and structured writing: the activities in which the method effectively operates.

Below are the case types with the greatest reduction in time:

\begin{figure}[H]
\centering
\includegraphics[width=0.80\linewidth,height=7cm,keepaspectratio]{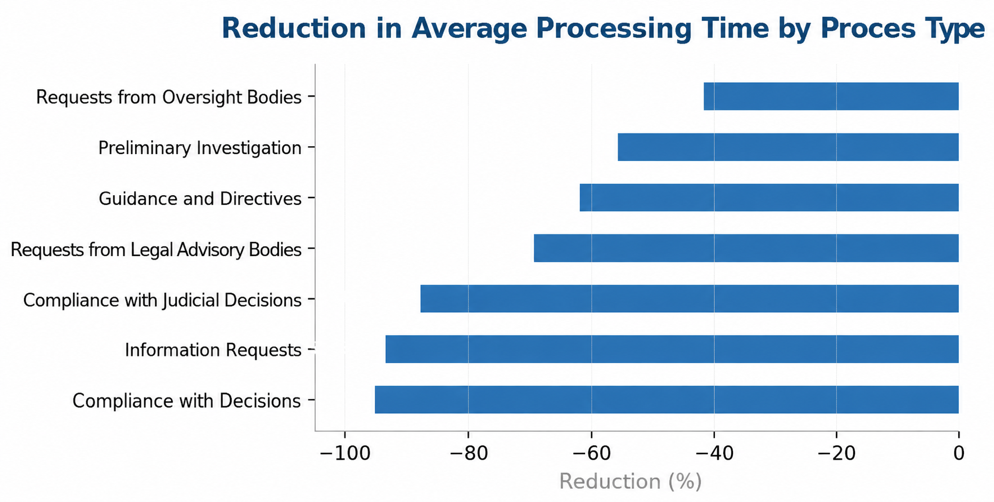}
\par\smallskip
\textit{Figure 3: Reduction in average time by case type (SES/CONT)}
\end{figure}

\begin{longtable}[]{@{}
  >{\centering\arraybackslash}p{(\linewidth - 6\tabcolsep) * \real{0.2499}}
  >{\centering\arraybackslash}p{(\linewidth - 6\tabcolsep) * \real{0.2499}}
  >{\centering\arraybackslash}p{(\linewidth - 6\tabcolsep) * \real{0.2499}}
  >{\centering\arraybackslash}p{(\linewidth - 6\tabcolsep) * \real{0.2502}}@{}}
\toprule\noalign{}
\begin{minipage}[b]{\linewidth}\centering
Case type
\end{minipage} & \begin{minipage}[b]{\linewidth}\centering
2023
\end{minipage} & \begin{minipage}[b]{\linewidth}\centering
2024
\end{minipage} & \begin{minipage}[b]{\linewidth}\centering
Reduction
\end{minipage} \\
\midrule\noalign{}
\endhead
\bottomrule\noalign{}
\endlastfoot
Decision Enforcement & 17.82 days & 0.81 days & \textbf{−95.5\%} \\
Information Request (Internal Control) & 22.81 days & 1.38 days & \textbf{−93.9\%} \\
Judicial Action: Enforcement & 24.96 days & 2.92 days & \textbf{−88.3\%} \\
Consultations from Legal Bodies & 9.27 days & 2.82 days & −69.6\% \\
Guidelines and Directives & 15.74 days & 5.92 days & −62.4\% \\
Preliminary Investigation & 7.39 days & 3.16 days & −57.2\% \\
Consultations from Oversight Bodies & 12.96 days & 7.00 days & −46.0\% \\
\end{longtable}

When these time gains are aggregated to the 1,581 cases effectively processed in 2024, a useful concept emerges to size the time gain: estimated released capacity. If the unit had operated with the productivity of 2023, the 1,581 cases would have consumed 28,332 cumulative processing days (1,581 × 17.92). With the 2024 gain, actual consumption was 23,177 days (1,581 × 14.66). The difference, 5,155 days, is equivalent to processing approximately 352 additional cases, a 22.3\% increase in the unit's operational capacity.

\begin{figure}[H]
\centering
\includegraphics[width=0.80\linewidth,height=7cm,keepaspectratio]{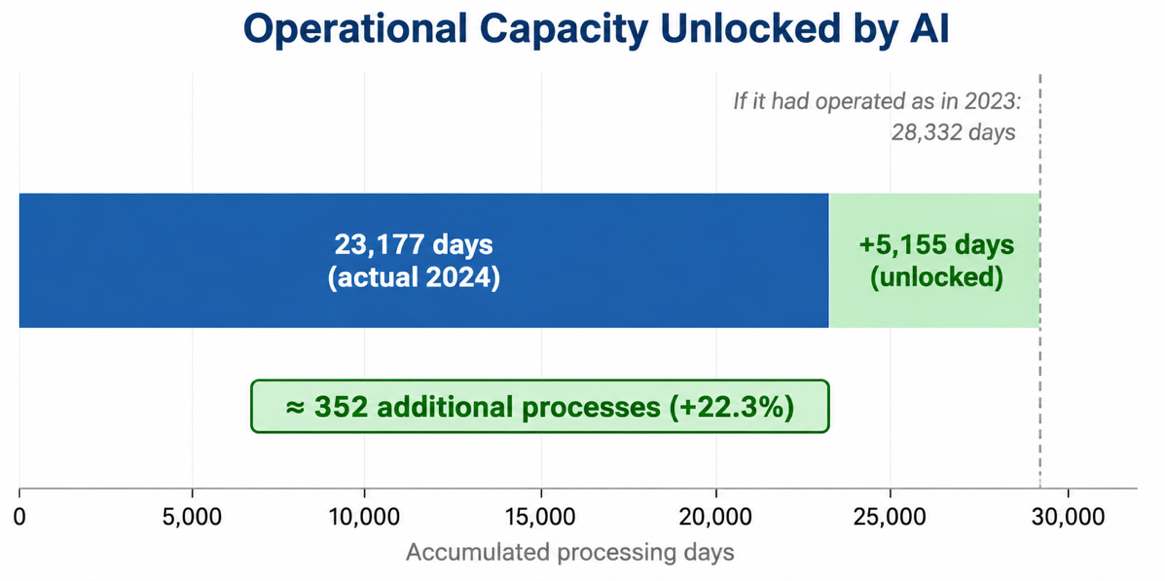}
\par\smallskip
\textit{Figure 4: Estimated released capacity at SES/CONT.}
\end{figure}

This released capacity materialized on three fronts. The first, and most visible, was the clearance of the accumulated backlog: at the end of 2024, the Internal Control Office had no case sitting more than 10 days in the inbox awaiting handling. The unit began to operate in continuous-flow mode. Report No.~12/2024 of SES/CONT/ASJULG records textually that ``decision drafts are produced within 10 days, avoiding backlogs.''

The second front was the stabilization of adjudications output. In 2023, the actual monthly average, excluding the atypical month described in Section II, was 30.36 adjudications per month. In 2024, the average was 31.00 adjudications per month, with the decisive advantage that the number was distributed much more regularly: the monthly amplitude fell from 56 adjudications (between minimum and maximum of 2023) to 25 (in 2024), a 55.4\% reduction. Monthly output was markedly more stable under the method: the same average volume, much more predictable distribution, without the need for emergency operation on weekends and holidays.

\begin{figure}[H]
\centering
\includegraphics[width=0.80\linewidth,height=7cm,keepaspectratio]{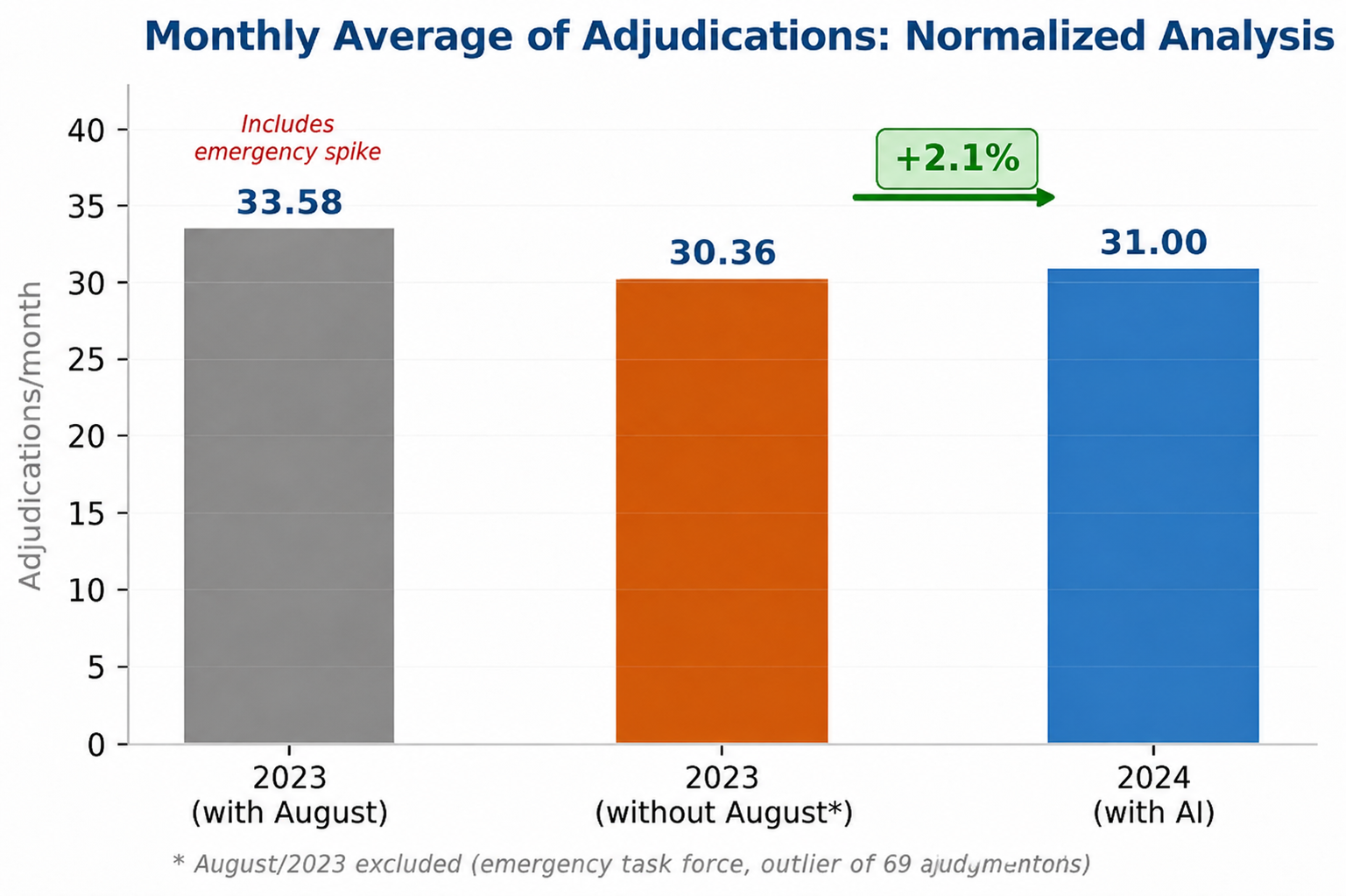}
\par\smallskip
\textit{Figure 5: Monthly average of Adjudications: normalized analysis (SES/CONT).}
\end{figure}

\begin{figure}[H]
\centering
\includegraphics[width=0.80\linewidth,height=7cm,keepaspectratio]{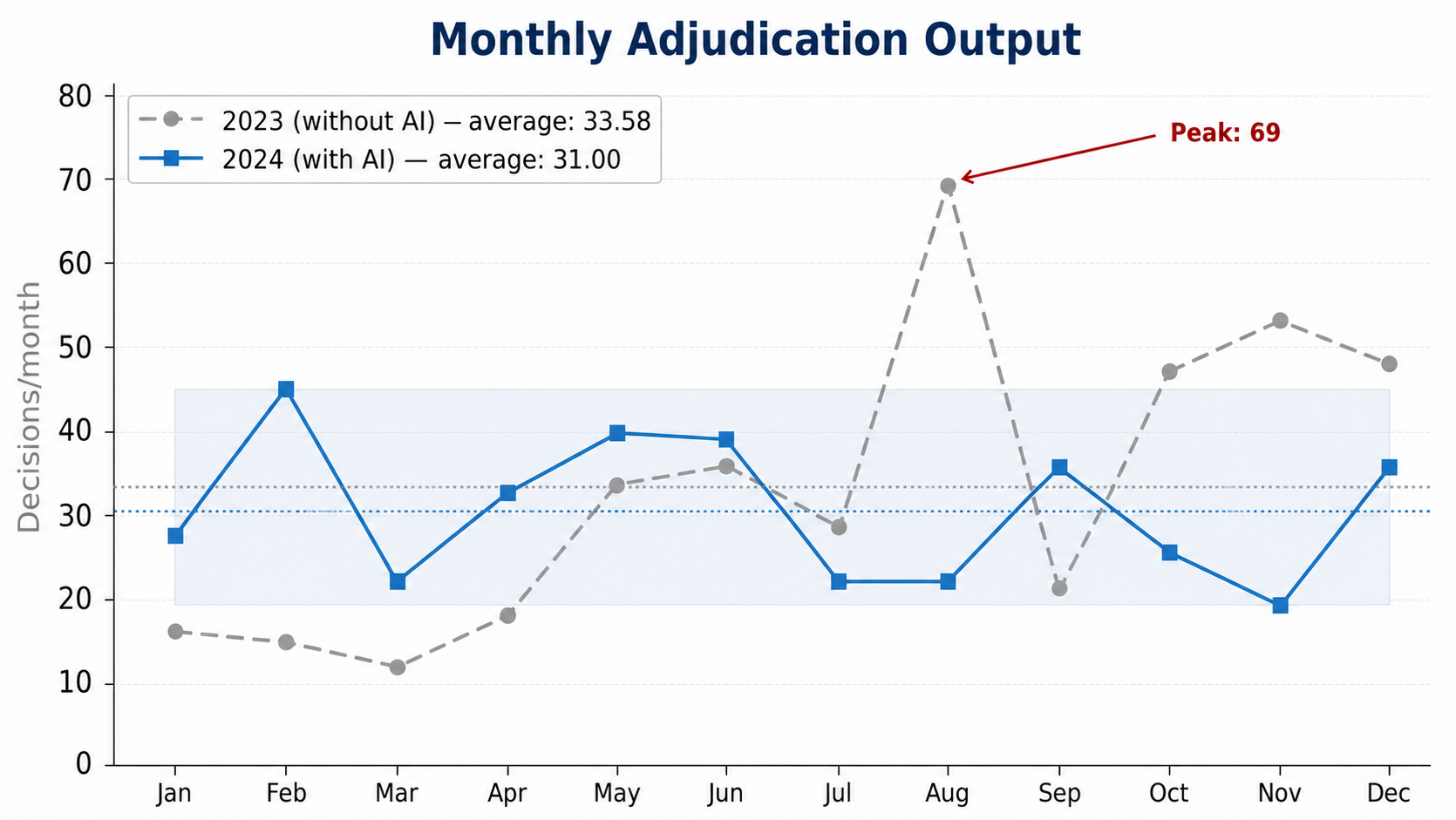}
\par\smallskip
\textit{Figure 6: Monthly Adjudication output (SES/CONT, 2023 vs.~2024).}
\end{figure}

The third front was the expansion of institutional articulation capacity. The output of Official Letters grew from 434 to 532 (+22.6\%), and that of Advisory Notices (the corrective recommendations addressed to the sectors responsible for problems identified in disciplinary cases) jumped from 28 to 214, a 664\% growth. This last figure marks a qualitative shift in the unit's posture: the Internal Control Office stopped operating only with punitive logic (instituting case, adjudicating, applying sanction) and began to operate also with preventive and corrective logic, signaling to the sectors the flows and practices that originated the irregularities so they could be remedied before the next incident.

\begin{figure}[H]
\centering
\includegraphics[width=0.80\linewidth,height=7cm,keepaspectratio]{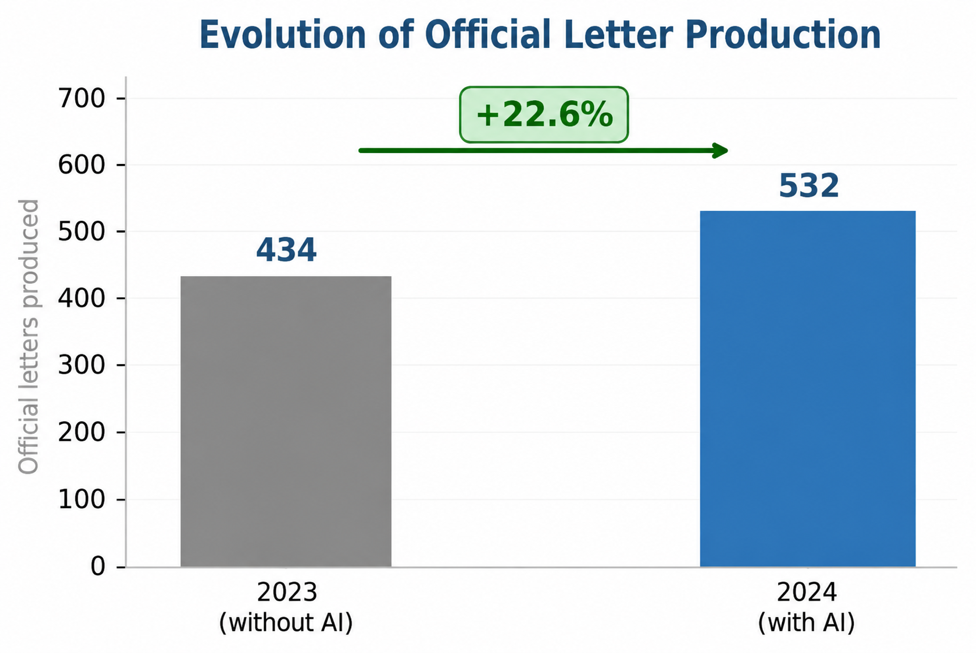}
\par\smallskip
\textit{Figure 7: Evolution of Official Letter output (SES/CONT).}
\end{figure}

\begin{figure}[H]
\centering
\includegraphics[width=0.80\linewidth,height=7cm,keepaspectratio]{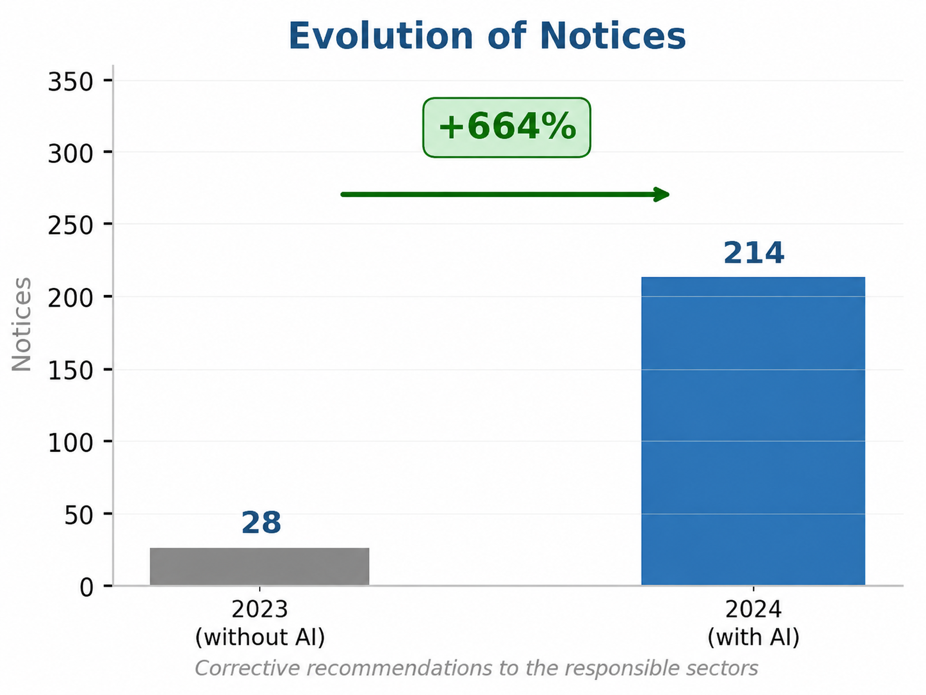}
\par\smallskip
\textit{Figure 8: Evolution of Advisory Notices (SES/CONT): shift from punitive to preventive posture.}
\end{figure}

The 2024 balance can be synthesized in three SEI-GDF-verifiable statements: the unit became 18.2\% faster without cutting deliveries; it gained capacity equivalent to 352 additional cases; and it changed the way it operates, expanding preventive work more than sevenfold. The institutional report stated that no AI-related problems were reported during the period.

\begin{figure}[H]
\centering
\includegraphics[width=0.80\linewidth,height=7cm,keepaspectratio]{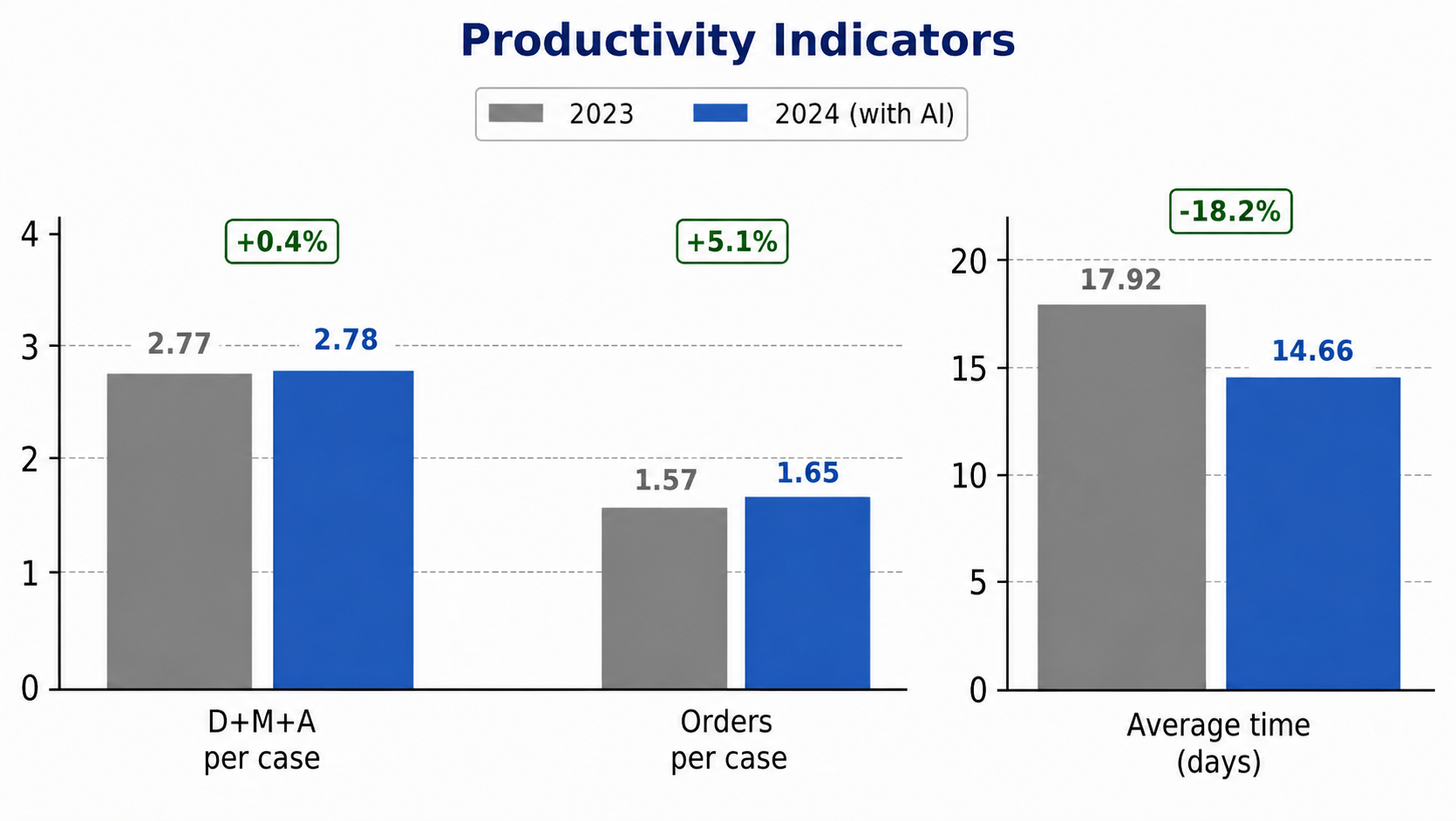}
\par\smallskip
\textit{Figure 9: Productivity indicators (SES/CONT, 2023 vs.~2024).}
\end{figure}

Section VII changes agencies and formulates the question that the second block of the paper must answer: does the method survive application in another agency, with another team and another subject matter?

\section{VII. Transfer to UCI/SEDET}\label{vii.-transfer-to-ucisedet}

The auditable results of the previous section make concrete a methodology that, until this point, could still be questioned as an isolated case. The remaining question, which defines the second block of this paper, is whether what worked at the Sectoral Internal Control Office of the Department of Health also works in another agency, with another team, and with another type of work.

This question stopped being hypothetical between 2024 and 2025, when the author was appointed Special Advisor of the Internal Control Unit of the Federal District Department of Economic Development, Labor and Income (UCI/SEDET-DF). The change was a functional transition within the Government and created precisely the replicability test that the methodology lacked.

Three variables changed in parallel. The first was the agency: a departure from a Department with a substantive mandate of public health into another with a substantive mandate of economic development, labor, and income. Each Department has its own administrative culture, its own internal workflows, and dynamics of relationship with external oversight bodies that do not transfer automatically.

The second variable was the subject matter. At SES/CONT, the axis was correctional control: disciplinary cases, inquiries, administrative adjudications, with a normative base centered on the legal regime of the public servant and on disciplinary procedures. At UCI/SEDET, the axis is general internal control: analysis of contracting, audit of financial execution, oversight of programs and agreements, with a normative base centered on the Public Procurement Law and on the constitutional principles of Public Administration. With the subject matter change, the type of document the AI must support, the analytical complexity expected of each piece, and the level of institutional risk embedded in each error all change together.

The third variable was the team. The method was originally installed in a team with predominantly Health training. At UCI/SEDET, the team was different, with a different trajectory, posted in another Department, with institutional ties built with another unit. The method had to be presented to this new team and produce, within a reasonable timeframe, gains comparable to those measured in the previous agency.

What remained constant was the method. The method, with the four layers described in Section III and the operational principles consolidated in Section IV, was transferred without structural reformulation. The specific adaptations that UCI required will be described in Section IX. The foundation, the pedagogical sequence, the use of free-tier versions of tools, and integral human supervision remained intact.

If the method were in fact a set of solutions designed for a single case, the transition would have revealed this quickly: Health and Economic Development have little in common in operational detail. If, on the contrary, the method was what it claimed to be, a replicable method based on transferable pedagogical principles, then UCI/SEDET would have to produce, in 2025, results comparable to those of SES/CONT in 2024, even if with its own profile. What happened is the object of Sections VIII, IX, and X.

\section{VIII. Diagnosis: UCI/SEDET (2024)}\label{viii.-diagnosis-ucisedet-2024}

The Internal Control Unit of the Federal District Department of Economic Development, Labor and Income (UCI/SEDET-DF) is directly linked to the Department's Cabinet. Its function is preventive and concomitant internal control: it examines administrative cases relevant to the Department, especially those with material financial impact, and produces technical reports with compliance analysis, risk identification, and recommendations directed to the responsible public managers.

There is an institutional trait of this unit that must be registered before the numbers, because it changes what counts as a result. Under Brazilian administrative law, recommendations issued by an Internal Control Unit are not binding: the public manager may comply with them or not. This means that the unit's principal asset is not only the quantity of recommendations issued, but the technical quality of each one. The more defensible the piece, in evidentiary rigor and legal grounding, the higher the probability that the manager will comply, and the greater the preventive function the unit effectively exercises over the Department.

In 2024, based on the official SEI-GDF data consolidated in Report No.~1/2026 of SEDET/GAB/UCI (SEI Doc. 193384267, case 04035-00000853/2026-24), these are the aggregate indicators of the unit's work.

The 34 days of average time express the internal duration of the procedure within the unit, from receipt of a case to production of the output technical document. In a unit whose main deliverable is the technical report, and whose work falls on cases with relevant financial volume, this time reflects directly on the cadence of contracting, payments, and administrative decisions of the Department.

\begin{longtable}[]{@{}
  >{\centering\arraybackslash}p{(\linewidth - 2\tabcolsep) * \real{0.5000}}
  >{\centering\arraybackslash}p{(\linewidth - 2\tabcolsep) * \real{0.5000}}@{}}
\toprule\noalign{}
\begin{minipage}[b]{\linewidth}\centering
Indicator
\end{minipage} & \begin{minipage}[b]{\linewidth}\centering
2024 (without AI)
\end{minipage} \\
\midrule\noalign{}
\endhead
\bottomrule\noalign{}
\endlastfoot
Average processing time & 34 days \\
Cases processed & 256 \\
Documents produced & 251 \\
Technical reports issued & 66 \\
\end{longtable}

The natural question, and the one that guided the arrival of the method throughout 2025, was whether the method of AI application could, at UCI, simultaneously increase the volume and depth of technical reports without compromising response time or the rigor demanded by the nature of the work. The answer began with the presentation of the EGOV-DF official course to the Unit's team, in the same format in which the public servants of the Sectoral Internal Control Office of the Department of Health had been trained, and continues with the methodological adaptations described in Section IX.

\section{IX. Adaptations of the method at UCI/SEDET}\label{ix.-adaptations-of-the-method-at-ucisedet}

The presentation of the method to the UCI/SEDET-DF team followed the same institutional path adopted at the Sectoral Internal Control Office of the Department of Health: public servants enrolled in the official Artificial Intelligence course of the Federal District School of Government, taught by the author, with the same program content. The continuity of the training source is a relevant record because it eliminates, for the purposes of the methodological comparison guiding this paper, the variable ``quality of training'' as a competing explanation for the results. The technical training offered was the same; what changed were the agency, the subject matter, and the team.

The transition from classes to actual operation, however, required specific adaptations. UCI does not work with disciplinary cases; it works with oversight of financial execution, contract analysis, and program auditing. The unit's central document, the technical report, has structure, complexity, and pace distinct from the Decision drafts of the Sectoral Internal Control Office of the Department of Health. The adaptations were distributed across three fronts.

The first front was the curation of applications in which AI would be used. The adoption strategy prioritized cases of high impact and low risk: manual and repetitive activities that historically consumed significant time from the technical team but that did not involve automated critical decision-making. The goal was to release the human workforce for higher-value work (critical and technical analysis, adjudications, and decision-making), leaving for AI the processing of information and the preliminary structuring of documents. The three applications that consolidated as routine were assistance in drafting Technical Reports, synthesis of long documents, and grammatical review and standardization of texts.

Starting in January 2025, the internal operation was reinforced with the construction of custom AIs proprietary to UCI, initially on the Gemini platform (Gems) and later migrated to Claude (Projects). The strategy was modular: for each recurring subject in cases analyzed by the unit (Funding Agreements under Law No.~13.019/2014 (MROSC), Public Procurement under Law No.~14.133/2021, Special Accounts Cases (Tomada de Contas Especial, TCE), Prior Year Expenditures (Despesas de Exercícios Anteriores, DEA), among others), a specific custom AI was created, with a knowledge base composed of the applicable federal legislation, regulatory decrees, and control norms. For the Public Procurement custom AI, for example, the base brings together Law No.~14.133/2021, Federal District Decree No.~32.598/2010, Federal District Decree No.~44.330/2023, Federal District Decree No.~45.933/2024, and Ordinance No.~29/2021 of the Federal District Comptroller General. The tooling enabled the drafting of technical reports, verification of adherence to applicable legislation, and standardization of the structure of opinions, always within the anonymization protocols formalized in UCI's AI Governance Framework.

The second front was the formal documentation of AI use in the unit. A specific AI Governance Framework was prepared for UCI (SEI Doc. 194251158), with operational guidelines on what can and cannot be submitted to AI tools, when human review is mandatory, and which data-anonymization protocols must be observed before any interaction. The Framework translates general governance principles into operational rules adherent to general internal-control matters, to the Data Protection Law, and to the principles of Public Administration.

The third front was incremental adoption. The introduction of AI into the routine did not happen in a single block but in waves: it began with the simplest and most stable tasks, advanced to activities of greater complexity as the team gained confidence and mastery, and consolidated internal standards as each good result became a reference for the next application. Four lessons guided the operation throughout 2025 and merit explicit mention, because they distinguish a mature method from enthusiastic adoption. The first is that the quality of the result depends directly on the quality of the command: well-structured prompts, with adequate context and clear instructions, accompanied by example technical reports, produce significantly better results. The second is that AI accelerates but does not replace the need for technical knowledge or qualified human judgment. Critical review by a qualified professional remained indispensable to ensure accuracy and adequacy of analyses. The third is that documenting best practices creates multipliers: building custom AIs and standardizing the most effective prompts allowed the entire team to benefit from individual discoveries. The fourth is that initial resistance, common to any process change, dissolved faster with practical results than with theoretical arguments.

With the course delivered, the Governance Framework consolidated, and the operational routine adapted, the unit operated under the method regime throughout 2025. What happened to the indicators compared to the previous year is the object of Section X.

\section{X. Results: UCI/SEDET (2024→2025)}\label{x.-results-ucisedet-20242025}

The year 2025 was UCI/SEDET's first year operating within the method. The numbers below are extracted from the official SEI-GDF statistics and consolidated in Report No.~1/2026 of SEDET/GAB/UCI (SEI Doc. 193384267, case 04035-00000853/2026-24), signed by the Unit Head, as well as from all the Documents and Technical Reports issued in 2025.

The 2024 → 2025 comparison brings substantial and simultaneous gains across all aggregate indicators. Average processing time fell from 34 days to 17 days (−50\%). Processing capacity rose from 256 to 336 cases processed (+31\%). Total documentary output grew from 251 to 419 documents (+67\%). The output of the densest piece, the technical report, rose from 66 to 122 (+85\%), per the total consolidated in the signed 2025 management report. Read month by month (Figure 14), this output stayed at a low baseline across 2023 and 2024 and stepped up in 2025, the first year under the method.

\begin{figure}[H]
\centering
\includegraphics[width=0.80\linewidth,height=7cm,keepaspectratio]{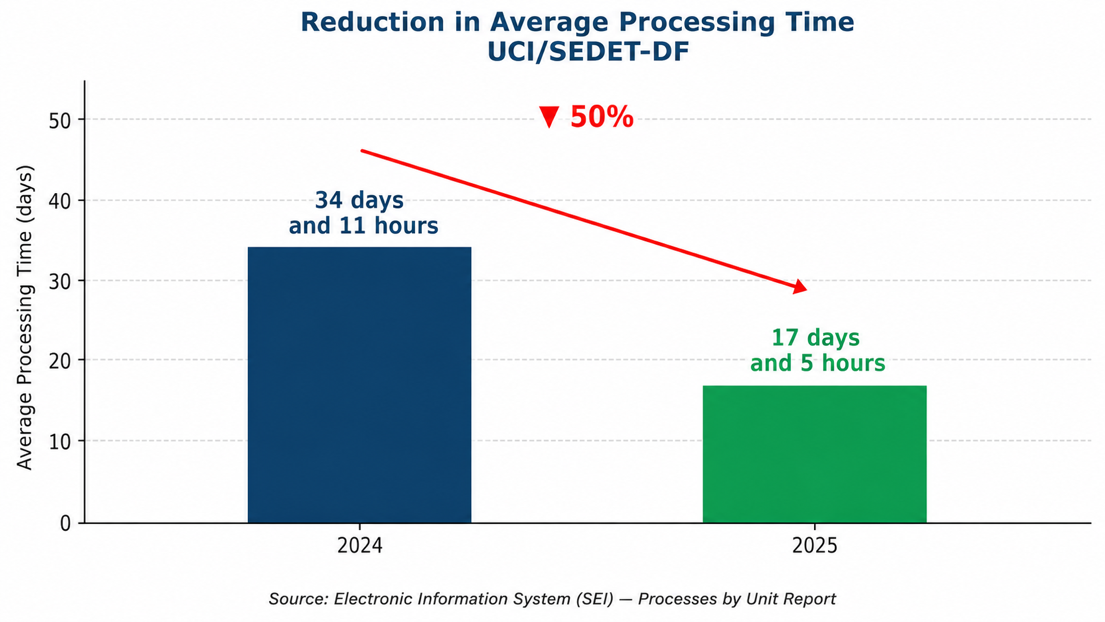}
\par\smallskip
\textit{Figure 10: Reduction in average processing time at UCI/SEDET (2024 vs.~2025).}
\end{figure}

\begin{figure}[H]
\centering
\includegraphics[width=0.80\linewidth,height=7cm,keepaspectratio]{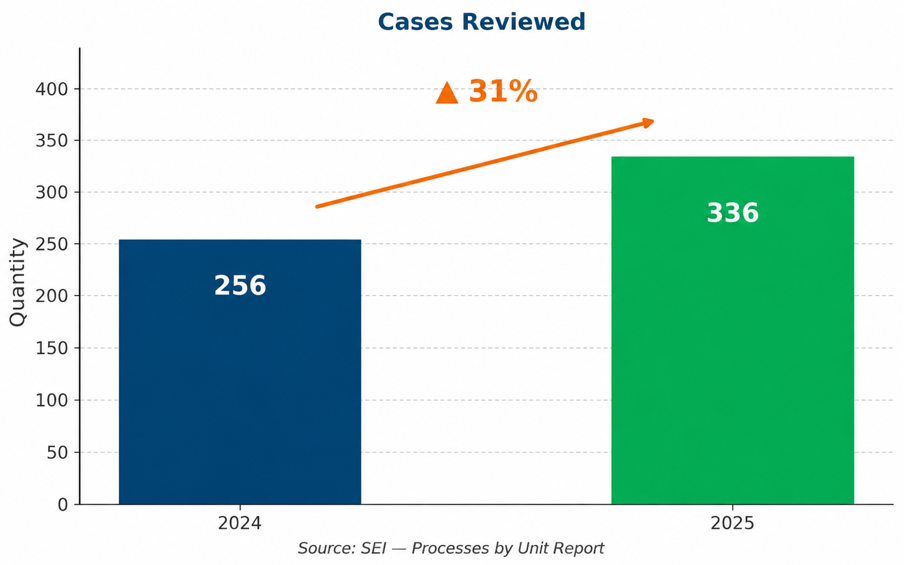}
\par\smallskip
\textit{Figure 11: Cases processed at UCI/SEDET (2024 vs.~2025).}
\end{figure}

\begin{figure}[H]
\centering
\includegraphics[width=0.80\linewidth,height=7cm,keepaspectratio]{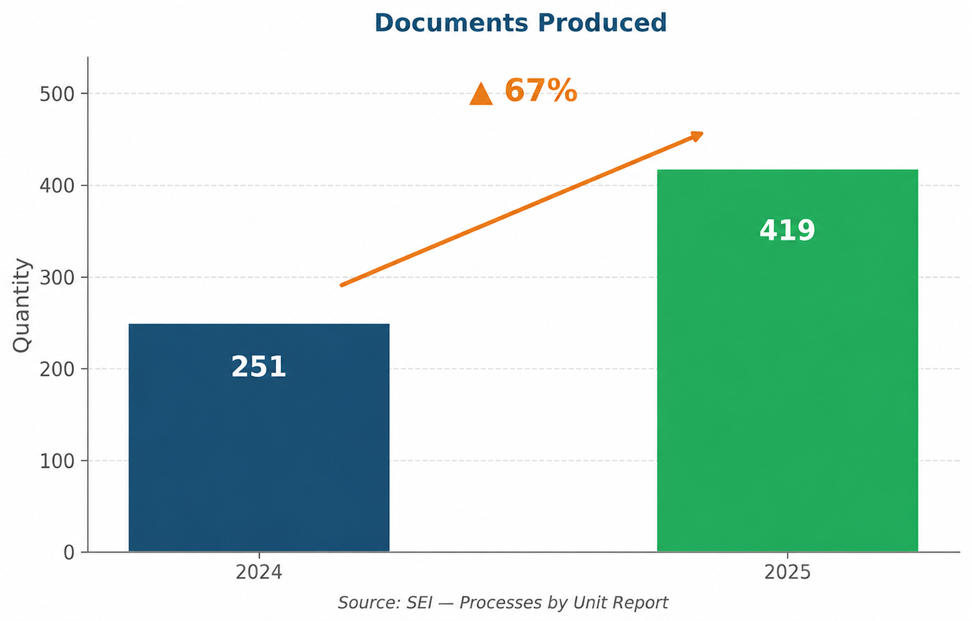}
\par\smallskip
\textit{Figure 12: Documents produced at UCI/SEDET (2024 vs.~2025).}
\end{figure}

\begin{figure}[H]
\centering
\includegraphics[width=0.80\linewidth,height=7cm,keepaspectratio]{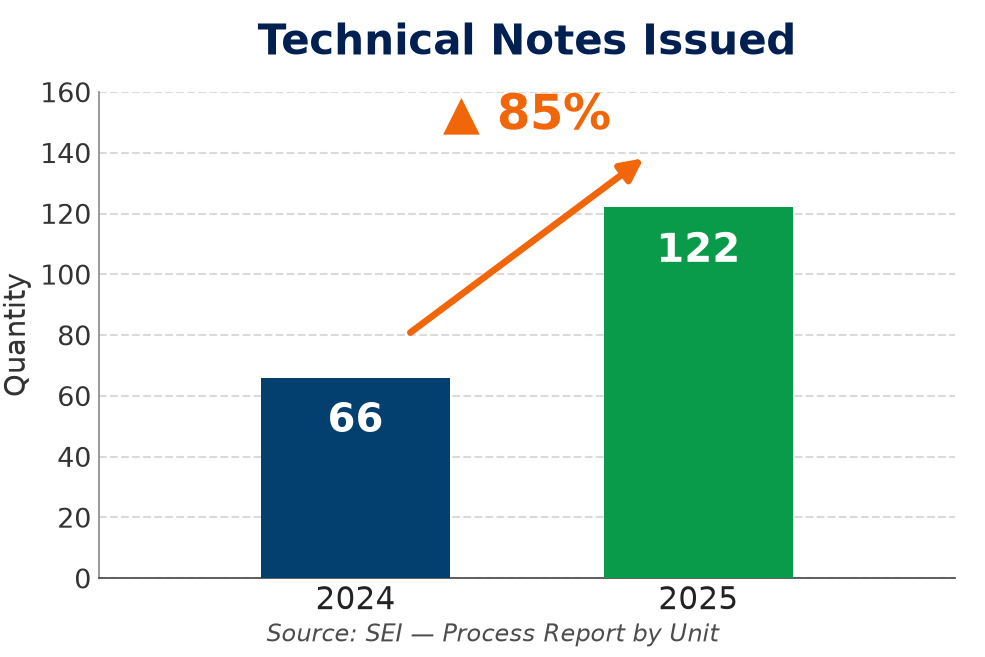}
\par\smallskip
\textit{Figure 13: Technical reports issued at UCI/SEDET: 66 in 2024 and 122 in 2025, an increase of approximately 85\%. Source: the 2024 figure is the raw SEI-GDF document statistics; the 2025 figure is the total consolidated in the signed UCI/SEDET 2025 management report (see the reconciliation note in Appendix A.0).}
\end{figure}

\begin{figure}[H]
\centering
\includegraphics[width=0.80\linewidth,height=7cm,keepaspectratio]{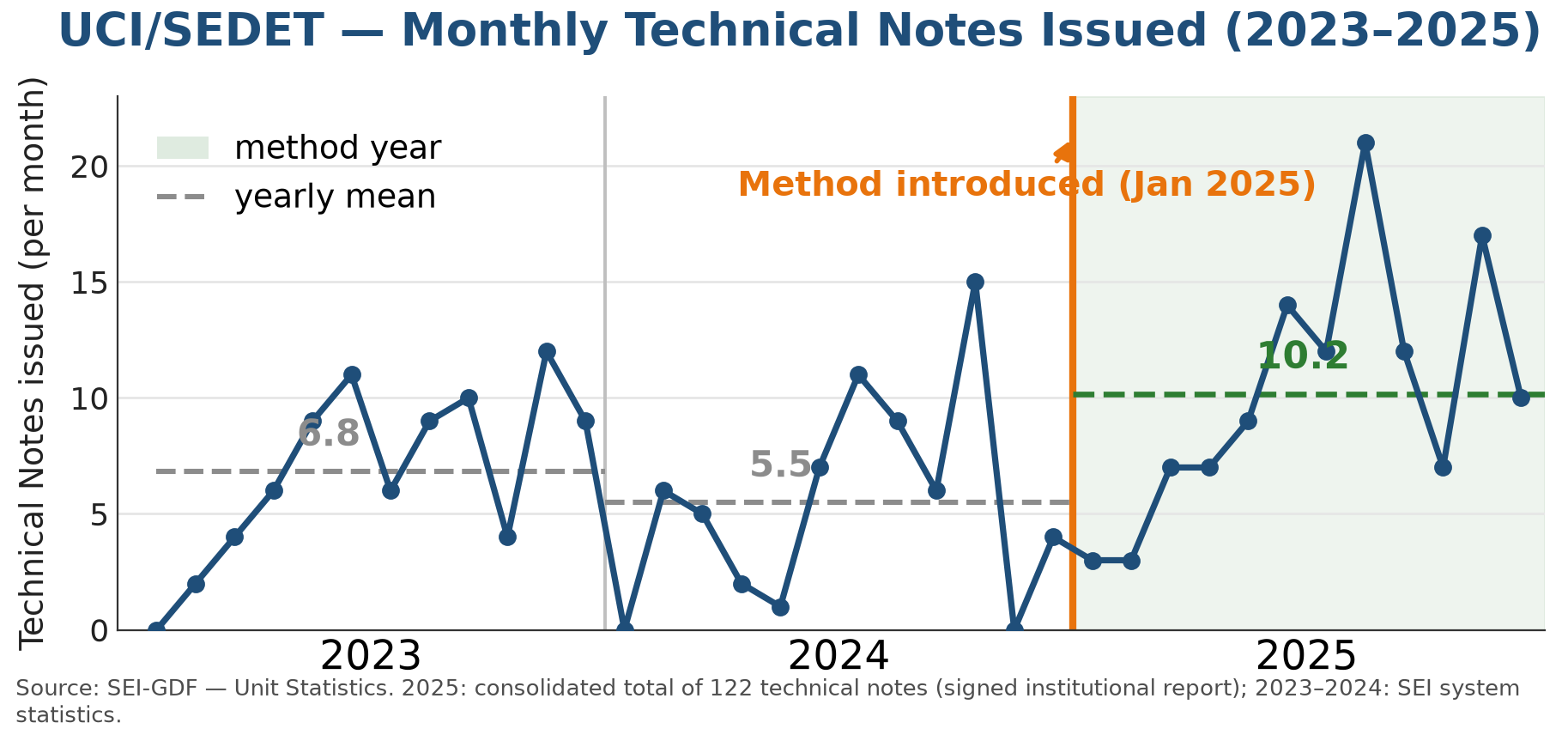}
\par\smallskip
\textit{Figure 14: Monthly technical reports issued at UCI/SEDET (2023--2025). The yearly mean rises from 6.8 (2023) and 5.5 (2024) to 10.2 in 2025, the first year under the method; the step-up is concentrated in the method year rather than across the two baseline years. Source: the 2023--2024 monthly figures are the raw SEI-GDF document statistics; the 2025 monthly series is the total consolidated in the signed UCI/SEDET 2025 management report, summing to 122 (see the reconciliation note in Appendix A.0).}
\end{figure}

\begin{longtable}[]{@{}
  >{\centering\arraybackslash}p{(\linewidth - 6\tabcolsep) * \real{0.3560}}
  >{\centering\arraybackslash}p{(\linewidth - 6\tabcolsep) * \real{0.2034}}
  >{\centering\arraybackslash}p{(\linewidth - 6\tabcolsep) * \real{0.2372}}
  >{\centering\arraybackslash}p{(\linewidth - 6\tabcolsep) * \real{0.2034}}@{}}
\toprule\noalign{}
\begin{minipage}[b]{\linewidth}\centering
Indicator
\end{minipage} & \begin{minipage}[b]{\linewidth}\centering
2024
\end{minipage} & \begin{minipage}[b]{\linewidth}\centering
2025 (with AI)
\end{minipage} & \begin{minipage}[b]{\linewidth}\centering
Variation
\end{minipage} \\
\midrule\noalign{}
\endhead
\bottomrule\noalign{}
\endlastfoot
Average processing time & 34 days & 17 days & \textbf{−50\%} \\
Cases processed & 256 & 336 & \textbf{+31\%} \\
Documents produced & 251 & 419 & \textbf{+67\%} \\
Technical reports issued & 66 & 122 & \textbf{+85\%} \\
Dispatches produced & 102 & 163 & \textbf{+60\%} \\
\end{longtable}

At the Internal Control Office, the reduction in time came with stable documentary output. At UCI, the reduction in time came accompanied by a simultaneous increase in documentary volume and in analytical intensity. Section XI takes up the difference in comparative perspective.

Analytical intensity deserves separate mention. The 122 technical reports consolidated in the signed 2025 management report resulted in the issuance of 286 formal recommendations to public managers. This figure is important because it means that the increase in output was not only quantitative: the unit produced more pieces and deepened the analysis within each piece, identifying more improvement points, more potential risks, and more opportunities for management enhancement per case examined.

The item-level review of the recommendations identified through the full text of each of the 122 technical reports consolidated in the signed 2025 management report, classified by the authors by nature and materiality (Appendix A.4), produced a modeled gross value at risk of US\$ 43.7 million. Applying a probability matrix calibrated with reference to international public-audit literature, the aggregate potential-mitigation estimate ranges from US\$ 1.1 million (conservative scenario) to US\$ 5.2 million (optimistic scenario), with a central estimate of US\$ 2.8 million: a modeled estimate, not realized or audited savings. The methodological detail, the classification criteria, and the probability matrix are described in Appendix A.4.

In financial terms, the 122 technical reports consolidated in 2025 concern cases and matters whose total value, per the UCI/SEDET official statistics signed in the 2025 management report (SEI Doc. 193436555), was US\$ 94.8 million: the value of the matters submitted to technical analysis. The temporal distribution of the amount was heterogeneous, with months of low volume and months of high load. Table A.3.2 of the appendix organizes the value analyzed in each month.

August deserves emphasis. It was the year's highest month both in the number of cases analyzed (21) and in financial volume: US\$ 19.1 million analyzed in a single month. Per the institutional report, the unit absorbed this peak without generating bottlenecks or delays and without compromising the depth or technical quality of the analyses produced.

The gain in efficiency was not restricted to one type of case. UCI itself records, in the official report, that the operational improvement was observed across all activities performed by the unit, from routine analysis to cases of greater technical and financial complexity. There was no trade-off of quality for speed in any segment of the work.

The 2025 balance of UCI/SEDET can be synthesized in four SEI-GDF-verifiable statements: the unit doubled the average processing speed; increased the volume of processed cases by 31\%; increased technical-note output by 85\% and dispatches (despachos, routine administrative orders that move a case forward in Brazilian administrative practice) by 60\%; and analyzed US\$ 94.8 million in financial volume, with a peak of US\$ 19.1 million absorbed without generating bottlenecks or delays. The Human-in-the-Loop methodology was maintained as a non-negotiable pillar, with express prohibition on issuing any document without integral, thorough technical human review, operationalized through a three-step Triple Review (drafting professional, direct supervisor, and signing authority).

Section XI discusses what the two cases, read together, teach about the method.

\section{XI. Integrated discussion}\label{xi.-integrated-discussion}

Read together, Sections VI and X bring an auditable set of evidence about the gains observed when the method was methodically installed in a public unit. The two cases are not identical repetitions. The Sectoral Internal Control Office of the Department of Health and the Internal Control Unit of SEDET-DF operate in distinct Departments, with distinct substantive mandates, distinct subject matters, and distinct teams. What remained constant between the two were the method, the instructor, the course through which the team was trained (the official EGOV-DF course), the exclusive use of free-tier versions of commercial AI platforms, and the non-negotiable principle of integral human supervision. The method and the institutional structure of training were constant; the agency, the subject matter, and the team varied. It is this configuration that allows the two sets of numbers to be read as convergent evidence.

The table below condenses, side by side, the main indicators of the two cases, to support the comparative reading in the following sections.

\begin{longtable}[]{@{}
  >{\centering\arraybackslash}p{(\linewidth - 4\tabcolsep) * \real{0.3243}}
  >{\centering\arraybackslash}p{(\linewidth - 4\tabcolsep) * \real{0.3537}}
  >{\centering\arraybackslash}p{(\linewidth - 4\tabcolsep) * \real{0.3220}}@{}}
\toprule\noalign{}
\begin{minipage}[b]{\linewidth}\centering
\textbf{Dimension}
\end{minipage} & \begin{minipage}[b]{\linewidth}\centering
\textbf{SES/CONT (2023 → 2024)}
\end{minipage} & \begin{minipage}[b]{\linewidth}\centering
\textbf{UCI/SEDET (2024 → 2025)}
\end{minipage} \\
\midrule\noalign{}
\endhead
\bottomrule\noalign{}
\endlastfoot
\textbf{Average processing time} & 17.92 → 14.66 days \textbf{(−18.2\%)} & 34 → 17 days \textbf{(−50.0\%)} \\
\textbf{Estimated released capacity} & +22.3\% \textbf{(≈352 additional cases)} & +31\% \textbf{(256 → 336 cases)} \\
\textbf{Monthly Production Stability Across the Unit's Most Important Documents} & CV reduced in 3 of 4 types (Decision, Adjudication, Official Letter) & CV reduced in both types (Dispatch, Technical Report) \\
\textbf{Dense analytical piece: monthly output} & Adjudications: 31.0/month vs. 30.4/month (+2.1\%, ex-atypical month) & Technical Reports: 66 → 122 (+84.8\%) \\
\textbf{Preventive function} & Advisory Notices: 28 → 214 (+664\%) & 286 formal recommendations issued \\
\textbf{Institutional articulation} & Official Letters: 434 → 532 (+22.6\%) & n/a (general internal control profile) \\
\textbf{Other types with greatest time improvement} & Decision Enforcement: −95.5\%; Information Request: −93.9\%; Judicial Action: −88.3\% & Improvement across all activities \\
\textbf{AI-related problems / controls documented in the institutional report} & None reported & Not addressed as an incident count; mandatory human review before issuance (Human-in-the-Loop, Triple Review) \\
\end{longtable}

The most visible convergence, and the one that most directly supports the paper's central argument, is the magnitude of the time gain. The Internal Control Office reduced average processing time by 18.2\%; UCI reduced it by 50\%. The percentages differ, but the direction is the same, in two agencies with opposite operational profiles. This convergence is not readily explained by the most obvious alternative hypotheses, which we consider explicitly below. The observed gains are temporally coincident with the introduction of the method and consistent with the hypothesis of a causal relationship: in both cases, the team remained the same between the base year and the AI year, documentary output per case did not decrease (stable at the Internal Control Office, growing at UCI), and the normative base of the two units was not altered.

Two such alternatives deserve to be named explicitly, because both accompanied the method and neither can be fully separated from it in an observational design. The first is a change of leadership: the author assumed direction of each unit at the start of its cycle, so new management coincided with the introduction of the method. The second is the reorganization of document workflows that took place as the method was adopted. We do not claim to isolate the method from these co-occurring factors.

Three features of the two-case design constrain these alternatives. The replication occurred on a different unit, a different subject matter, and a different baseline, a pattern hard to reconcile with a one-off artifact such as reversion from an atypical year. Documentary output per case held stable or grew, which is inconsistent with a speed-for-quality trade-off. And the gains concentrated in exactly the document types the method targets, rather than spreading uniformly as a generic, tool-driven improvement would. The evidence remains observational; these features narrow rather than eliminate alternative explanations, but together they make the training-based account the most parsimonious reading of the two cases.

The most visible divergence between the two cases lies in the output profile. At the Internal Control Office, there was a reduction in time with documentary output per case practically stable (2.77 to 2.78 D+M+A per case) after adoption of the method, and the released capacity was converted into clearance of the accumulated backlog, stabilization of monthly output, and expansion of the preventive function (Advisory Notices +664\%). At UCI, the reduction in time came accompanied by a parallel increase in total documentary volume (+67\%) and in the densest analytical piece (Technical Reports +85\%, with 286 formal recommendations issued). The difference is institutional. The Internal Control Office operated at the limit of capacity, with backlog, and the post-adoption indicators are consistent with the release of that pent-up capacity. UCI operated with technical capacity underused due to lack of time for the dense analytical piece, and the post-adoption indicators are consistent with the expression of that capacity in more pieces and deeper analysis. The gain observed, in each case, is consistent with the performance constraint of the starting point. There is, moreover, a qualitative gain common to both cases: both units came to operate simultaneously in repressive (ex post control) and preventive and corrective registers. The Advisory Notices issued by SES/CONT (+664\%) and the 286 formal recommendations of UCI/SEDET are the documented manifestation of this turn, in which internal control begins to signal to managers the flows and practices that originated the irregularities so they can be remedied before the next incident.

The month of August, in each of the two cycles, illustrates the structural change particularly clearly. In August 2023, the Internal Control Office absorbed an extraordinary peak of 69 Adjudications by means of an emergency task force: public servants working beyond normal capacity, part of the output produced on non-working days, real risk of losing cases to the statute of limitations. In August 2025, UCI absorbed the year's peak in financial volume (US\$ 19.1 million in a single month), and the institutional report states this occurred without generating bottlenecks or delays and without compromising the depth of technical analysis. Same calendar; different operational geometry.

What the method does not do also deserves recording, based on the lessons consolidated in the UCI Report. The method does not replace technical knowledge: AI accelerates output, but critical review by a qualified professional remained indispensable in both cases to ensure accuracy and adequacy. The method does not waive the effort of commanding it: well-structured prompts, with adequate context, clear instructions, and pertinent examples, produce significantly better results than generic prompts, and that difference does not resolve itself; it requires training and practice. The method is not instantaneous: adoption was incremental, from simple to complex tasks, and initial resistance yielded faster to practical results than to theoretical arguments. And the method does not operate in a vacuum: each unit required a formally documented protection layer (the custom AI with legal base at the Internal Control Office, the specific AI Governance Framework at UCI), adherent to the matter that unit handles.

There is also the result the two cases produced jointly: no incident identified. In both annual cycles, the internal control mechanisms of the two units did not identify sensitive-data leakage, administrative decisions voided by failures stemming from AI use, or formal questioning by external oversight bodies regarding the method's compliance with the Data Protection Law or with the principles of Public Administration. The reading consistent with the evidence is that this result followed from concern with governance being built from the outset, with the foundation built explaining the reasons why AI can be wrong, for example, and from integral human supervision being treated as a functioning condition, not as an additional measure.

The stability of output, mentioned qualitatively in Sections VI and X, is quantifiable by the coefficient of variation (CV) of monthly output by document type, as shown in Tables A.2.4 and A.3.4 of the appendix. The table below summarizes the CV variation between the base year and the year of method adoption in the two units for the most important documents:

\begin{longtable}[]{@{}
  >{\raggedright\arraybackslash}p{(\linewidth - 12\tabcolsep) * \real{0.1693}}
  >{\raggedright\arraybackslash}p{(\linewidth - 12\tabcolsep) * \real{0.1739}}
  >{\raggedright\arraybackslash}p{(\linewidth - 12\tabcolsep) * \real{0.1224}}
  >{\raggedright\arraybackslash}p{(\linewidth - 12\tabcolsep) * \real{0.1588}}
  >{\raggedright\arraybackslash}p{(\linewidth - 12\tabcolsep) * \real{0.1197}}
  >{\raggedright\arraybackslash}p{(\linewidth - 12\tabcolsep) * \real{0.1409}}
  >{\raggedright\arraybackslash}p{(\linewidth - 12\tabcolsep) * \real{0.1150}}@{}}
\toprule\noalign{}
\begin{minipage}[b]{\linewidth}\raggedright
Unit
\end{minipage} & \begin{minipage}[b]{\linewidth}\raggedright
Type
\end{minipage} & \begin{minipage}[b]{\linewidth}\raggedright
Base year
\end{minipage} & \begin{minipage}[b]{\linewidth}\raggedright
CV base year
\end{minipage} & \begin{minipage}[b]{\linewidth}\raggedright
AI year
\end{minipage} & \begin{minipage}[b]{\linewidth}\raggedright
CV AI year
\end{minipage} & \begin{minipage}[b]{\linewidth}\raggedright
Variation
\end{minipage} \\
\midrule\noalign{}
\endhead
\bottomrule\noalign{}
\endlastfoot
SES/CONT & Decision & 2023 & 39.8\% & 2024 & 17.3\% & \textbf{−56.5\%} \\
SES/CONT & Adjudication & 2023 & 52.4\% & 2024 & 26.5\% & −49.4\% \\
SES/CONT & Official Letter & 2023 & 19.8\% & 2024 & 13.6\% & −31.3\% \\
UCI/SEDET & Dispatch & 2024 & 73.3\% & 2025 & 49.3\% & −32.7\% \\
UCI/SEDET & Technical Report & 2024 & 83.3\% & 2025 (consolidated) & 53.0\% & \textbf{−36.4\%} \\
\end{longtable}

\section{XII. Conclusions}\label{xii.-conclusions}

The synthesis, therefore, has four verifiable fronts. The first concerns portability: efficiency gains accompanied the method in two agencies with distinct institutional profiles, distinct teams, and distinct subject matters, without structural reformulation between the two cases. The second is that the method operated, within the scope of the study, without internal control mechanisms identifying any incident, with governance documented formally in both units. The third is that the method is accessible: results were obtained with free-tier versions of commercial platforms, which opens the possibility of replication to other public entities with budget constraints. The fourth concerns the method's potential financial effect, which we model rather than measure directly: under a probability matrix built on explicit, literature-anchored assumptions, the item-level review of the recommendations identified across the 122 technical reports consolidated in the signed 2025 management report, classified by the authors for this exercise (Appendix A.4), places potential mitigation in a range of US\$ 1.1 million (conservative) to US\$ 5.2 million (optimistic), with a central estimate of US\$ 2.8 million. This is a modeled range, not realized or audited savings, and is separate from the US\$ 94.8 million in financial volume the unit's signed statistics record as submitted to technical analysis.

\textbf{APPENDIX}

\section{Appendix A: Documentary References and Tables of Official Data}\label{appendix-a-documentary-references-and-tables-of-official-data}

The data and analyses presented in this paper are anchored in institutional documents of the Electronic Information System of the Federal District Government (SEI-GDF), all identified by unique document number and corresponding case, and therefore verifiable by third parties. This appendix consolidates the documentary references used and replicates, in auditable format, the tables of official data that support the main text.

\subsection{A.0 Methodological note on data extraction}\label{a.0-methodological-note-on-data-extraction}

The quantitative data presented in this paper were extracted from the Electronic Information System of the Federal District Government (SEI-GDF), the official case-processing system of the Federal District Public Administration.

Extraction was performed via the native ``Estatísticas da Unidade'' functionality, with filtering by issuing unit (SES/CONT or UCI/SEDET) and by full calendar year. The functionality generates, in PDF format, three blocks of information per year: (i) cases generated in the period, by type; (ii) documents generated in the period, by type (Decision, Memorandum, Adjudication, Official Letter, Advisory Notice, Technical Report, and other types available to the unit's profile); (iii) average processing times by case type. The average-time indicator corresponds to gross time, in continuous hours, between receipt of the case at the unit and the conclusion of the case that closes the processing stage in that unit, according to the SEI-GDF's internal methodology.

The comparison between the two comparative cycles (SES/CONT 2023×2024; UCI/SEDET 2024×2025) was made over extractions performed with the same procedure, with no known changes in the system's calculation methodology between the years compared. The issuing unit of the reports is the same in each pair of years, preserving the comparison base. No additional filters were applied to the extracted PDFs.

The statistical post-processing of the monthly-output tables (means, medians, standard deviations, quartiles, IQR, and coefficients of variation reported in Tables A.2.4 and A.3.4) was performed by a deterministic pipeline maintained by the author, with cross-audit confirming, for all reported series, the identity between the sum of the 12 monthly extracted values and the total declared in the SEI PDFs. The pipeline is available upon request to the author.

\textbf{Reconciliation of the UCI/SEDET 2025 technical-note count (122 versus 127).} The official spreadsheet on page 48 of the signed UCI/SEDET 2025 management report (``Estatística Notas Técnicas Unidade de Controle Interno, UCI/SEDET, 2025,'' SEI Doc. 193436555) consolidates technical notes that analyze procurement, contract-payment, or partnership processes, typified as CM, PM, IN, CL, or PL, and reports a total of 122 such notes for 2025, with a combined value of US\$ 94.8 million. A separate, raw query of the SEI-GDF ``Estatísticas da Unidade'' functionality (p.~59 of the same report) identified 127 documents labeled ``Nota Técnica'' issued by the unit in 2025. The signed 2025 UCI management report consolidated 122 technical notes, compared with 66 in 2024, an increase of approximately 85\%. We reconciled the two universes item by item and use the report's consolidated total of 122 as the official performance measure. The five-document difference reflects technical notes that were not analyses of procurement, contract-payment, or partnership processes: one follow-up note on the implementation of 2024 recommendations, one review of a draft normative instrument, one response to an external correctional inquiry, and two special accounts proceedings (\emph{tomadas de contas especial}), as identified in the item-level reconciliation. The same report recorded 286 formal recommendations (p.~5). Accordingly, this paper uses 122 technical notes and 286 recommendations throughout, while retaining the raw SEI count of 127 only to document the reconciliation procedure.

Currency conversion: All monetary values in this English version are approximate conversions of the official figures in Brazilian reais (official total of the analyzed universe: R\$ 521,358,798.09) at the Central Bank of Brazil PTAX selling rate of 31 December 2025 (R\$ 5.5024 per US\$ 1.00).

\subsection{A.1 SEI-GDF documents cited in the paper}\label{a.1-sei-gdf-documents-cited-in-the-paper}

\subsubsection{A.1.1 Sectoral Internal Control Office of the Department of Health (SES/CONT)}\label{a.1.1-sectoral-internal-control-office-of-the-department-of-health-sescont}

SEI Doc. 146621014: Memorandum No.~4/2024, SES/CONT/ASJULG

\begin{longtable}[]{@{}
  >{\raggedright\arraybackslash}p{(\linewidth - 2\tabcolsep) * \real{0.5000}}
  >{\raggedright\arraybackslash}p{(\linewidth - 2\tabcolsep) * \real{0.5000}}@{}}
\toprule\noalign{}
\begin{minipage}[b]{\linewidth}\raggedright
Field
\end{minipage} & \begin{minipage}[b]{\linewidth}\raggedright
Content
\end{minipage} \\
\midrule\noalign{}
\endhead
\bottomrule\noalign{}
\endlastfoot
SEI case & 00060-00314546/2024-15 \\
Date & June 25, 2024 \\
Subject & Proposal for the creation of the course ``Artificial Intelligence for Public Servants'' \\
\end{longtable}

SEI Doc. 144321884: Course Lesson Plan

\begin{longtable}[]{@{}
  >{\raggedright\arraybackslash}p{(\linewidth - 2\tabcolsep) * \real{0.5000}}
  >{\raggedright\arraybackslash}p{(\linewidth - 2\tabcolsep) * \real{0.5000}}@{}}
\toprule\noalign{}
\begin{minipage}[b]{\linewidth}\raggedright
Field
\end{minipage} & \begin{minipage}[b]{\linewidth}\raggedright
Content
\end{minipage} \\
\midrule\noalign{}
\endhead
\bottomrule\noalign{}
\endlastfoot
SEI case & 00060-00314546/2024-15 \\
Content & Complete program structure of the five in-person classes of the course Artificial Intelligence in the public sector: techniques, risks, and applications (20 hours) \\
\end{longtable}

SEI Doc. 197403428: Executive Management Report: Impact of Artificial Intelligence on Operational Efficiency

\begin{longtable}[]{@{}
  >{\raggedright\arraybackslash}p{(\linewidth - 2\tabcolsep) * \real{0.5000}}
  >{\raggedright\arraybackslash}p{(\linewidth - 2\tabcolsep) * \real{0.5000}}@{}}
\toprule\noalign{}
\begin{minipage}[b]{\linewidth}\raggedright
Field
\end{minipage} & \begin{minipage}[b]{\linewidth}\raggedright
Content
\end{minipage} \\
\midrule\noalign{}
\endhead
\bottomrule\noalign{}
\endlastfoot
SEI case & 00060-00582291/2024-11 \\
Date & January 2, 2025 \\
Issuing unit & Sectoral Internal Control Office of the Department of Health (SES/CONT) \\
Content & Comparative analysis 2023 vs.~2024 of the unit's operational indicators after implementation of the method \\
\end{longtable}

Report No.~12/2024: SES/CONT/ASJULG

\begin{longtable}[]{@{}
  >{\raggedright\arraybackslash}p{(\linewidth - 2\tabcolsep) * \real{0.5000}}
  >{\raggedright\arraybackslash}p{(\linewidth - 2\tabcolsep) * \real{0.5000}}@{}}
\toprule\noalign{}
\begin{minipage}[b]{\linewidth}\raggedright
Field
\end{minipage} & \begin{minipage}[b]{\linewidth}\raggedright
Content
\end{minipage} \\
\midrule\noalign{}
\endhead
\bottomrule\noalign{}
\endlastfoot
Date & December 14, 2024 \\
Content & Institutional record that the unit was not facing case backlog, with a maximum 10-day deadline for drafting decisions \\
\end{longtable}

\subsubsection{A.1.2 Internal Control Unit (UCI/SEDET-DF)}\label{a.1.2-internal-control-unit-ucisedet-df}

SEI Doc. 194251158: AI Governance Framework, UCI

\begin{longtable}[]{@{}
  >{\raggedright\arraybackslash}p{(\linewidth - 2\tabcolsep) * \real{0.5000}}
  >{\raggedright\arraybackslash}p{(\linewidth - 2\tabcolsep) * \real{0.5000}}@{}}
\toprule\noalign{}
\begin{minipage}[b]{\linewidth}\raggedright
Field
\end{minipage} & \begin{minipage}[b]{\linewidth}\raggedright
Content
\end{minipage} \\
\midrule\noalign{}
\endhead
\bottomrule\noalign{}
\endlastfoot
Issuing unit & Internal Control Unit, SEDET-DF \\
Content & Institutional document with operational guidelines for the use of AI tools at UCI \\
\end{longtable}

SEI Doc. 193384267, Report No.~1/2026: Management Results Report 2025, UCI/GAB/SEDET

\begin{longtable}[]{@{}
  >{\raggedright\arraybackslash}p{(\linewidth - 2\tabcolsep) * \real{0.5000}}
  >{\raggedright\arraybackslash}p{(\linewidth - 2\tabcolsep) * \real{0.5000}}@{}}
\toprule\noalign{}
\begin{minipage}[b]{\linewidth}\raggedright
Field
\end{minipage} & \begin{minipage}[b]{\linewidth}\raggedright
Content
\end{minipage} \\
\midrule\noalign{}
\endhead
\bottomrule\noalign{}
\endlastfoot
SEI case & 04035-00000853/2026-24 \\
Date & January 28, 2026 \\
Issuing unit & Internal Control Unit, SEDET-DF \\
Content & Comparative analysis 2024 vs.~2025 of the unit's operational indicators after implementation of the method \\
\end{longtable}

\subsubsection{\texorpdfstring{ A.1.3 Federal District School of Government (EGOV-DF)}{ A.1.3 Federal District School of Government (EGOV-DF)}}\label{a.1.3-federal-district-school-of-government-egov-df}

\textbf{SEI Doc. 146526272: Memorandum No.~166/2024, SEEC/SEGEA/EGOV}

\begin{longtable}[]{@{}
  >{\raggedright\arraybackslash}p{(\linewidth - 2\tabcolsep) * \real{0.5000}}
  >{\raggedright\arraybackslash}p{(\linewidth - 2\tabcolsep) * \real{0.5000}}@{}}
\toprule\noalign{}
\begin{minipage}[b]{\linewidth}\raggedright
Field
\end{minipage} & \begin{minipage}[b]{\linewidth}\raggedright
Content
\end{minipage} \\
\midrule\noalign{}
\endhead
\bottomrule\noalign{}
\endlastfoot
SEI case & 04044-00021535/2024-26 \\
Date & July 22, 2024 \\
Subject & Institutional approval of the course Artificial Intelligence in the public sector: techniques, risks, and applications, with funding, through internal instruction, in in-person mode, directed at public agents of the agencies and entities of the Direct and Indirect Administration of the Federal District \\
\end{longtable}

\subsubsection{A.1.4 Normative base of the SES/CONT custom AI}\label{a.1.4-normative-base-of-the-sescont-custom-ai}

The custom AI of the Sectoral Internal Control Office of the Department of Health (described in Section IV) operated on a knowledge base composed of ten legal instruments directly applicable to the unit's correctional matter. These ten instruments are not isolated in the Reference List because they are not invoked argumentatively in the body of the paper; they appear in this subsection exclusively as documentary traceability of the operational material loaded into the custom AI:

\begin{itemize}
\item
  Organic Law of the Federal District. Federal District Legislative Chamber, 1993.
\item
  Decree No.~37.297, of April 29, 2016: Code of Ethics of Public Servants and Civil Employees of the Federal District Executive Branch.
\item
  Federal District Comptroller General. Normative Instruction No.~2, of July 25, 2016: mediation procedures within the Federal District public administration.
\item
  Federal District Comptroller General. Normative Instruction No.~1, of March 12, 2021: Conduct Adjustment Term (TAC).
\item
  Federal District Comptroller General. Normative Instruction No.~2, of October 19, 2021: Admissibility Assessment and Preliminary Investigation Procedure (PIP).
\item
  Federal District Comptroller General. Theoretical Manual of Disciplinary Procedures. Brasília, DF.
\item
  Federal District Comptroller General. Practical Manual of Disciplinary Procedures. Brasília, DF.
\item
  Complementary Law No.~840, of December 23, 2011: Legal regime of the civil public servants of the Federal District, of autarchies and of public foundations of the Federal District.
\item
  Law No.~4.938, of September 19, 2012: Establishes the Federal District Correctional System (SICOR).
\item
  Decree No.~39.701, of March 7, 2019: Standards of disciplinary procedures in the Federal District.
\end{itemize}

\subsubsection{A.1.5 Normative base of the UCI/SEDET custom AI (Public Procurement)}\label{a.1.5-normative-base-of-the-ucisedet-custom-ai-public-procurement}

UCI/SEDET-DF operated, from January 2025, with modular custom AIs by subject matter (described in Section IX). For the Public Procurement custom AI, the knowledge base gathered the applicable federal and Federal District legislation. These instruments appear in this subsection as documentary traceability of the operational material loaded, in parallel to A.1.4 (normative base of the SES/CONT custom AI):

\begin{itemize}
\item
  Law No.~13.019, of July 31, 2014: Regulatory Framework for Civil Society Organizations (MROSC), Funding Agreements.
\item
  Law No.~14.133, of April 1, 2021: Public Procurement and Administrative Contracts Law.
\item
  Federal District Decree No.~32.598, of December 15, 2010: rules of planning, budget, finance, patrimony, and accounting of the Federal District.
\item
  Federal District Decree No.~44.330, of April 7, 2023: regulates Law No.~14.133/2021 within the Federal District.
\item
  Federal District Decree No.~45.933, of June 18, 2024: update of the public procurement and contracts regulation of the Federal District.
\item
  Federal District Comptroller General. Ordinance No.~29, of 2021: guidelines applicable to internal control of Federal District public procurement.
\end{itemize}

\subsection{A.2 Tables of official data: Sectoral Internal Control Office of the Department of Health (SES/CONT)}\label{a.2-tables-of-official-data-sectoral-internal-control-office-of-the-department-of-health-sescont}

The tables in this section were extracted from the Executive Management Report (SEI Doc. 197403428).

\subsubsection{A.2.1 General indicators 2023 vs.~2024}\label{a.2.1-general-indicators-2023-vs.-2024}

\begin{longtable}[]{@{}
  >{\centering\arraybackslash}p{(\linewidth - 6\tabcolsep) * \real{0.2499}}
  >{\centering\arraybackslash}p{(\linewidth - 6\tabcolsep) * \real{0.2499}}
  >{\centering\arraybackslash}p{(\linewidth - 6\tabcolsep) * \real{0.2499}}
  >{\centering\arraybackslash}p{(\linewidth - 6\tabcolsep) * \real{0.2502}}@{}}
\toprule\noalign{}
\begin{minipage}[b]{\linewidth}\centering
Indicator
\end{minipage} & \begin{minipage}[b]{\linewidth}\centering
2023
\end{minipage} & \begin{minipage}[b]{\linewidth}\centering
2024 (with AI)
\end{minipage} & \begin{minipage}[b]{\linewidth}\centering
Variation
\end{minipage} \\
\midrule\noalign{}
\endhead
\bottomrule\noalign{}
\endlastfoot
Average processing time & 17.92 days & 14.66 days & \textbf{−18.2\%} \\
Cases processed & 1,752 & 1,581 & −9.8\% \\
Total D+M+A produced & 4,846 & 4,393 & −9.3\% \\
D+M+A per case & 2.77 & 2.78 & +0.4\% \\
Memoranda per case & 1.57 & 1.65 & +5.1\% \\
\end{longtable}

\subsubsection{A.2.2 Case types with greatest time reductions (high analytical complexity)}\label{a.2.2-case-types-with-greatest-time-reductions-high-analytical-complexity}

\begin{longtable}[]{@{}
  >{\centering\arraybackslash}p{(\linewidth - 6\tabcolsep) * \real{0.2499}}
  >{\centering\arraybackslash}p{(\linewidth - 6\tabcolsep) * \real{0.2499}}
  >{\centering\arraybackslash}p{(\linewidth - 6\tabcolsep) * \real{0.2499}}
  >{\centering\arraybackslash}p{(\linewidth - 6\tabcolsep) * \real{0.2502}}@{}}
\toprule\noalign{}
\begin{minipage}[b]{\linewidth}\centering
Case type
\end{minipage} & \begin{minipage}[b]{\linewidth}\centering
2023
\end{minipage} & \begin{minipage}[b]{\linewidth}\centering
2024
\end{minipage} & \begin{minipage}[b]{\linewidth}\centering
Reduction
\end{minipage} \\
\midrule\noalign{}
\endhead
\bottomrule\noalign{}
\endlastfoot
Decision Enforcement & 17.82 days & 0.81 days & −95.5\% \\
Information Request (Internal Control) & 22.81 days & 1.38 days & −93.9\% \\
Judicial Action: Enforcement & 24.96 days & 2.92 days & −88.3\% \\
Consultations from Legal Bodies & 9.27 days & 2.82 days & −69.6\% \\
Guidelines and Directives & 15.74 days & 5.92 days & −62.4\% \\
Preliminary Investigation & 7.39 days & 3.16 days & −57.2\% \\
Consultations from Oversight Bodies & 12.96 days & 7.00 days & −46.0\% \\
\end{longtable}

\subsubsection{A.2.3 Adjudication output: normalized analysis}\label{a.2.3-adjudication-output-normalized-analysis}

\begin{longtable}[]{@{}
  >{\centering\arraybackslash}p{(\linewidth - 6\tabcolsep) * \real{0.2499}}
  >{\centering\arraybackslash}p{(\linewidth - 6\tabcolsep) * \real{0.2499}}
  >{\centering\arraybackslash}p{(\linewidth - 6\tabcolsep) * \real{0.2499}}
  >{\centering\arraybackslash}p{(\linewidth - 6\tabcolsep) * \real{0.2502}}@{}}
\toprule\noalign{}
\begin{minipage}[b]{\linewidth}\centering
Scenario
\end{minipage} & \begin{minipage}[b]{\linewidth}\centering
Total
\end{minipage} & \begin{minipage}[b]{\linewidth}\centering
Months
\end{minipage} & \begin{minipage}[b]{\linewidth}\centering
Monthly average
\end{minipage} \\
\midrule\noalign{}
\endhead
\bottomrule\noalign{}
\endlastfoot
2023 (August included, atypical month) & 403 & 12 & 33.58 \\
2023 (August excluded, atypical month)¹ & 334 & 11 & 30.36 \\
2024 (all months, with AI) & 372 & 12 & 31.00 \\
\end{longtable}

¹ August 2023 disregarded as it was atypical volume arising from an emergency task force, not representative of regular operation. See detail in Section II.

\subsubsection{A.2.4 Descriptive statistics of monthly output by document type: SES/CONT}\label{a.2.4-descriptive-statistics-of-monthly-output-by-document-type-sescont}

Base: 12 monthly values per type/year, extracted from the unit's official SEI-GDF statistics. SD = sample standard deviation; CV = coefficient of variation (SD/mean × 100); IQR = interquartile range (Q3 − Q1).

\begin{longtable}[]{@{}
  >{\raggedright\arraybackslash}p{(\linewidth - 22\tabcolsep) * \real{0.1497}}
  >{\raggedright\arraybackslash}p{(\linewidth - 22\tabcolsep) * \real{0.0711}}
  >{\raggedright\arraybackslash}p{(\linewidth - 22\tabcolsep) * \real{0.0740}}
  >{\raggedright\arraybackslash}p{(\linewidth - 22\tabcolsep) * \real{0.0849}}
  >{\raggedright\arraybackslash}p{(\linewidth - 22\tabcolsep) * \real{0.0916}}
  >{\raggedright\arraybackslash}p{(\linewidth - 22\tabcolsep) * \real{0.0740}}
  >{\raggedright\arraybackslash}p{(\linewidth - 22\tabcolsep) * \real{0.0615}}
  >{\raggedright\arraybackslash}p{(\linewidth - 22\tabcolsep) * \real{0.0849}}
  >{\raggedright\arraybackslash}p{(\linewidth - 22\tabcolsep) * \real{0.0849}}
  >{\raggedright\arraybackslash}p{(\linewidth - 22\tabcolsep) * \real{0.0651}}
  >{\raggedright\arraybackslash}p{(\linewidth - 22\tabcolsep) * \real{0.0849}}
  >{\raggedright\arraybackslash}p{(\linewidth - 22\tabcolsep) * \real{0.0735}}@{}}
\toprule\noalign{}
\begin{minipage}[b]{\linewidth}\raggedright
Type
\end{minipage} & \begin{minipage}[b]{\linewidth}\raggedright
Year
\end{minipage} & \begin{minipage}[b]{\linewidth}\raggedright
Total
\end{minipage} & \begin{minipage}[b]{\linewidth}\raggedright
Mean
\end{minipage} & \begin{minipage}[b]{\linewidth}\raggedright
Median
\end{minipage} & \begin{minipage}[b]{\linewidth}\raggedright
SD
\end{minipage} & \begin{minipage}[b]{\linewidth}\raggedright
Min
\end{minipage} & \begin{minipage}[b]{\linewidth}\raggedright
Q1
\end{minipage} & \begin{minipage}[b]{\linewidth}\raggedright
Q3
\end{minipage} & \begin{minipage}[b]{\linewidth}\raggedright
Max
\end{minipage} & \begin{minipage}[b]{\linewidth}\raggedright
IQR
\end{minipage} & \begin{minipage}[b]{\linewidth}\raggedright
CV\%
\end{minipage} \\
\midrule\noalign{}
\endhead
\bottomrule\noalign{}
\endlastfoot
Decision & 2022 & 1,885 & 157.08 & 122.0 & 72.68 & 91 & 102.25 & 222.50 & 291 & 120.25 & 46.3 \\
& 2023 & 1,684 & 140.33 & 120.0 & 55.92 & 91 & 104.00 & 151.25 & 266 & 47.25 & 39.8 \\
& 2024 (with AI) & 1,408 & 117.33 & 124.0 & 20.34 & 87 & 100.50 & 135.25 & 142 & 34.75 & 17.3 \\
Memorandum & 2022 & 2,856 & 238.00 & 238.5 & 45.42 & 164 & 206.50 & 267.25 & 315 & 60.75 & 19.1 \\
& 2023 & 2,759 & 229.92 & 231.5 & 20.37 & 193 & 217.25 & 244.75 & 259 & 27.50 & 8.9 \\
& 2024 (with AI) & 2,613 & 217.75 & 226.5 & 33.19 & 152 & 202.00 & 238.25 & 258 & 36.25 & 15.2 \\
Adjudication & 2022 & 180 & 15.00 & 10.0 & 14.82 & 1 & 7.75 & 17.75 & 56 & 10.00 & 98.8 \\
& 2023 & 403 & 33.58 & 31.5 & 17.58 & 13 & 18.50 & 47.25 & 69 & 28.75 & 52.4 \\
& 2024 (with AI) & 372 & 31.00 & 30.5 & 8.21 & 20 & 23.00 & 36.75 & 45 & 13.75 & 26.5 \\
Official Letter & 2022 & 357 & 29.75 & 27.0 & 8.81 & 18 & 24.00 & 36.00 & 45 & 12.00 & 29.6 \\
& 2023 & 434 & 36.17 & 35.5 & 7.16 & 22 & 32.75 & 38.50 & 50 & 5.75 & 19.8 \\
& 2024 (with AI) & 532 & 44.33 & 43.5 & 6.04 & 30 & 41.75 & 49.00 & 52 & 7.25 & 13.6 \\
\end{longtable}

Note on Advisory Notices (SES/CONT): the Advisory Notice type had residual volume in 2022 (n=8) and 2023 (n=28), and grew to 214 in 2024, reflecting the shift in preventive function discussed in Section VI. For this type, the coefficient of variation is not a meaningful metric for comparison across years of the series and was omitted from this table.

\subsubsection{A.2.5 Evolution of Official Letter output}\label{a.2.5-evolution-of-official-letter-output}

\begin{longtable}[]{@{}
  >{\centering\arraybackslash}p{(\linewidth - 6\tabcolsep) * \real{0.2499}}
  >{\centering\arraybackslash}p{(\linewidth - 6\tabcolsep) * \real{0.2499}}
  >{\centering\arraybackslash}p{(\linewidth - 6\tabcolsep) * \real{0.2499}}
  >{\centering\arraybackslash}p{(\linewidth - 6\tabcolsep) * \real{0.2502}}@{}}
\toprule\noalign{}
\begin{minipage}[b]{\linewidth}\centering
Indicator
\end{minipage} & \begin{minipage}[b]{\linewidth}\centering
2023
\end{minipage} & \begin{minipage}[b]{\linewidth}\centering
2024
\end{minipage} & \begin{minipage}[b]{\linewidth}\centering
Variation
\end{minipage} \\
\midrule\noalign{}
\endhead
\bottomrule\noalign{}
\endlastfoot
Official Letters produced & 434 & 532 & +22.6\% \\
Cases processed (reference) & 1,752 & 1,581 & −9.8\% \\
Official Letters per case processed & 0.25 & 0.34 & +36.0\% \\
\end{longtable}

\subsubsection{A.2.6 Synthesis of SES/CONT evidence}\label{a.2.6-synthesis-of-sescont-evidence}

\begin{longtable}[]{@{}
  >{\centering\arraybackslash}p{(\linewidth - 4\tabcolsep) * \real{0.3597}}
  >{\centering\arraybackslash}p{(\linewidth - 4\tabcolsep) * \real{0.1126}}
  >{\centering\arraybackslash}p{(\linewidth - 4\tabcolsep) * \real{0.5277}}@{}}
\toprule\noalign{}
\begin{minipage}[b]{\linewidth}\centering
Evidence
\end{minipage} & \begin{minipage}[b]{\linewidth}\centering
Result
\end{minipage} & \begin{minipage}[b]{\linewidth}\centering
Meaning
\end{minipage} \\
\midrule\noalign{}
\endhead
\bottomrule\noalign{}
\endlastfoot
Overall average time & −18.2\% & From 17.92 to 14.66 days per case \\
Estimated released capacity & +22.3\% & Equivalent to 352 additional cases (5,155 days) \\
Case backlog & Cleared & Inbox without cases over 10 days at end of 2024 \\
High-complexity cases & Up to −95.5\% & 70\% of types analyzed improved (7 of 10) \\
Adjudications (normalized analysis) & +2.1\% & Average of 31.00/month exceeded 30.36/month in 2023 without task force \\
Adjudications (stability) & −55.4\% amplitude & More regular output, without extraordinary peaks \\
Productivity per case & +0.4\% & Same documentary delivery in 18.2\% less time \\
Official Letters & +22.6\% & Greater institutional articulation capacity \\
Advisory Notices & +664\% & Shift from punitive to preventive posture \\
AI-related problems reported & None & Institutional report \\
\end{longtable}

\subsection{A.3 Tables of official data: Internal Control Unit (UCI/SEDET-DF)}\label{a.3-tables-of-official-data-internal-control-unit-ucisedet-df}

The 2024--2025 data in the tables of this section were extracted from Report No.~1/2026 (SEI Doc. 193384267); the 2023 series in Table A.3.4 comes from the unit's official SEI-GDF statistics (see the note to that table).

\subsubsection{A.3.1 General indicators 2024 vs.~2025}\label{a.3.1-general-indicators-2024-vs.-2025}

\begin{longtable}[]{@{}
  >{\centering\arraybackslash}p{(\linewidth - 6\tabcolsep) * \real{0.2499}}
  >{\centering\arraybackslash}p{(\linewidth - 6\tabcolsep) * \real{0.2499}}
  >{\centering\arraybackslash}p{(\linewidth - 6\tabcolsep) * \real{0.2499}}
  >{\centering\arraybackslash}p{(\linewidth - 6\tabcolsep) * \real{0.2502}}@{}}
\toprule\noalign{}
\begin{minipage}[b]{\linewidth}\centering
Indicator
\end{minipage} & \begin{minipage}[b]{\linewidth}\centering
2024
\end{minipage} & \begin{minipage}[b]{\linewidth}\centering
2025 (with AI)
\end{minipage} & \begin{minipage}[b]{\linewidth}\centering
Variation
\end{minipage} \\
\midrule\noalign{}
\endhead
\bottomrule\noalign{}
\endlastfoot
Average processing time & 34 days & 17 days & −50\% \\
Cases processed & 256 & 336 & +31\% \\
Documents produced & 251 & 419 & +67\% \\
Technical reports issued & 66 & 122 & +85\% \\
Dispatches produced & 102 & 163 & +60\% \\
\end{longtable}

\subsubsection{A.3.2 Monthly distribution of financial volume analyzed in 2025}\label{a.3.2-monthly-distribution-of-cases-and-financial-volume-analyzed-in-2025}

\begin{longtable}[]{@{}
  >{\centering\arraybackslash}p{(\linewidth - 2\tabcolsep) * \real{0.5000}}
  >{\centering\arraybackslash}p{(\linewidth - 2\tabcolsep) * \real{0.5000}}@{}}
\toprule\noalign{}
\begin{minipage}[b]{\linewidth}\centering
Month
\end{minipage} & \begin{minipage}[b]{\linewidth}\centering
Value analyzed
\end{minipage} \\
\midrule\noalign{}
\endhead
\bottomrule\noalign{}
\endlastfoot
January & US\$ 0.9 mi \\
February & US\$ 13.1 mi \\
March & US\$ 6.9 mi \\
April & US\$ 3.7 mi \\
May & US\$ 8.0 mi \\
June & US\$ 11.7 mi \\
July & US\$ 4.3 mi \\
August & US\$ 19.1 mi \\
September & US\$ 4.2 mi \\
October & US\$ 11.7 mi \\
November & US\$ 4.9 mi \\
December & US\$ 6.3 mi \\
Total 2025 & US\$ 94.8 million \\
\end{longtable}

Note: monthly values are rounded to US\$ 0.1 million (converted from the official reais figures at the rate declared in Appendix A.0); the sum of the rounded monthly figures may therefore differ slightly from the annual total, which is US\$ 94.8 million (an approximate conversion of the unrounded source data; see Appendix A.0 for the official figure in reais and the exchange rate used). Source: official spreadsheet ``Estatística Notas Técnicas Unidade de Controle Interno, UCI/SEDET, 2025,'' p.~48 of the signed UCI/SEDET 2025 management report (SEI Doc. 193436555).

\subsubsection{A.3.3 Analytical intensity and Governance}\label{a.3.3-analytical-intensity-and-governance}

\begin{longtable}[]{@{}
  >{\centering\arraybackslash}p{(\linewidth - 2\tabcolsep) * \real{0.5000}}
  >{\centering\arraybackslash}p{(\linewidth - 2\tabcolsep) * \real{0.5000}}@{}}
\toprule\noalign{}
\begin{minipage}[b]{\linewidth}\centering
Indicator
\end{minipage} & \begin{minipage}[b]{\linewidth}\centering
2025
\end{minipage} \\
\midrule\noalign{}
\endhead
\bottomrule\noalign{}
\endlastfoot
Technical reports issued & 122 \\
Formal recommendations issued to public managers & 286 \\
Financial volume analyzed & US\$ 94.8 million \\
Human review procedure before issuance & Mandatory (Human-in-the-Loop policy; three-step Triple Review) \\
\end{longtable}

\subsubsection{A.3.4 Descriptive statistics of monthly output by document type: UCI/SEDET}\label{a.3.4-descriptive-statistics-of-monthly-output-by-document-type-ucisedet}

Base: 12 monthly values per type/year. Unless otherwise noted, series are the raw SEI-GDF unit statistics (the native ``Estatísticas da Unidade'' extraction described in Appendix A.0). The 2025 Technical Report series is the exception: it is the total consolidated in the signed management report (122; see Appendix A.0), not the raw SEI-GDF count.

\begin{longtable}[]{@{}
  >{\raggedright\arraybackslash}p{(\linewidth - 22\tabcolsep) * \real{0.1496}}
  >{\raggedright\arraybackslash}p{(\linewidth - 22\tabcolsep) * \real{0.0769}}
  >{\raggedright\arraybackslash}p{(\linewidth - 22\tabcolsep) * \real{0.0769}}
  >{\raggedright\arraybackslash}p{(\linewidth - 22\tabcolsep) * \real{0.0788}}
  >{\raggedright\arraybackslash}p{(\linewidth - 22\tabcolsep) * \real{0.0916}}
  >{\raggedright\arraybackslash}p{(\linewidth - 22\tabcolsep) * \real{0.0728}}
  >{\raggedright\arraybackslash}p{(\linewidth - 22\tabcolsep) * \real{0.0719}}
  >{\raggedright\arraybackslash}p{(\linewidth - 22\tabcolsep) * \real{0.0728}}
  >{\raggedright\arraybackslash}p{(\linewidth - 22\tabcolsep) * \real{0.0784}}
  >{\raggedright\arraybackslash}p{(\linewidth - 22\tabcolsep) * \real{0.0737}}
  >{\raggedright\arraybackslash}p{(\linewidth - 22\tabcolsep) * \real{0.0784}}
  >{\raggedright\arraybackslash}p{(\linewidth - 22\tabcolsep) * \real{0.0782}}@{}}
\toprule\noalign{}
\begin{minipage}[b]{\linewidth}\raggedright
Type
\end{minipage} & \begin{minipage}[b]{\linewidth}\raggedright
Year
\end{minipage} & \begin{minipage}[b]{\linewidth}\raggedright
Total
\end{minipage} & \begin{minipage}[b]{\linewidth}\raggedright
Mean
\end{minipage} & \begin{minipage}[b]{\linewidth}\raggedright
Median
\end{minipage} & \begin{minipage}[b]{\linewidth}\raggedright
SD
\end{minipage} & \begin{minipage}[b]{\linewidth}\raggedright
Min
\end{minipage} & \begin{minipage}[b]{\linewidth}\raggedright
Q1
\end{minipage} & \begin{minipage}[b]{\linewidth}\raggedright
Q3
\end{minipage} & \begin{minipage}[b]{\linewidth}\raggedright
Max
\end{minipage} & \begin{minipage}[b]{\linewidth}\raggedright
IQR
\end{minipage} & \begin{minipage}[b]{\linewidth}\raggedright
CV\%
\end{minipage} \\
\midrule\noalign{}
\endhead
\bottomrule\noalign{}
\endlastfoot
Dispatch & 2023 & 85 & 7.08 & 6.5 & 4.56 & 1 & 4.25 & 10.50 & 14 & 6.25 & 64.4 \\
& 2024 & 102 & 8.50 & 9.0 & 6.23 & 0 & 3.25 & 13.25 & 18 & 10.00 & 73.3 \\
& 2025 (with AI) & 163 & 13.58 & 12.5 & 6.69 & 5 & 8.00 & 19.25 & 25 & 11.25 & 49.3 \\
Technical Report & 2023 & 82 & 6.83 & 7.5 & 3.76 & 0 & 4.00 & 9.25 & 12 & 5.25 & 55.1 \\
& 2024 & 66 & 5.50 & 5.5 & 4.58 & 0 & 1.75 & 7.50 & 15 & 5.75 & 83.3 \\
& 2025 (with AI, consolidated) & 122 & 10.17 & 9.5 & 5.39 & 3 & 7.00 & 12.50 & 21 & 5.50 & 53.0 \\
\end{longtable}

\subsection{A.4 Assessment of hypothetical savings from UCI/SEDET 2025 technical reports}\label{a.4-assessment-of-hypothetical-savings-from-ucisedet-2025-technical-reports}

\subsubsection{A.4.1 Universe and classification principle}\label{a.4.1-universe-and-classification-principle}

This assessment covers the 122 technical reports consolidated in the signed UCI/SEDET 2025 management report (SEI Doc. 193384267), the same official universe adopted throughout this paper (see the reconciliation note in Appendix A.0). Each of these 122 reports has an official value registered in the report's own spreadsheet (SEI Doc. 193436555, p.~48), the sum of which is US\$ 94.8 million: the value of the matters submitted to technical analysis.

For the purpose of this appendix's financial model, values at risk were attributed recommendation by recommendation, from the full text of each of the 122 technical reports, following the moderate criterion described in A.4.3, and audited against the original documents in the SEI.

Classification was made by the content of each recommendation, not by the formal verdict of the report. In a non-binding internal-control system, it is common for reports with a favorable conclusion to contain material recommendations whose adoption would prevent future loss; classifying by formal verdict would systematically underestimate the effectiveness of the control.

The granularity of the analysis is the individual recommendation, not the report as a unit. The recommendations identified and classified by the authors from the full text of each of the 122 technical reports feed this financial model (see A.4.7). This classification is the authors' own, built strictly for modeling purposes, and does not replace the institutional total of 286 formal recommendations stated in the signed 2025 management report (p.~5), which governs every non-modeling statement of this paper (see Appendix A.0). The difference reflects how individual textual passages were segmented into discrete, separately assessable recommendations for the purpose of attributing value at risk, not a disagreement with the official count.

\subsubsection{A.4.2 Applied typology}\label{a.4.2-applied-typology}

Each recommendation was classified along two dimensions. The first is nature:

\begin{longtable}[]{@{}
  >{\raggedright\arraybackslash}p{(\linewidth - 2\tabcolsep) * \real{0.5000}}
  >{\raggedright\arraybackslash}p{(\linewidth - 2\tabcolsep) * \real{0.5000}}@{}}
\toprule\noalign{}
\begin{minipage}[b]{\linewidth}\raggedright
Nature
\end{minipage} & \begin{minipage}[b]{\linewidth}\raggedright
Definition
\end{minipage} \\
\midrule\noalign{}
\endhead
\bottomrule\noalign{}
\endlastfoot
I: Procedural/documentary & Recommendations on documentation, description, workflow, standardization, formal internal controls, or prospective improvements. No direct financial implication for the act under analysis. \\
II: Legal compliance & Failure to comply with a specific legal requirement that may cause nullity or external-control questioning (expired certificate, absence of budgetary commitment, defect in a mandatory clause, documentary divergence without quantified value). \\
III: Quantitative anomaly & Divergence in value, overpricing, undue payment, duplication, specific disallowance. Measurable monetary value at risk. \\
IV: Material irregularity & Indications of fraud, collusion, favoritism, or explicit recommendation of administrative disciplinary proceedings/inquiry/communication to the Public Prosecutor's Office/TCDF/MPC. Rare cases in preventive control; none of the 122 technical reports of the fiscal year fell under this category. \\
\end{longtable}

The second dimension is materiality:

\begin{longtable}[]{@{}
  >{\raggedright\arraybackslash}p{(\linewidth - 2\tabcolsep) * \real{0.5000}}
  >{\raggedright\arraybackslash}p{(\linewidth - 2\tabcolsep) * \real{0.5000}}@{}}
\toprule\noalign{}
\begin{minipage}[b]{\linewidth}\raggedright
Materiality
\end{minipage} & \begin{minipage}[b]{\linewidth}\raggedright
Definition
\end{minipage} \\
\midrule\noalign{}
\endhead
\bottomrule\noalign{}
\endlastfoot
High & Nature IV; nature III with explicit value; nature II with validity defect affecting payment/contract of significant value. \\
Medium & Nature III without explicit direct value; nature II with a clause that may generate contractual nullity. \\
Low & Nature II in isolation (minor curable legal non-compliance). \\
Null & Only nature I (pure procedural). \\
\end{longtable}

\subsubsection{A.4.3 Moderate criterion for attribution of value at risk}\label{a.4.3-moderate-criterion-for-attribution-of-value-at-risk}

The value at risk of each recommendation was attributed according to six exhaustive cases:

\begin{enumerate}
\def\labelenumi{\arabic{enumi}.}
\item
  Explicit value mentioned in the recommendation (specific disallowance, refund, quantified reimbursement): uses that value.
\item
  Defect affecting the validity of the act (expired certificate, absence of budgetary commitment, divergence between receipt acknowledgment and invoice, absence of formal designation of fiscal monitor by published act, contractual clause inconsistent with the public call, absence of mandatory clause, non-compliance with legal qualification requirement): uses the principal value of the payment/contract under analysis.
\item
  Quantitative divergence without specified value: uses the principal value as the maximum exposure ceiling, with explicit justification that the actual risk may be lower.
\item
  Pure procedural recommendation (standardization, detailing, report improvement, preventive designation for future procurements): zero value.
\item
  Recommendation of future improvement, monitoring, or maintenance of good practices: zero value.
\item
  Generic observance of legality without identification of specific failure: zero value.
\end{enumerate}

Additionally, when multiple recommendations from the same report identify different defects but affect the same payment/contract, the value is attributed to the most serious recommendation; the others receive zero value with the justification ``ancillary defect to recommendation X.Y, avoiding double counting.'' This rule is necessary because the actual loss on a given financial object can only occur once, regardless of how many formal defects UCI identifies simultaneously.

\subsubsection{A.4.4 Probability matrix}\label{a.4.4-probability-matrix}

Each recommendation is assigned a probability of risk materialization, modulated by nature and materiality:

\begin{longtable}[]{@{}
  >{\raggedright\arraybackslash}p{(\linewidth - 6\tabcolsep) * \real{0.2499}}
  >{\raggedright\arraybackslash}p{(\linewidth - 6\tabcolsep) * \real{0.2499}}
  >{\raggedright\arraybackslash}p{(\linewidth - 6\tabcolsep) * \real{0.2499}}
  >{\raggedright\arraybackslash}p{(\linewidth - 6\tabcolsep) * \real{0.2502}}@{}}
\toprule\noalign{}
\begin{minipage}[b]{\linewidth}\raggedright
Nature
\end{minipage} & \begin{minipage}[b]{\linewidth}\raggedright
Conservative
\end{minipage} & \begin{minipage}[b]{\linewidth}\raggedright
Central
\end{minipage} & \begin{minipage}[b]{\linewidth}\raggedright
Optimistic
\end{minipage} \\
\midrule\noalign{}
\endhead
\bottomrule\noalign{}
\endlastfoot
I: Procedural & 0\% & 0\% & 0\% \\
II: Legal compliance & 2\% & 5\% & 10\% \\
III: Quantitative anomaly & 10\% & 25\% & 40\% \\
IV: Material irregularity & 30\% & 50\% & 70\% \\
\end{longtable}

The materiality modifier adjusts each probability: high = 1.00; medium = 0.75; low = 0.50; null = 0.00.

The estimate of potential mitigation per recommendation is calculated as: value at risk × probability (nature, scenario) × modifier (materiality).

The aggregate estimate is the sum across the recommendations classified in this exercise (Appendix A.4.7), for each of the three scenarios.

\subsubsection{A.4.5 Probability calibration: sources used}\label{a.4.5-probability-calibration-sources-used}

The probability matrix (Table A.4.4) was calibrated on a prudential basis: it proceeds from the principle that the advisory prior control characteristic of UCI does not replace the manager's decision, which limits the fraction of recommendations that effectively prevents the loss. The probabilities were set in three scenarios (conservative, central, and optimistic), anchored in the following reference literature:

(i) U.S. federal public auditing. The U.S. Government Accountability Office (GAO), the federal external-audit body and functional equivalent of the Brazilian TCU, publishes annual estimates of improper payments (US\$ 162 billion in FY 2024, US\$ 186 billion in FY 2025; approximately US\$ 3 trillion accumulated since 2003), and the associated payment-integrity literature documents typical effective recovery rates considerably lower than the volume identified.

(ii) Occupational fraud studies. The ACFE Report to the Nations (1,921 cases in 138 countries, 2024 edition) documents that organizations with structured internal controls (particularly surprise audits, financial statement audits, whistleblower hotlines, and proactive data analysis) present substantial reductions (up to 50\%) in fraud losses and duration compared with organizations without such controls.

Drawing on this empirical evidence on improper payments and occupational fraud, which shows both substantial residual losses and significant, but far from complete, loss reductions in organizations with strong internal controls, we define three scenarios (conservative, central, optimistic). In the central scenario, we set the probability of materialization at 5\% for nature II (legal compliance), 25\% for nature III (quantitative anomaly), and 50\% for nature IV (material irregularity). These values are not directly estimated from the literature; rather, they represent prudential, judgment-based assumptions that are consistent with the ranges suggested by aggregate studies. The optimistic scenario reflects the upper bound of effectiveness implied by the literature, whereas the conservative one represents a prudential floor.

\subsubsection{A.4.6 Validation process}\label{a.4.6-validation-process}

The result went through four sequential validation phases. The first was full note-by-note reading. The second was focused reconciliation: reports with extensive text and low initial recommendation count were re-examined to incorporate additional recommendations identified. The third was cross-audit: a stratified sample covering all processing blocks and all natures was reviewed, with correction of divergences identified in nature, materiality, and value at risk. The fourth was final validation: a sample of the main contributors to the financial aggregate was reviewed by the author, with application of the observations.

\subsubsection{A.4.7 Consolidated results}\label{a.4.7-consolidated-results}

Tier 1: Official value of the universe analyzed:

US\$ 94.8 million, the sum of the official value of each of the 122 technical reports registered in the signed management report's own spreadsheet (SEI Doc. 193436555, p.~48; see Appendix A.0).

Tier 2: Recommendations by nature (authors' own classification for modeling purposes; see Appendix A.0 for the institutional total of 286 recommendations):

\begin{longtable}[]{@{}
  >{\raggedright\arraybackslash}p{(\linewidth - 2\tabcolsep) * \real{0.5000}}
  >{\raggedright\arraybackslash}p{(\linewidth - 2\tabcolsep) * \real{0.5000}}@{}}
\toprule\noalign{}
\begin{minipage}[b]{\linewidth}\raggedright
Nature
\end{minipage} & \begin{minipage}[b]{\linewidth}\raggedright
Gross value at risk
\end{minipage} \\
\midrule\noalign{}
\endhead
\bottomrule\noalign{}
\endlastfoot
IV: Material irregularity & N/A \\
III: Quantitative anomaly & US\$ 3.4 million \\
II: Legal compliance & US\$ 40.4 million \\
I: Procedural & N/A \\
Total & US\$ 43.7 million \\
\end{longtable}

\textbf{Tier 3: Potential mitigation estimate:}

\begin{longtable}[]{@{}
  >{\raggedright\arraybackslash}p{(\linewidth - 2\tabcolsep) * \real{0.5000}}
  >{\raggedright\arraybackslash}p{(\linewidth - 2\tabcolsep) * \real{0.5000}}@{}}
\toprule\noalign{}
\begin{minipage}[b]{\linewidth}\raggedright
Scenario
\end{minipage} & \begin{minipage}[b]{\linewidth}\raggedright
Total
\end{minipage} \\
\midrule\noalign{}
\endhead
\bottomrule\noalign{}
\endlastfoot
Conservative & US\$ 1.1 million \\
Central & US\$ 2.8 million \\
Optimistic & US\$ 5.2 million \\
\end{longtable}

These estimates are not realized or audited savings: they are a modeled range of potential mitigation, conditional on the probability matrix of A.4.4 and on managers' effective adoption of the recommendations, which this study does not track.

\subsubsection{A.4.8 Limitations of the estimate}\label{a.4.8-limitations-of-the-estimate}

\begin{enumerate}
\def\labelenumi{\arabic{enumi}.}
\item
  Non-binding nature of UCI recommendations. UCI issues advisory prior control; there is no systematic record of which recommendations were effectively complied with by managers. The estimate quantifies mitigation potential, not documented realization.
\item
  Non-observable counterfactual. It is not possible to establish with certainty the alternative scenario without UCI control. The probabilities applied are derived from international literature, not calibrated with GDF-specific data.
\item
  Underestimation by conservative classification. Recommendations classified as Nature I (procedural) receive zero probability, even when they may prevent indirect losses. This is a deliberate choice for methodological rigor.
\item
  Non-monetary benefits not captured. Gains in transparency, procedural compliance, documentary standardization, and team technical capacity are not quantified by the financial estimate.
\item
  Moderate criterion is interpretive. The attribution of the entire payment/contract value to validity defects is defensible but represents a methodological choice that could be presented in more or less generous form under alternative interpretations.
\end{enumerate}

\section{REFERENCES}\label{references}

SEI-GDF Institutional Documents

{[}1{]} FEDERAL DISTRICT. Department of Health. Sectoral Internal Control Office (SES/CONT). Executive Management Report: Impact of Artificial Intelligence on Operational Efficiency. SEI Doc. No.~197403428, SEI case 00060-00582291/2024-11. Brasília, DF, January 2, 2025.

{[}2{]} FEDERAL DISTRICT. Department of Economic Development, Labor and Income. Internal Control Unit (UCI/SEDET-DF). Report No.~1/2026: Management Results Report 2025: UCI/GAB/SEDET. SEI Doc. No.~193384267, SEI case 04035-00000853/2026-24. Brasília, DF, January 28, 2026.

{[}3{]} FEDERAL DISTRICT. Department of Economic Development, Labor and Income. Internal Control Unit (UCI/SEDET-DF). AI Governance Framework: UCI. SEI Doc. No.~194251158. Brasília, DF.

{[}4{]} FEDERAL DISTRICT. Department of Health. Sectoral Internal Control Office, Advisory Office for the Adjudication of Administrative Proceedings (SES/CONT/ASJULG). Memorandum No.~4/2024: Proposal for the creation of the course ``Artificial Intelligence in the public sector: techniques, risks, and applications''. SEI Doc. No.~146621014, SEI case 00060-00314546/2024-15. Signatory: Vinicius Santana Gomes. Brasília, DF, June 25, 2024.

{[}5{]} FEDERAL DISTRICT. Executive Secretariat of the Civil House, Sub-secretariat of Administrative Management, School of Government (SEEC/SEGEA/EGOV). Memorandum No.~166/2024: Institutional approval of the course ``Artificial Intelligence in the public sector: techniques, risks, and applications''. SEI Doc. No.~146526272, SEI case 04044-00021535/2024-26. Signatory: Raquel Aben Athar de Sousa (Acting Executive Director). Brasília, DF, July 22, 2024.

{[}6{]} FEDERAL DISTRICT. Department of Health. Sectoral Internal Control Office, Advisory Office for the Adjudication of Administrative Proceedings (SES/CONT/ASJULG). Report No.~12/2024: Institutional record of the clearance of the unit's case backlog. Brasília, DF, December 14, 2024.

Brazilian Legislation

{[}7{]} BRAZIL. Law No.~13.709, of August 14, 2018. Brazilian General Data Protection Law (LGPD). Official Gazette of the Union, Brasília, DF.

Public Auditing

{[}8{]} U.S. GOVERNMENT ACCOUNTABILITY OFFICE (GAO). Improper Payments: Information on Agencies' Fiscal Year 2024 Estimates. GAO-25-107753. Washington, DC: GAO, 2025. Available at: \url{https://www.gao.gov/products/gao-25-107753}.

{[}9{]} U.S. GOVERNMENT ACCOUNTABILITY OFFICE (GAO). Payment Integrity: Agencies' Estimated Improper Payments Increased to \$186 Billion in Fiscal Year 2025. GAO-26-108694. Washington, DC: GAO, 2026. Available at: \url{https://www.gao.gov/products/gao-26-108694}.

{[}10{]} ASSOCIATION OF CERTIFIED FRAUD EXAMINERS (ACFE). Occupational Fraud 2024: A Report to the Nations. Austin, TX: ACFE, 2024. Available at: \url{https://www.acfe.com/-/media/files/acfe/pdfs/rttn/2024/2024-report-to-the-nations.pdf}.

International AI Governance Frameworks

{[}11{]} NATIONAL INSTITUTE OF STANDARDS AND TECHNOLOGY (NIST). Artificial Intelligence Risk Management Framework (AI RMF 1.0). NIST AI 100-1. Gaithersburg, MD: U.S. Department of Commerce, 2023. Available at: \url{https://www.nist.gov/itl/ai-risk-management-framework}.

{[}12{]} U.S. GOVERNMENT ACCOUNTABILITY OFFICE (GAO). Artificial Intelligence: An Accountability Framework for Federal Agencies and Other Entities. GAO-21-519SP. Washington, DC: GAO, 2021. Available at: \url{https://www.gao.gov/products/gao-21-519sp}.

{[}13{]} EUROPEAN PARLIAMENT AND COUNCIL OF THE EUROPEAN UNION. Regulation (EU) 2024/1689 of 13 June 2024 laying down harmonised rules on artificial intelligence (Artificial Intelligence Act). Official Journal of the European Union, Brussels: European Union, 2024. Available at: \url{https://eur-lex.europa.eu/eli/reg/2024/1689/oj}.

{[}14{]} ORGANISATION FOR ECONOMIC CO-OPERATION AND DEVELOPMENT (OECD). Recommendation of the Council on Artificial Intelligence. OECD/LEGAL/0449. Paris: OECD, adopted 22 May 2019, amended 8 November 2023 and 3 May 2024. Available at: \url{https://oecd.ai/en/ai-principles}.

Methodological References

{[}15{]} WEI, Jason; WANG, Xuezhi; SCHUURMANS, Dale; BOSMA, Maarten; ICHTER, Brian; XIA, Fei; CHI, Ed H.; LE, Quoc V.; ZHOU, Denny. Chain-of-Thought Prompting Elicits Reasoning in Large Language Models. Advances in Neural Information Processing Systems (NeurIPS), 2022. Available at: \url{https://arxiv.org/abs/2201.11903} (DOI: \url{https://doi.org/10.48550/arXiv.2201.11903)}.

{[}16{]} BROWN, Tom B.; MANN, Benjamin; RYDER, Nick et al.~Language Models are Few-Shot Learners. Advances in Neural Information Processing Systems (NeurIPS), 2020. Available at: \url{https://arxiv.org/abs/2005.14165} (DOI: \url{https://doi.org/10.48550/arXiv.2005.14165)}.

Technology Adoption and AI Training

{[}17{]} COHEN, Wesley M.; LEVINTHAL, Daniel A. Absorptive Capacity: A New Perspective on Learning and Innovation. Administrative Science Quarterly, v. 35, n.~1, p.~128--152, 1990. Available at: \url{https://doi.org/10.2307/2393553}.

{[}18{]} DUNLEAVY, Patrick; MARGETTS, Helen; BASTOW, Simon; TINKLER, Jane. New Public Management Is Dead: Long Live Digital-Era Governance. Journal of Public Administration Research and Theory, v. 16, n.~3, p.~467--494, 2006. Available at: \url{https://doi.org/10.1093/jopart/mui057}.

{[}19{]} HEEKS, Richard. Most eGovernment-for-development projects fail: How can risks be reduced? Manchester: Institute for Development Policy and Management, University of Manchester, 2003. Available at: \url{https://ssrn.com/abstract=3540052}.

{[}20{]} LONG, Duri; MAGERKO, Brian. What is AI literacy? Competencies and design considerations. Proceedings of the 2020 CHI Conference on Human Factors in Computing Systems, p.~1--16, 2020. Available at: \url{https://doi.org/10.1145/3313831.3376727}.

\clearpage
{\centering
{\LARGE\bfseries A Principal Barreira à Adoção de IA no Setor Público É a Falta de Capacitação:\par}
\vspace{0.35em}
{\large Como um Método Estruturado Esteve Associado a Ganhos de Produtividade em Dois Casos do Governo Brasileiro\footnotemark[2]\par}
\vspace{0.9em}
{\large Vinicius Santana Gomes\footnotemark[1]\par}
\vspace{1.2em}
\par}
\footnotetext[1]{Assessor Especial, Unidade de Controle Interno, Secretaria de Estado de Desenvolvimento Econômico, Trabalho e Renda do Governo do Distrito Federal; Instrutor Oficial em Inteligência Artificial, Escola de Governo do Distrito Federal. Contato: vinicius.gomes@sedet.df.gov.br; gomes.vs@gmail.com.}
\footnotetext[2]{Versão em português, traduzida do original em inglês, ambas publicadas no arXiv: \url{https://arxiv.org/abs/2606.01517}.}

\textbf{Abstract}

A adoção de inteligência artificial generativa no setor público tem sido
tratada predominantemente como problema tecnológico, com a expectativa
de que ganhos de produtividade decorram da disponibilidade de modelos
cada vez mais capazes. Este \emph{paper} sustenta, a partir de dois
casos auditáveis no Serviço Público Brasileiro, que o gargalo
determinante da adoção observado nessas unidades não foi tecnológico,
mas de capacitação, e descreve a metodologia pedagógica estruturada em
quatro camadas, desenvolvida pelo autor deste \emph{paper}. O método foi
aplicado em duas unidades com perfis institucionais distintos: a
Controladoria Setorial da Saúde do Governo do Distrito Federal
(SES/CONT) ao longo do ano de 2024, e a Unidade de Controle Interno da
Secretaria de Estado de Desenvolvimento Econômico, Trabalho e Renda do
Governo do Distrito Federal (UCI/SEDET) ao longo de 2025. Em ambos os
casos, os indicadores oficiais do Sistema Eletrônico de Informações
(SEI-GDF), verificáveis por terceiros, registraram ganhos que acompanharam
a implantação do método: redução do tempo médio de tramitação de 18,2\% na SES/CONT
e de 50\% na UCI/SEDET, com a UCI ainda registrando aumento de 85\% na
produção de notas técnicas, a expedição de 286 recomendações formais aos
gestores e a análise de processos e matérias cujo valor total, segundo a estatística oficial assinada pela própria unidade, somou R\$ 521,3 milhões, o valor das matérias submetidas à análise técnica. Na SES/CONT, o relatório institucional informou que não foram relatados problemas relacionados à IA no período analisado. Na UCI/SEDET, o relatório institucional registra que toda minuta produzida com apoio de IA esteve sujeita a procedimento obrigatório de revisão humana antes da expedição, sob política de Human-in-the-Loop operacionalizada por meio de Tripla Revisão em três etapas. A análise é compatível
com a hipótese de que o método é replicável entre órgãos com mandatos
distintos, opera dentro de protocolos desenhados para atender à
legislação nacional e internacional de proteção de dados e aos
princípios da Administração Pública, e é acessível a entes públicos com
restrições orçamentárias, já que utilizou modelos de IA gratuitos.

\textbf{Palavras-chave:} Inteligência Artificial generativa; setor
público; governança de IA; metodologia de capacitação; replicabilidade
institucional; Proteção de dados.

\hypertarget{i.-introduuxe7uxe3o}{%
\section{I. Introdução}\label{i.-introduuxe7uxe3o}}

Existe uma frase recorrente em apresentações sobre inteligência
artificial no setor público: os ganhos da IA chegarão quando a
tecnologia amadurecer. A frase adia a discussão. Os modelos já existem,
estão em uso doméstico e institucional, e seguem sendo aprimorados em
ciclos de meses. O atraso está em outro lugar.

Ao assumir a chefia da Assessoria da Controladoria Setorial da Saúde do
Distrito Federal, o autor deste \emph{paper} encontrou uma equipe
multidisciplinar, sem formação jurídica homogênea, processando
manualmente um volume de processos cujo tempo médio de tramitação o
SEI-GDF registrava em 17 dias, 22 horas e 11 minutos. Eram dados
públicos, disponíveis para qualquer pessoa.

A pergunta que foi feita à equipe naquele momento não foi se a IA
poderia ajudar. Foi outra: se a ferramenta existe, é gratuita e qualquer
servidor consegue abrir uma aba do navegador e usar, por que ela não
estava sendo utilizada na unidade para aumentar a produtividade, a
qualidade do trabalho e diminuir o tempo de tramitação dos processos,
com a diminuição do passivo acumulado?

A resposta dessa pergunta, e a metodologia construída ao longo dos dois
anos seguintes para respondê-la na prática, é o objeto deste
\emph{paper}.

A primeira evidência que o trabalho tornou visível, nos dois casos
estudados, é que o gargalo entre a oferta de IA e a entrega de valor não
é tecnológico. É de capacitação. Modelos cada vez mais capazes chegam à
mesa de um operador sem repertório para usá-los com segurança ou com
método. A segunda é que capacitação genérica em IA não resolve esse
gargalo no serviço público. O que resolveu foi um método estruturado,
com governança jurídica embutida, que separa o que a ferramenta faz, o
que o operador precisa fazer, e o que o cargo público exige que seja
feito antes que qualquer documento saia da unidade.

O método aqui descrito tem evidência empírica auditável em dois órgãos
do Governo do Distrito Federal: a Controladoria Setorial da Saúde
(SES/CONT, ano de 2024) e a Unidade de Controle Interno da SEDET-DF
(UCI/SEDET, ano de 2025). Os dois conjuntos de dados saem do SEI-GDF
(sistema oficial do Governo do Distrito Federal do Brasil), são
verificáveis por terceiros, e estão sintetizados em relatórios
institucionais formais, já consolidados e assinados pelas respectivas
chefias. Igualmente relevante: o método foi executado na versão gratuita
das plataformas comerciais de IA, pois os Órgãos não tinham a IA
contratada no momento da utilização.

O texto principal segue a ordem em que o método foi efetivamente
construído e validado. As Seções II a VI cobrem a Controladoria Setorial
da Saúde: o diagnóstico operacional encontrado em 2023, a construção do
método em parceria com a Escola de Governo do Distrito Federal
(EGOV-DF), a capacitação dos servidores da unidade matriculados como
alunos no mesmo curso oficial aberto a todo o Governo do Distrito
Federal, e a medição do ano de 2024 contra os do ano anterior. As Seções
VII a X cobrem a transferência do método para a Unidade de Controle
Interno da SEDET-DF ao longo de 2025, sob a pergunta que o segundo bloco
precisa responder: o método sobrevive à mudança de órgão, de temática
processual e de equipe? A Seção XI discute o que os dois casos ensinam
juntos, com atenção também ao que não funcionou e aos limites
identificados durante a implementação. A Seção XII apresenta as
conclusões do estudo.

O autor declara, para fins de transparência, que cumulou no período
analisado os papéis de proponente do método, instrutor do curso na
Escola de Governo, gestor das duas unidades estudadas e signatário do
relatório institucional que serve de fonte primária no caso da
Controladoria Setorial da Saúde. O relatório institucional do segundo
caso (UCI/SEDET) foi assinado pelo superior hierárquico do autor na
estrutura da unidade.

O apêndice técnico ao final do paper reúne as referências documentais e
as tabelas de dados oficiais que subsidiaram a análise.

\hypertarget{i.1.-trabalhos-relacionados}{%
\subsection{I.1. Trabalhos
relacionados}\label{i.1.-trabalhos-relacionados}}

O presente paper se insere em três conjuntos de literatura que raramente
conversam entre si: frameworks regulatórios de governança de
inteligência artificial, estudos de adoção de tecnologia no setor
público, e pesquisas sobre capacitação no uso de modelos generativos. O
método é uma tentativa de integrar essas três frentes em uma única
arquitetura pedagógica, e a finalidade desta seção é situar a
contribuição em relação a cada uma.

No plano regulatório, frameworks como o NIST AI Risk Management
Framework (NIST, 2023) e o GAO AI Accountability Framework (GAO, 2021)
estabelecem estruturas de governança baseadas em ciclos de
identificação, mensuração e tratamento de riscos. O EU AI Act (EU, 2024)
traduz princípios em categorias jurídicas de risco, e os Princípios da
OCDE (OECD, 2019) consolidam uma base normativa internacional. Esses
instrumentos descrevem o que deve ser observado e como avaliar o
cumprimento. O método, apresentado na Seção III, opera no plano
complementar: traduz esses princípios em conduta operacional do servidor
público, com aderência ao sigilo processual e à Legislação de Proteção
de Dados.

No plano organizacional, a literatura sobre adoção tecnológica no setor
público enfatiza, desde a década de 1990, que a disponibilidade da
tecnologia não é condição suficiente para a entrega de valor: o que
determina o desempenho efetivo é a capacidade absortiva da organização
(Cohen \& Levinthal, 1990), entendida como a capacidade de reconhecer o
valor de nova informação externa, assimilá-la e aplicá-la às suas
operações. A literatura sobre digital-era governance (Dunleavy et al.,
2006) consolidou a tese de que as transformações do setor público
dependem cada vez mais de reintegração, holismo orientado a necessidades
e digitalização. Estudos sobre implementação de governo eletrônico
(Heeks, 2003) mostram que falhas de adoção decorrem, com frequência, de
lacunas entre o desenho do sistema e a realidade organizacional
(design--reality gaps). O método aqui detalhado se posiciona como
intervenção nessa fronteira na área de Inteligência Artificial: trabalha
o repertório da equipe antes da introdução da ferramenta e mede o
resultado nos indicadores oficiais da unidade.

No plano pedagógico, a literatura recente sobre AI literacy (Long \&
Magerko, 2020) identifica as competências necessárias para que usuários
compreendam e se engajem criticamente com sistemas de IA, enquanto os
trabalhos sobre modelos generativos de linguagem e engenharia de prompts
(Brown et al., 2020; Wei et al., 2022) caracterizam padrões técnicos
como formulação de prompts, few-shot prompting e chain-of-thought
prompting. Essas contribuições, no entanto, tratam o usuário em abstrato
e não abordam o regime jurídico específico em que ele opera. O método
aqui descrito integra esses elementos técnicos a uma camada de
governança jurídica embutida desde a primeira aula, e mede o resultado
no setor público.

A contribuição diferencial do método, à luz desses três conjuntos de
literatura, é a integração simultânea dos três planos em uma única
arquitetura aplicada. Frameworks regulatórios, sem operacionalização,
permanecem como instrumento normativo; capacidade absortiva, sem método
pedagógico, permanece como diagnóstico; pesquisas de AI literacy, sem
ancoragem jurídica e institucional, permanecem como recomendação
técnica. O método abaixo descrito une os três e oferece evidência
empírica auditável do efeito dessa integração em duas unidades do
serviço público.

\hypertarget{ii.-diagnuxf3stico-sescont-2023}{%
\section{II. Diagnóstico: SES/CONT
(2023)}\label{ii.-diagnuxf3stico-sescont-2023}}

A Controladoria Setorial da Saúde do Distrito Federal (SES/CONT) foi
instituída pelo Decreto 39.546, de 19 de dezembro de 2018 (Regimento
Interno da Secretaria de Estado de Saúde, Art. 38). O Decreto 45.128, de
31 de outubro de 2023, criou no interior dessa estrutura a Assessoria de
Apoio aos Julgamentos de Processos Administrativos (ASJULG, Art. 39-B),
regulamentada pela Portaria 1.290/2023, SES/DF, posteriormente
alterada pela Portaria 1.166/2024. A unidade, portanto, operava dentro
de um arcabouço normativo formal, com competências regulamentadas para
produzir decisões, despachos, julgamentos e demais documentos técnicos
que sustentem a atuação correcional da Secretaria.

A primeira característica do panorama encontrado em 2023, e que define
todo o problema que viria a ser tratado pelo método, é o perfil da
equipe. Como a Secretaria de Estado de Saúde do Distrito Federal não
possui contratação específica para a função de análise correcional, com
formação jurídica, é comum que os servidores que compõem a assessoria
tenham formação na área da Saúde, e não em Direito. A equipe atua, no
entanto, na produção de peças cuja exigência é jurídica e
administrativa: fundamentação normativa, análise de admissibilidade,
interpretação de processos disciplinares, manejo de prazos e prescrição.
O descompasso entre a formação predominante da equipe e a natureza da
entrega esperada é o que torna a unidade um caso paradigmático do
gargalo de capacitação descrito na seção anterior.

O segundo elemento do panorama é o volume. Em 2023, a unidade tramitou
1.752 processos. O tempo médio de tramitação consolidado pelo SEI-GDF
foi de 17 dias, 22 horas e 11 minutos (17,92 dias). A produção
documental somou 4.846 documentos entre Decisões, Despachos e
Julgamentos, com 434 Ofícios e 28 Comunicados expedidos.

Em agosto de 2023, a unidade produziu 69 Julgamentos em um único mês.
Esse número, sozinho, representou 17,1\% de toda a produção anual de
Julgamentos. Não foi resultado de operação regular: tratava-se de um
lote de processos com risco iminente de prescrição, encaminhados
simultaneamente à Controladoria após atraso acumulado em outro setor.
Para absorver a demanda dentro do prazo legal, foi necessário mobilizar
uma força-tarefa emergencial. Diversos servidores trabalharam acima da
capacidade normal naquela semana, e parte da produção foi feita em dias
não úteis para evitar a perda de processos por prescrição. Nenhum
processo foi perdido naquele lote. O episódio, no entanto, deixou claro
o que o tempo médio agregado não dizia: a unidade operava no limite, e
qualquer pico de demanda externa empurrava o trabalho regular para o
feriado e dias de folga dos servidores.

Aos elementos anteriores somava-se uma restrição estrutural específica à
matéria tratada pela Controladoria. Processos disciplinares envolvem
dados sigilosos, identificação de pessoas, e informações cuja exposição
a sistemas externos é vedada por sigilo processual e pela Lei de
Proteção de Dados. Qualquer ferramenta de inteligência artificial a ser
introduzida na rotina precisaria ser capaz de auxiliar a equipe sem
expor dado protegido. Essa restrição, à primeira vista uma limitação,
viria a se mostrar elemento decisivo do desenho metodológico, e está
descrita em detalhe nos Seções III e IV.

O diagnóstico de 2023, portanto, não era o de uma unidade mal funcional.
Era o de uma unidade que entregava acima da própria capacidade
sustentável, com equipe disposta, mas pedagogicamente desassistida
quanto às ferramentas disponíveis, e dentro de um perímetro jurídico que
tornava soluções genéricas de mercado inviáveis. O método precisava
nascer dessa interseção.

\hypertarget{iii.-o-muxe9todo}{%
\section{III. O Método}\label{iii.-o-muxe9todo}}

O diagnóstico da seção anterior deixou três frentes simultâneas para
resolver: uma equipe sem repertório jurídico homogêneo precisando
produzir peças que exigem fundamentação técnica; uma operação cuja
capacidade efetiva era constantemente ultrapassada pela demanda; e um
perímetro de sigilo e proteção de dados que vedava o uso direto de
ferramentas externas. Não havia, no mercado brasileiro de capacitação em
IA naquele momento, programa que abordasse simultaneamente esses três
vetores. Cursos de prompt engineering ensinavam técnica sem contexto
jurídico; cursos de governança de IA discutiam frameworks regulatórios
sem aterrissar na rotina do servidor; cursos genéricos de ``IA para
gestores'' eram superficiais demais para sustentar produção documental
real.

A escolha foi montar o repertório à mão, integrando trilhas que
normalmente não conversam entre si: fundamentos técnicos de IA
generativa, engenharia de prompts, ética e risco, governança
regulatória, e IA aplicada. O percurso de estudo dirigido, ao longo de
pouco mais de um ano, articulou cursos e certificações de instituições
internacionais cobrindo quatro eixos: (i) base conceitual e gerencial em
IA aplicada a negócios e à administração pública; (ii) técnica
operacional de uso de modelos generativos e engenharia de prompts; (iii)
governança regulatória e ética da IA; e (iv) gestão de pessoas no
contexto da adoção tecnológica. Nenhuma trilha isolada bastava; o
repertório útil era a interseção.

Esse repertório, organizado pela exigência do problema e não pela
facilidade do conteúdo, precisava ser transferível a uma equipe que não
teria tempo nem mandato para refazer o mesmo percurso. Foi nessa
exigência de transferibilidade que nasceu a estrutura metodológica que
viria a sustentar tanto o curso oficial da Escola de Governo do Distrito
Federal quanto a operação interna das duas unidades: a \textbf{Casa da
IA}, método pedagógico em quatro camadas para adoção de Inteligência
Artificial no setor público.

\begin{figure}[H]
\centering
\includegraphics[width=0.85\linewidth]{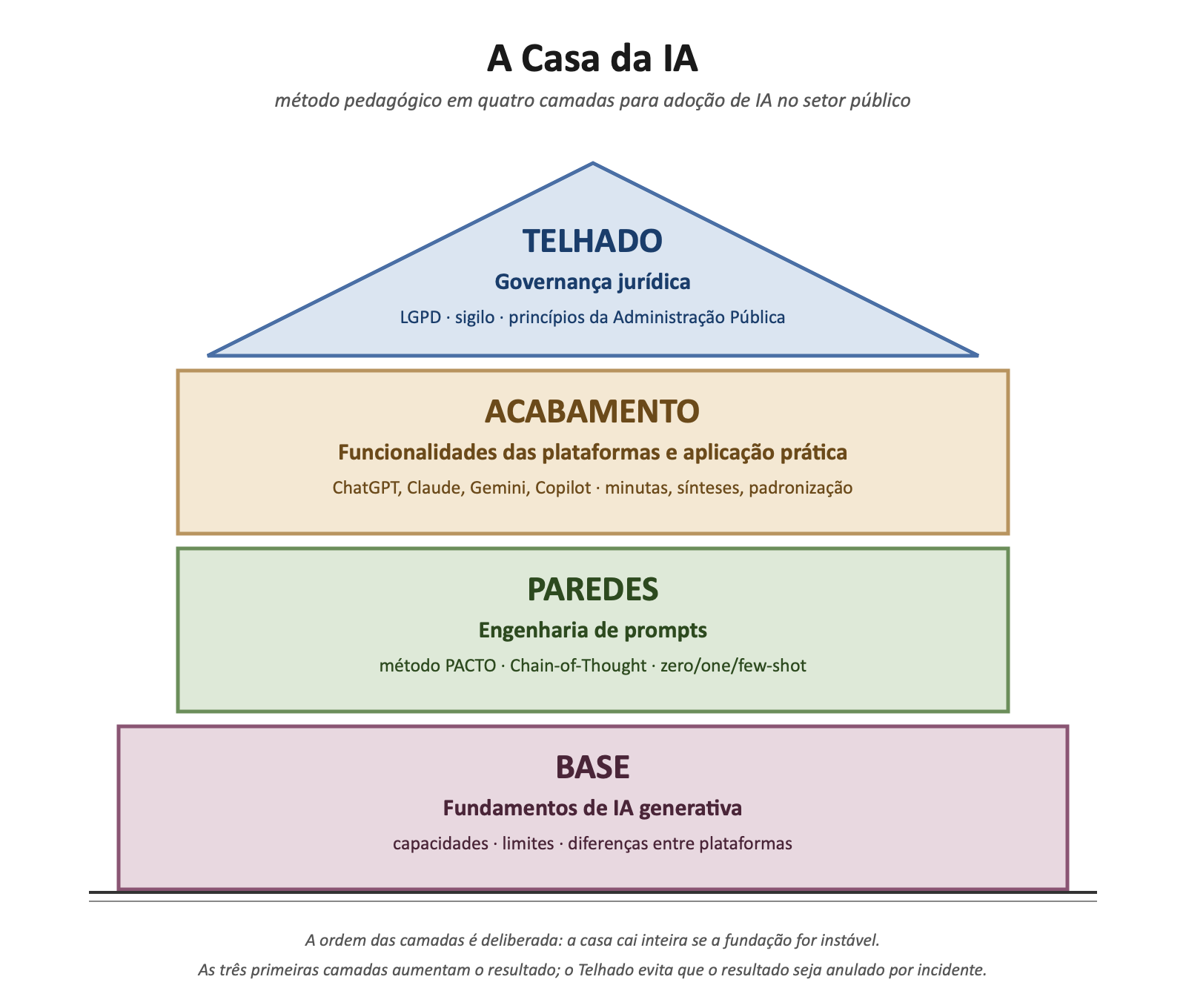}
\par\smallskip
\textit{Figura 1: A Casa da IA: estrutura das quatro camadas, com exemplos de conteúdo aplicado em cada uma.}
\end{figure}

Cada camada cumpre uma função no aprendizado do servidor, e pular
qualquer uma gera um modo específico de falha previsível. A
\textbf{Base} ensina o que é a inteligência artificial generativa e como
o modelo produz texto. É a fundação do modelo mental que o servidor
precisa ter sobre a ferramenta antes de operá-la. As \textbf{Paredes}
ensinam a técnica correta de construção do prompt: método PACTO
(Persona, Ação, Contexto, Tom, Observações, com Exemplo opcional), Chain
of Thought, e o uso adequado dos modos zero-shot, one-shot e few-shot.
Sem prompt bem estruturado, mesmo a funcionalidade mais avançada da
ferramenta deixa de produzir resultado utilizável. O \textbf{Acabamento}
traz as funcionalidades da IA aplicadas às atividades concretas do posto
de trabalho do servidor (estruturação de minutas, síntese de documentos
extensos, padronização de redação técnica, sumarização de
jurisprudência) e as diferenças funcionais entre as principais
plataformas disponíveis ao servidor. É nessa camada que cada
funcionalidade é amarrada ao tipo de documento que o servidor produz,
com identificação clara da parte que a IA pode preparar, da parte que
cabe ao servidor escrever, e do ponto em que a revisão humana é
obrigatória. O \textbf{Telhado} integra desde o início Lei de Proteção
de Dados, sigilo processual, princípios da Administração Pública,
controle externo, e os frameworks internacionais de governança de IA
aplicáveis ao setor público (NIST AI Risk Management Framework, GAO AI
Accountability Framework, EU AI Act, recomendações da OCDE).

A ordem das camadas é deliberada, e a casa cai inteira se a fundação for
instável. Sem Base, o servidor usa a ferramenta como motor de busca e
replica eventuais alucinações em documento oficial. Sem Paredes, escreve
prompts vagos e recebe respostas igualmente vagas, gastando tempo em
tentativa-e-erro sem chegar a resultado utilizável. Sem Acabamento, o
conhecimento fica abstrato e não desce até a tarefa concreta do
servidor, sabe-se o que a IA pode fazer em tese, mas não onde ela
encaixa no documento que precisa ser produzido até o final do dia. Sem
Telhado, as três camadas anteriores produzem incidentes de Lei de
Proteção de Dados, decisões administrativas mal fundamentadas, e
responsabilização da Administração e do servidor, exatamente o tipo de
adoção de IA que paralisa projetos públicos. As três primeiras camadas
aumentam o resultado; o Telhado é o que evita que o resultado seja
anulado por incidente.

Uma escolha de projeto precisa ser nomeada aqui. O método foi montado
para funcionar com as plataformas comerciais de IA disponíveis na versão
gratuita (ChatGPT, Claude, Gemini, Copilot, NotebookLM, Perplexity,
POE). Em nenhum momento o método pressupôs licenciamento corporativo,
plataforma customizada, ou infraestrutura de IA própria. A razão é
técnica, estratégica e motivacional ao mesmo tempo. Técnica porque a
evidência empírica apresentada nos Seções VI e X mostra que as
ferramentas gratuitas são suficientes para a maioria das tarefas
analíticas e redacionais do servidor médio, desde que ele saiba
operá-las dentro do Telhado de governança. Estratégica porque vincular a
metodologia a uma licença corporativa específica criaria uma barreira
inicial de adoção que excluiria a maior parte dos órgãos públicos,
especialmente aqueles cujo orçamento não comporta esse tipo de
aquisição. Motivacional porque resultados concretos obtidos com as
versões gratuitas oferecem aos gestores institucionais a base empírica
para justificar, quando e se necessário, a aquisição posterior das
versões corporativas pagas das mesmas plataformas. O método precisava
ser replicável em qualquer órgão público, independentemente da escala ou
da jurisdição. A acessibilidade do método foi parâmetro de projeto.

Há um elemento operacional nesse desenho que define boa parte da sua
praticidade. A \textbf{funcionalidade de IA personalizada}, disponível
nas principais plataformas comerciais como Projetos no ChatGPT e no
Claude, e Gems no Gemini, por exemplo, ainda nas versões gratuitas,
viabiliza o uso eficaz da IA generativa em órgãos públicos sem
necessidade de licenciamento corporativo ou plataforma própria.
Carregando à IA personalizada a base normativa e dos documentos corretos
pertinente à matéria da unidade, o servidor obtém respostas ancoradas
nessa base, com persona e tom configurados para o contexto
institucional. Em ambas as unidades estudadas neste paper, SES/CONT e
UCI/SEDET, a construção de IAs personalizadas foi instrumento
operacional-chave do método.

Com o repertório consolidado e a estrutura pedagógica desenhada, o passo
seguinte foi materializar o método em um curso oficial, aberto a
qualquer servidor do Governo do Distrito Federal. É o objeto da Seção
IV.

\hypertarget{iv.-curso-oficial-na-escola-de-governo}{%
\section{IV. Curso oficial na Escola de
Governo}\label{iv.-curso-oficial-na-escola-de-governo}}

A Escola de Governo do Distrito Federal (EGOV-DF) é a instituição
oficial de formação dos servidores do Governo do Distrito Federal.
Construir o método como curso da EGOV-DF, e não como treinamento interno
teve três efeitos diretos. O primeiro, institucional: o curso passa a
integrar o catálogo oficial de formação do GDF, com mandato pedagógico
formal, e os servidores que o concluem recebem certificação de
instituição reconhecida. O segundo, escalar: a turma deixa de ser
limitada ao quadro da Unidade, e qualquer servidor do GDF, de qualquer
Secretaria, pode se matricular, abrindo a formação a um público muito
mais amplo do que o de uma única unidade. O terceiro, evidencial: ao
aplicar o curso na própria Unidade depois de ele já existir
externamente, a equipe se torna campo de prova de algo que existe
independentemente dela, e os resultados de produtividade ganham validade
interna que treinamento construído ad-hoc dentro da unidade jamais
teria.

O curso foi estruturado em \textbf{cinco aulas presenciais, totalizando
20 horas de instrução} (quatro horas por aula), sob o título oficial
\emph{Inteligência Artificial no setor público: técnicas, riscos e
aplicações}. A carga horária de 20 horas foi uma decisão deliberada:
cursos mais longos competem com a rotina operacional do servidor,
dispersam o foco e elevam a evasão antes da conclusão. A experiência com
turmas em formatos de 12, 16 e 20 horas mostrou que esse intervalo é o
mais efetivo, justamente por não competir com a rotina do servidor e por
preservar a profundidade pedagógica das aulas. A proposta foi
protocolada pelo autor deste paper junto à Escola de Governo (Memorando
Nº 4/2024 da SES/CONT/ASJULG) e aprovada pela Escola de Governo
(Memorando Nº 166/2024 da SEEC/SEGEA/EGOV, processo SEI
04044-00021535/2024-26). A modalidade foi presencial e destinada a
agentes públicos dos órgãos e entidades da Administração Direta e
Indireta do Distrito Federal, nas carreiras civis ou militares.

\begin{longtable}[]{@{}
  >{\raggedright\arraybackslash}p{(\linewidth - 4\tabcolsep) * \real{0.1007}}
  >{\raggedright\arraybackslash}p{(\linewidth - 4\tabcolsep) * \real{0.7987}}
  >{\raggedright\arraybackslash}p{(\linewidth - 4\tabcolsep) * \real{0.1007}}@{}}
\toprule
Aula & Tema & Carga\tabularnewline
\midrule
\endhead
1 & Introdução e Aplicações da Inteligência Artificial &
4h\tabularnewline
2 & Engenharia de Prompts para Inteligência Artificial: Técnicas e
Abordagens & 4h\tabularnewline
3 & Plataformas de Inteligência Artificial & 4h\tabularnewline
4 & Limitações, Riscos e Questões Éticas da Inteligência Artificial &
4h\tabularnewline
5 & Inteligência Artificial no Serviço Público & 4h\tabularnewline
\textbf{Total} & \textbf{20h} &\tabularnewline
\bottomrule
\end{longtable}

A correspondência entre as aulas e a estrutura do método é direta: a
Aula 1 sustenta a Base; a Aula 2 forma as Paredes; as Aulas 3 e 5
compõem o Acabamento; e a Aula 4 concentra o Telhado, conforme detalhado
na Seção III.

A modalidade combina aulas expositivas em sala convencional com aulas
práticas em laboratório de informática, permitindo que o servidor saia
da formação tendo já operado as ferramentas em exercícios e projetos
diretamente aplicáveis à sua rotina.

Para o caso específico da Controladoria Setorial da Saúde, foi
construída uma extensão prática do método que merece registro separado:
uma IA personalizada, com base de conhecimento composta por dez
normativos jurídicos diretamente aplicáveis à unidade: Decreto
39.701/2019 (procedimentos disciplinares), Instruções Normativas 1/2021
(Termo de Ajustamento de Conduta), 2/2021 (Juízo de Admissibilidade e
Procedimento de Investigação Preliminar) e 2/2016 (Mediação), Lei
4.938/2012 (SICOR, Sistema de Correição do Distrito Federal), Lei
Complementar 840/2011 (Regime Jurídico do servidor do DF), Lei Orgânica
do DF, Decreto 37.297/2016 (Código de Ética), e os Manuais Teórico e
Prático de Procedimentos Disciplinares da Controladoria-Geral do DF. A
IA personalizada operava como assistente jurídico interno da equipe:
respondia dúvidas sobre legislação, sugeria estruturas para
fundamentações, e apontava precedentes administrativos relevantes,
sempre a partir da base documental carregada, sem necessidade de
consulta a sistemas externos com dados do caso concreto.

Essa arquitetura resolve a restrição estrutural identificada na Seção
II. As consultas à IA personalizada eram realizadas de forma
despersonalizada: o servidor descrevia a dúvida normativa sem informar
nomes, números de processo ou qualquer dado identificável da situação
real. A IA respondia a partir do conteúdo jurídico carregado.

Três considerações técnico-jurídicas sustentam o desenho do método.
Primeira: o tratamento de dados pelo Protocolo de Anonimização de Dados
(PAD), adotado antes de qualquer envio à plataforma, faz com que o
conteúdo encaminhado deixe de ser dado pessoal nos termos do art. 12 da
Legislação de Proteção de Dados (Lei nº 13.709/2018), que retira do
escopo da Lei os dados anonimizados. Segunda: para mitigar o risco
residual nas versões gratuitas das plataformas, o protocolo orienta
sobre a desativação das funcionalidades de autorização de treinamento
com as conversas e o uso dos modos de conversa momentânea ou temporária
dos fornecedores, que impedem o uso dos dados enviados para treinamento
dos modelos e estabelecem janela limitada de retenção de logs. Terceira:
a responsabilidade administrativa pelo documento expedido permanece
integralmente do servidor que o assina, conforme princípio
Human-in-the-Loop adotado em ambas as unidades e formalizado no
Framework de Governança de IA da UCI (Doc. SEI 194251158).

O último elemento da modelagem do curso, presente desde a primeira aula,
é o princípio da \textbf{supervisão humana integral}. Todo conteúdo
gerado por IA passa por revisão técnica do criador do documento e
validação do chefe da unidade antes de qualquer formalização. A IA é
assistente; o servidor permanece responsável. Sob esse princípio,
durante todo o ciclo da
Controladoria, nenhuma decisão administrativa, julgamento ou nota
técnica foi expedida sem revisão humana integral. O relatório institucional informou que não foram relatados problemas relacionados à IA no período observado.

Com o curso desenhado, o catálogo da EGOV-DF receptivo à oferta, e a IA
personalizada construída para o ambiente jurídico da Controladoria,
faltava o passo que viria a tornar tudo isso mensurável: aplicar o
método aos servidores que efetivamente produzem os documentos da
unidade. É o objeto da Seção V.

\hypertarget{v.-aplicauxe7uxe3o-na-sescont}{%
\section{V. Aplicação na SES/CONT}\label{v.-aplicauxe7uxe3o-na-sescont}}

O curso \emph{Inteligência Artificial no setor público: técnicas, riscos
e aplicações} foi aberto a qualquer servidor do Distrito Federal. Os
servidores da Controladoria Setorial da Saúde se matricularam como
alunos do curso, no mesmo curso oferecido a colegas de outras
Secretarias, sob a instrutoria oficial da EGOV-DF, e com certificação ao
final. Os servidores que viriam a operar a IA na unidade aprenderam o
método junto com servidores de outras pastas, em ambiente pedagógico
igualitário e auditável.

A operação interna foi montada com cinco elementos, todos provenientes
das aulas do curso da EGOV-DF e traduzidos para a rotina da unidade na
sequência da formação. O primeiro, fundamentos de IA generativa e seus
limites, garantiu que cada servidor entendesse o que a ferramenta podia
e o que não podia entregar. O segundo, técnicas de engenharia de prompt
aplicadas à redação de documentos oficiais, transformou prompts
genéricos em comandos com persona, ação, contexto, tom, observações e
exemplo quando necessário, conforme o método PACTO já descrito. O
terceiro, uso da IA personalizada com a base legal da Controladoria. O
quarto, protocolos de segurança e proteção de dados, foi treinado como
reflexo: nenhum dado identificável, nenhum número de processo, nenhuma
matéria sigilosa entra na ferramenta sem despersonalização prévia. O
quinto, revisão crítica e validação humana, transformou a IA em
assistente formal: todo documento gerado passava por revisão técnica do
servidor responsável e validação do chefe da unidade antes de qualquer
formalização.

Na prática diária, o método se materializou em padrões claros de uso.
Para a elaboração de uma minuta de Decisão, o servidor formulava a
dúvida normativa de forma despersonalizada na IA personalizada, obtinha
o esboço de fundamentação a partir da base legal da unidade, e então
retomava o texto integrando os dados específicos do processo concreto
fora da ferramenta. Em nenhum momento a IA era a autora do documento.
Era assistente operacional que ampliava a velocidade e a consistência do
trabalho humano.

A unidade operou nesse regime ao longo do ano de 2024. O que aconteceu
com os indicadores operacionais durante esse ano, comparados aos do ano
anterior, é o objeto da Seção VI.

\hypertarget{vi.-resultados-sescont-20232024}{%
\section{VI. Resultados: SES/CONT
(2023→2024)}\label{vi.-resultados-sescont-20232024}}

O ano de 2024 foi o primeiro ano da unidade operando dentro do método.
Os números abaixo, integralmente extraídos das estatísticas oficiais do
SEI-GDF e consolidados no Relatório Gerencial Executivo da SES/CONT
(Doc. SEI 197403428, processo 00060-00582291/2024-11), comparam o
desempenho de 2024 com o de 2023.

O indicador central da unidade, o tempo médio de tramitação, caiu de 17
dias, 22 horas e 11 minutos (17,92 dias) em 2023 para 14 dias, 15 horas
e 44 minutos (14,66 dias) em 2024. A redução foi de 3,26 dias por
processo, ou 18,2\% no agregado.

\begin{figure}[H]
\centering
\includegraphics[width=0.80\linewidth,height=7cm,keepaspectratio]{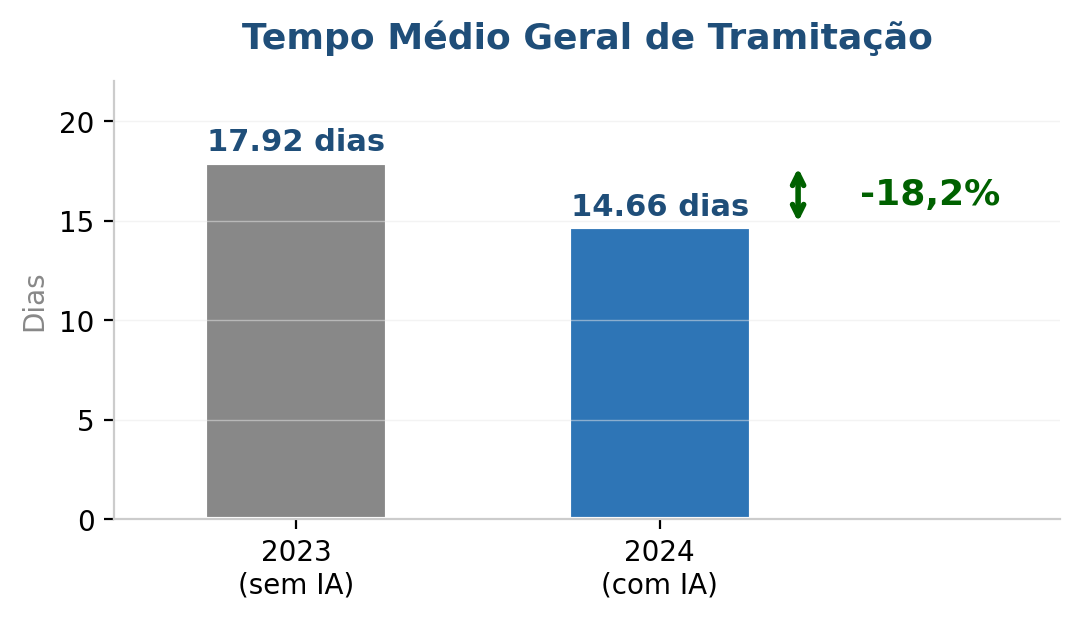}
\par\smallskip
\textit{Figura 2: Tempo médio geral de tramitação SES/CONT: 2023 vs. 2024.}
\end{figure}

\begin{longtable}[]{@{}
  >{\raggedright\arraybackslash}p{(\linewidth - 6\tabcolsep) * \real{0.5879}}
  >{\raggedright\arraybackslash}p{(\linewidth - 6\tabcolsep) * \real{0.1241}}
  >{\raggedright\arraybackslash}p{(\linewidth - 6\tabcolsep) * \real{0.1448}}
  >{\raggedright\arraybackslash}p{(\linewidth - 6\tabcolsep) * \real{0.1433}}@{}}
\toprule
Indicador & 2023 & 2024 (com IA) & Variação\tabularnewline
\midrule
\endhead
Tempo médio de tramitação dos processos da Unidade & 17,92 dias & 14,66
dias & \textbf{−18,2\%}\tabularnewline
D+D+J (Decisões + Despachos + Julgamentos) produzidos por processo &
2,77 & 2,78 & +0,4\%\tabularnewline
\bottomrule
\end{longtable}

A produção documental por processo permaneceu estável (2,78 contra 2,77
em 2023), o que significa que a unidade não cortou entregas para
entregar mais rápido. Cada processo continuou gerando aproximadamente o
mesmo número de Decisões, Despachos e Julgamentos, em 18,2\% menos
tempo.

A redução de tempo não se distribuiu uniformemente. Ela se concentrou
exatamente nos tipos de processo que exigem mais análise técnica,
fundamentação normativa e redação estruturada, ou seja, as atividades em
que o método efetivamente atua.

\begin{figure}[H]
\centering
\includegraphics[width=0.80\linewidth,height=7cm,keepaspectratio]{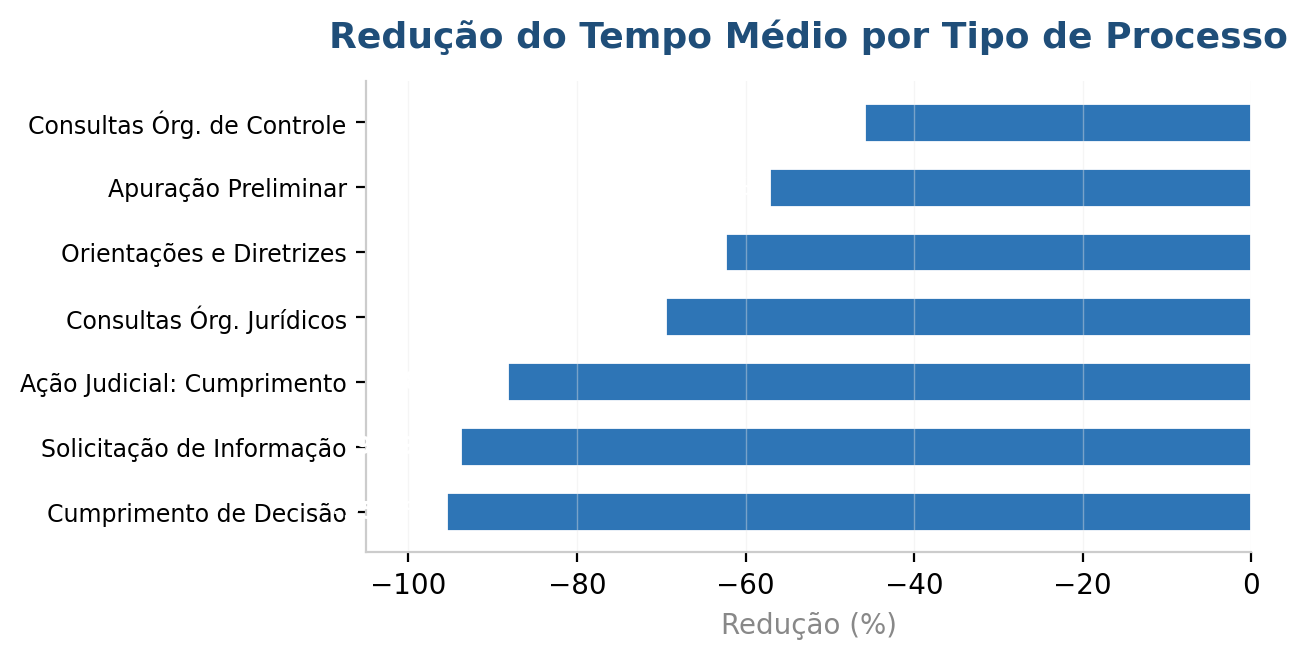}
\par\smallskip
\textit{Figura 3: Redução do tempo médio por tipo de processo (SES/CONT)}
\end{figure}

\begin{longtable}[]{@{}
  >{\raggedright\arraybackslash}p{(\linewidth - 6\tabcolsep) * \real{0.5629}}
  >{\raggedright\arraybackslash}p{(\linewidth - 6\tabcolsep) * \real{0.1554}}
  >{\raggedright\arraybackslash}p{(\linewidth - 6\tabcolsep) * \real{0.1344}}
  >{\raggedright\arraybackslash}p{(\linewidth - 6\tabcolsep) * \real{0.1473}}@{}}
\toprule
Tipo de Processo & 2023 & 2024 & Redução\tabularnewline
\midrule
\endhead
Cumprimento de Decisão & 17,82 dias & 0,81 dias &
\textbf{−95,5\%}\tabularnewline
Solicitação de Informação (Controle Interno) & 22,81 dias & 1,38 dias &
\textbf{−93,9\%}\tabularnewline
Ação Judicial: Cumprimento & 24,96 dias & 2,92 dias &
\textbf{−88,3\%}\tabularnewline
Consultas Órgãos Jurídicos & 9,27 dias & 2,82 dias &
−69,6\%\tabularnewline
Orientações e Diretrizes & 15,74 dias & 5,92 dias &
−62,4\%\tabularnewline
Apuração Preliminar & 7,39 dias & 3,16 dias & −57,2\%\tabularnewline
Consultas Órgãos de Controle & 12,96 dias & 7,00 dias &
−46,0\%\tabularnewline
\bottomrule
\end{longtable}

Quando se agregam esses ganhos de tempo aos 1.581 processos efetivamente
tramitados em 2024, surge um conceito útil para dimensionar o ganho de tempo: capacidade liberada estimada. Se a unidade tivesse operado com a
produtividade de 2023, os 1.581 processos teriam consumido 28.332 dias
de tramitação acumulada (1.581 × 17,92). Com o ganho de 2024, o consumo
real foi de 23.177 dias (1.581 × 14,66). A diferença, 5.155 dias,
equivale à tramitação de cerca de 352 processos adicionais, um aumento
de 22,3\% na capacidade operacional da unidade.

\begin{figure}[H]
\centering
\includegraphics[width=0.80\linewidth,height=7cm,keepaspectratio]{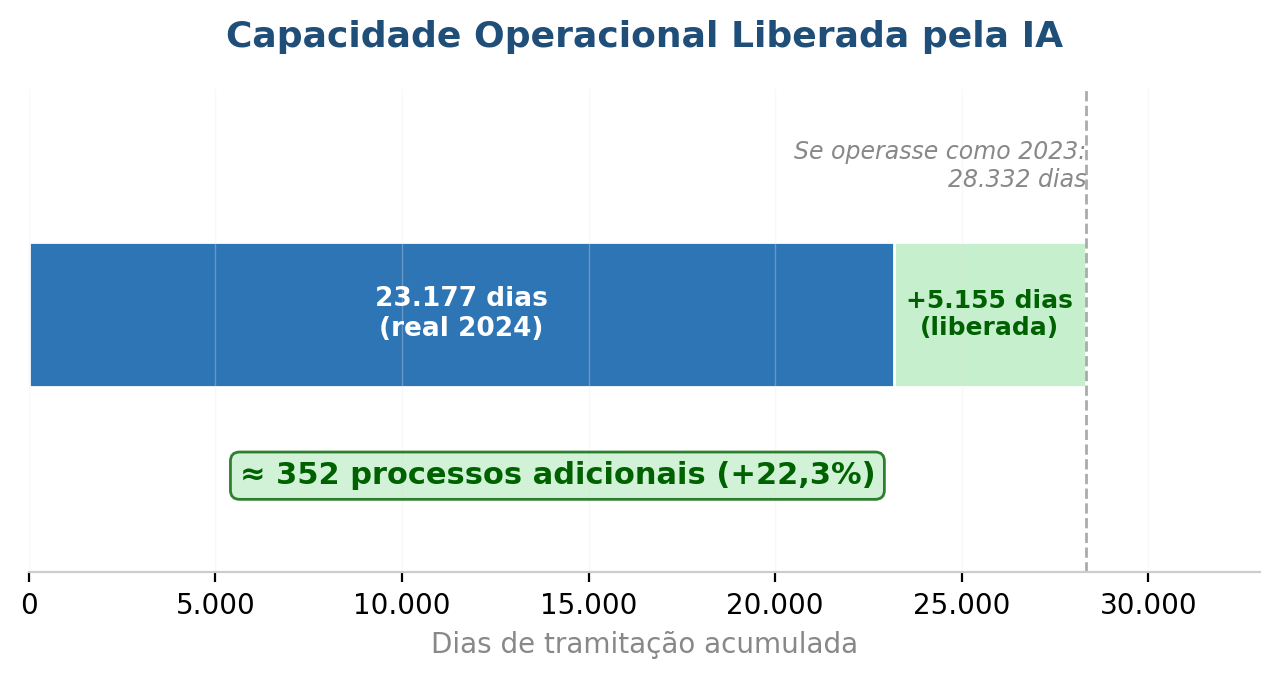}
\par\smallskip
\textit{Figura 4: Capacidade liberada estimada na SES/CONT.}
\end{figure}

Essa capacidade liberada se materializou em três frentes. A primeira, e
mais visível, foi o saneamento do passivo represado: ao final de 2024, a
Controladoria não tinha nenhum processo com mais de 10 dias na caixa
aguardando andamento. A unidade passou a operar em regime de fluxo
contínuo. O Relatório Nº 12/2024 da SES/CONT/ASJULG registra
textualmente que ``as minutas de decisão são elaboradas em até 10 dias,
evitando represamentos''.

A segunda frente foi a estabilização da produção de Julgamentos. Em
2023, a média mensal real, descontado o mês atípico descrito na Seção
II, foi de 30,36 julgamentos por mês. Em 2024, a média foi de 31,00
julgamentos por mês, com a vantagem decisiva de que o número se
distribuiu de forma muito mais regular: a amplitude mensal caiu de 56
julgamentos (entre o mínimo e o máximo de 2023) para 25 (em 2024), uma
redução de 55,4\%. A produção mensal ficou bem mais estável sob o método: mesmo volume médio, distribuição muito mais previsível, sem
necessidade de operação emergencial em finais de semana e feriados.

\begin{figure}[H]
\centering
\includegraphics[width=0.80\linewidth,height=7cm,keepaspectratio]{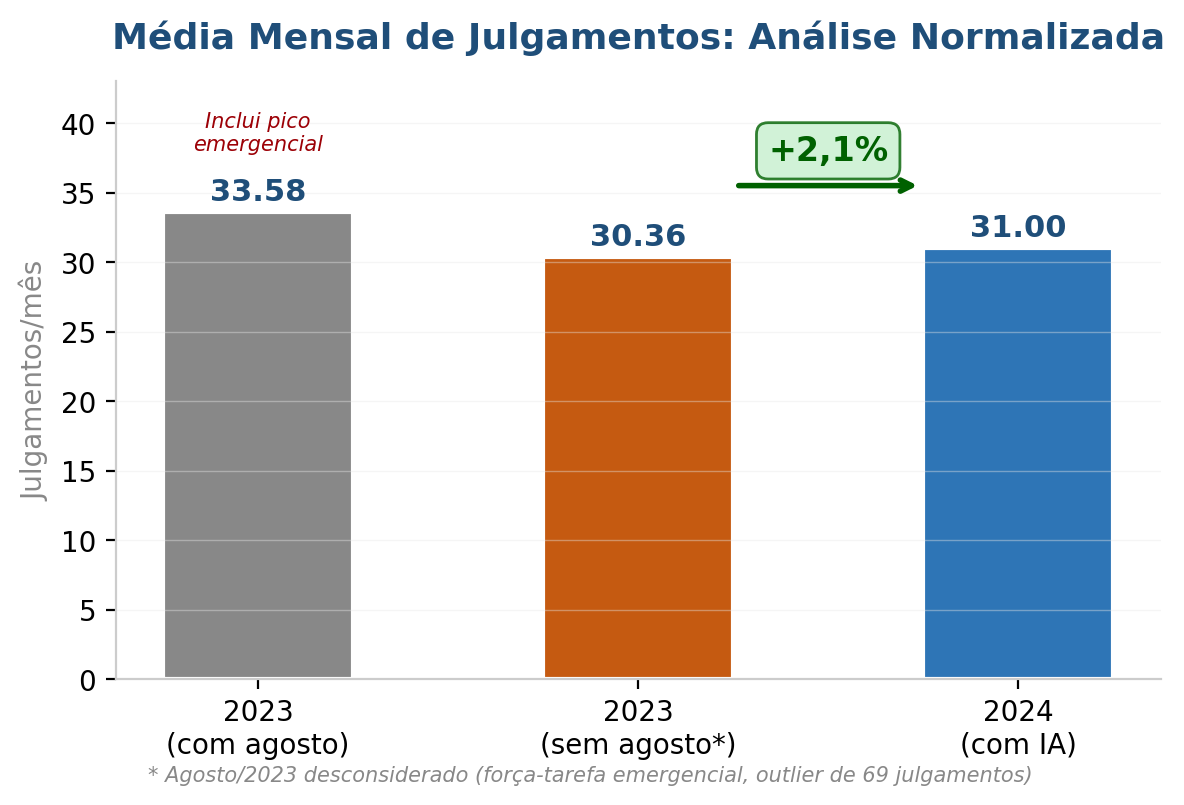}
\par\smallskip
\textit{Figura 5: Média mensal de Julgamentos: análise normalizada (SES/CONT).}
\end{figure}

\begin{figure}[H]
\centering
\includegraphics[width=0.80\linewidth,height=7cm,keepaspectratio]{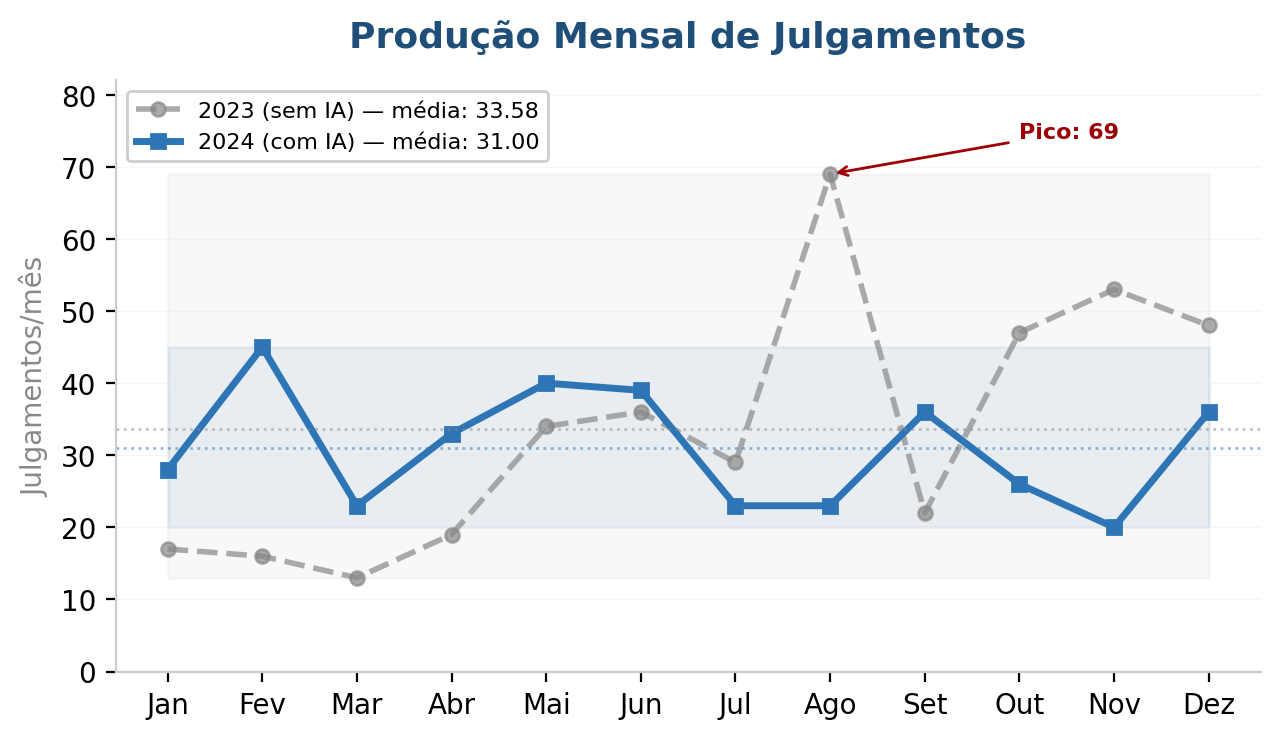}
\par\smallskip
\textit{Figura 6: Produção mensal de Julgamentos (SES/CONT, 2023 vs. 2024).}
\end{figure}

A terceira frente foi a expansão da capacidade de articulação
institucional. A produção de Ofícios cresceu de 434 para 532 (+22,6\%),
e a de Comunicados, que são as recomendações corretivas dirigidas aos
setores responsáveis por problemas identificados em processos
disciplinares, saltou de 28 para 214, um crescimento de 664\%. Esse
último número marca uma mudança qualitativa de postura da unidade: a
Controladoria deixou de operar apenas com a lógica punitiva (instaurar
processo, julgar, aplicar sanção) e passou a operar também com lógica
preventiva e corretiva, sinalizando aos setores os fluxos e práticas que
originaram as irregularidades, para que possam ser saneados antes do
próximo incidente.

\begin{figure}[H]
\centering
\includegraphics[width=0.80\linewidth,height=7cm,keepaspectratio]{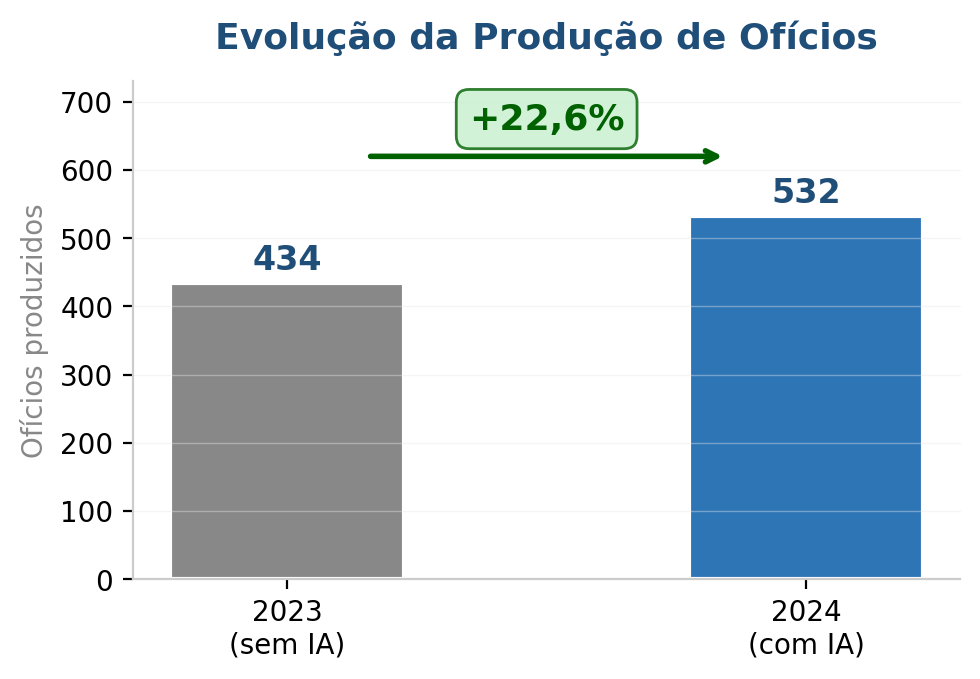}
\par\smallskip
\textit{Figura 7: Evolução da produção de Ofícios (SES/CONT).}
\end{figure}

\begin{figure}[H]
\centering
\includegraphics[width=0.80\linewidth,height=7cm,keepaspectratio]{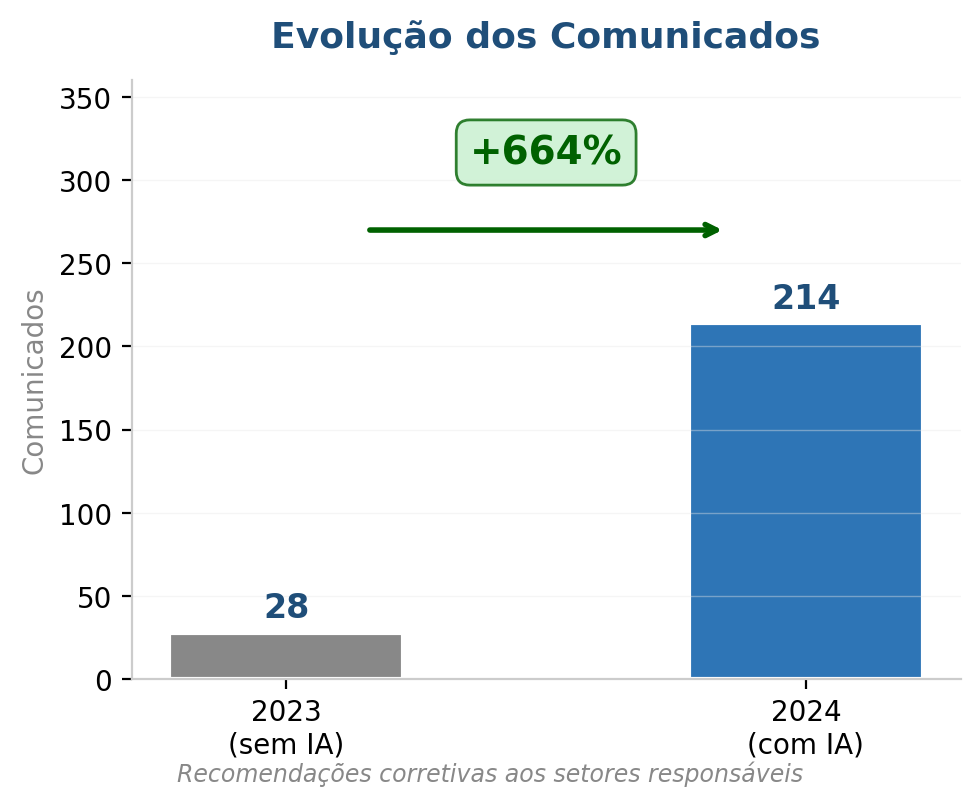}
\par\smallskip
\textit{Figura 8: Evolução dos Comunicados (SES/CONT): virada de postura punitiva para preventiva.}
\end{figure}

O balanço de 2024 pode ser sintetizado em três afirmações verificáveis
pelo SEI-GDF: a unidade ficou 18,2\% mais rápida sem cortar entregas;
ganhou capacidade equivalente a 352 processos adicionais; e mudou o
próprio modo como atua, expandindo o trabalho preventivo em mais de sete
vezes. O relatório institucional informou que não foram relatados problemas relacionados à IA no período.

\begin{figure}[H]
\centering
\includegraphics[width=0.80\linewidth,height=7cm,keepaspectratio]{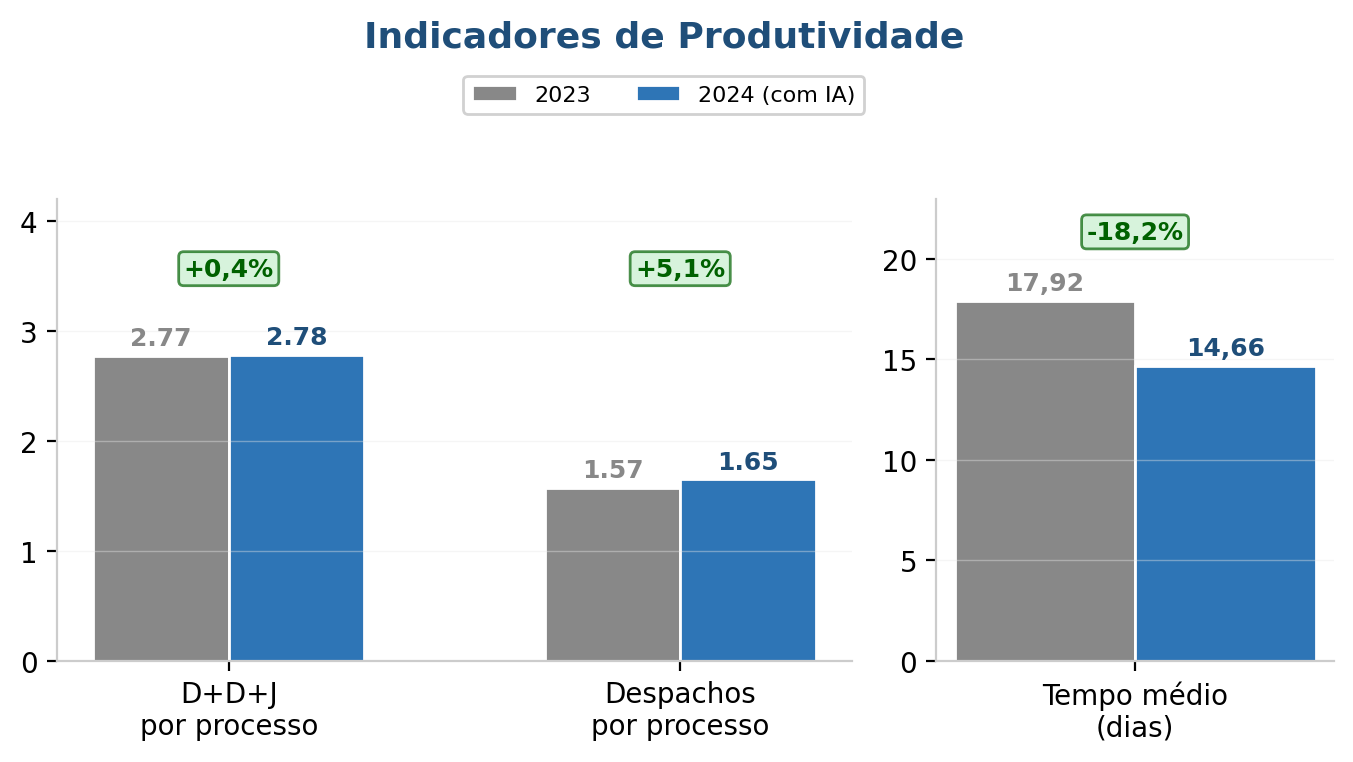}
\par\smallskip
\textit{Figura 9: Indicadores de produtividade (SES/CONT, 2023 vs. 2024).}
\end{figure}

A Seção VII muda de unidade e formula a pergunta que o segundo bloco do
paper precisa responder: o método sobrevive à aplicação em outro órgão,
com outra equipe e outra temática?

\hypertarget{vii.-transferuxeancia-para-ucisedet}{%
\section{VII. Transferência para
UCI/SEDET}\label{vii.-transferuxeancia-para-ucisedet}}

Os resultados auditáveis da seção anterior tornam concreta uma
metodologia que, até esse ponto, podia ainda ser questionada como caso
isolado. A pergunta que sobra, e que define o segundo bloco deste paper,
é se o que funcionou na Controladoria Setorial da Saúde funciona também
em outro órgão, com outra equipe, e com outro tipo de trabalho.

Essa pergunta deixou de ser hipotética entre 2024 e 2025, quando o autor
foi designado Assessor Especial da Unidade de Controle Interno da
Secretaria de Estado de Desenvolvimento Econômico, Trabalho e Renda do
Distrito Federal (UCI/SEDET-DF). A mudança foi uma transição funcional
dentro do Governo e criou exatamente o teste de replicabilidade que
faltava para a metodologia.

Três variáveis mudaram simultaneamente. A primeira foi o órgão: saída de
uma Secretaria com mandato finalístico de saúde pública e entrada em
outra com mandato finalístico de desenvolvimento econômico, trabalho e
renda. Cada Secretaria tem cultura administrativa própria, fluxos
internos próprios, e dinâmicas de relacionamento com os órgãos de
controle externo que não se transferem automaticamente.

A segunda variável foi a temática processual. Na SES/CONT, o eixo era
controle correcional: processos disciplinares, sindicâncias, julgamentos
administrativos, com base normativa centrada no regime jurídico do
servidor e nos procedimentos disciplinares. Na UCI/SEDET, o eixo é
controle interno geral: análise de contratações, auditoria de execução
financeira, fiscalização de programas e convênios, com base normativa
centrada na Lei de Licitações e nos princípios constitucionais da
Administração Pública. Mudam juntos, com a temática, o tipo de documento
que a IA precisa apoiar, a complexidade analítica esperada de cada peça,
e o nível de risco institucional embutido em cada erro.

A terceira variável foi a equipe. O método foi originalmente instalado
em uma equipe com formação predominante em Saúde. Na UCI/SEDET, a equipe
era outra, com trajetória diferente, lotada em outra Secretaria, com
vínculo institucional construído com outra unidade. O método precisava
ser apresentado a essa equipe nova e produzir, em prazo razoável, ganhos
comparáveis aos medidos no órgão anterior.

O que permaneceu constante foi o método. O método, com as quatro camadas
descritas na Seção III e com os princípios operacionais consolidados na
Seção IV, foi transferido sem reformulação estrutural. As adaptações
específicas que a UCI demandou serão descritas na Seção IX. A fundação,
a sequência pedagógica, o uso de versões gratuitas das ferramentas, e a
supervisão humana integral permaneceram intactos.

Se o método fosse, na verdade, um conjunto de soluções desenhadas sob
medida para um único caso, a transição revelaria isso rapidamente: a
Saúde e o Desenvolvimento Econômico têm pouco em comum no detalhe
operacional. Se, pelo contrário, o método fosse de fato replicável entre
órgãos com perfis distintos, então a UCI/SEDET teria que produzir, em
2025, resultados comparáveis aos da SES/CONT em 2024, ainda que com
perfil próprio. As Seções VIII, IX e X respondem a essa pergunta.

\hypertarget{viii.-diagnuxf3stico-ucisedet-2024}{%
\section{VIII. Diagnóstico: UCI/SEDET
(2024)}\label{viii.-diagnuxf3stico-ucisedet-2024}}

A Unidade de Controle Interno da Secretaria de Estado de Desenvolvimento
Econômico, Trabalho e Renda do Governo do Distrito Federal
(UCI/SEDET-DF) está vinculada diretamente ao Gabinete da Secretaria. Sua
função é o controle interno preventivo e concomitante: examina processos
administrativos relevantes da pasta, em especial aqueles com impacto
financeiro material, e produz notas técnicas com análise de
conformidade, identificação de risco e recomendações dirigidas aos
gestores responsáveis.

Há um traço institucional dessa unidade que precisa ser registrado antes
dos números, porque ele muda o que conta como resultado. Pelo direito
administrativo brasileiro, as recomendações expedidas por uma Unidade de
Controle Interno não são vinculantes: o gestor público pode acatá-las ou
não. Isso significa que o ativo principal do trabalho da UCI não é a
quantidade de recomendações expedidas, e sim a qualidade técnica de cada
uma. Quanto mais defensável a peça, em rigor probatório e fundamentação
jurídica, maior a probabilidade de que o gestor acate, e maior a função
preventiva que a unidade efetivamente exerce sobre a pasta.

Em 2024, com base nos dados oficiais do SEI-GDF consolidados no
Relatório Nº 1/2026 da SEDET/GAB/UCI (Doc. SEI 193384267, processo
04035-00000853/2026-24), esses são os indicadores agregados do trabalho
da unidade.

\begin{longtable}[]{@{}
  >{\raggedright\arraybackslash}p{(\linewidth - 2\tabcolsep) * \real{0.6666}}
  >{\raggedright\arraybackslash}p{(\linewidth - 2\tabcolsep) * \real{0.3334}}@{}}
\toprule
Indicador & 2024 (sem IA)\tabularnewline
\midrule
\endhead
Tempo médio de tramitação & 34 dias\tabularnewline
Processos tramitados & 256\tabularnewline
Documentos produzidos & 251\tabularnewline
Notas técnicas elaboradas & 66\tabularnewline
\bottomrule
\end{longtable}

Os 34 dias de tempo médio expressam a duração interna do trâmite na
unidade, do recebimento de um processo até a produção da peça técnica de
saída. Em uma unidade cujo principal entregável é a nota técnica, e cujo
trabalho recai sobre processos com volume financeiro relevante, esse
tempo se reflete diretamente na cadência de contratações, pagamentos e
decisões administrativas da Secretaria.

A pergunta natural, e que orientou a chegada do método ao longo de 2025,
era se o método de aplicação de IA poderia, na UCI, simultaneamente
aumentar o volume e a profundidade das notas técnicas, sem comprometer o
tempo de resposta nem o rigor exigido pela natureza do trabalho. A
resposta começou com a apresentação do mesmo curso oficial da EGOV-DF à
equipe da Unidade, no mesmo formato em que os servidores da
Controladoria Setorial da Saúde haviam sido capacitados, e prossegue com
as adaptações de método descritas na Seção IX.

\hypertarget{ix.-adaptauxe7uxf5es-do-muxe9todo-na-ucisedet}{%
\section{IX. Adaptações do método na
UCI/SEDET}\label{ix.-adaptauxe7uxf5es-do-muxe9todo-na-ucisedet}}

A apresentação do método à equipe da UCI/SEDET-DF seguiu o mesmo caminho
institucional adotado na Controladoria Setorial da Saúde: os servidores
se matricularam no curso oficial \emph{Inteligência Artificial} da
Escola de Governo do Distrito Federal, ministrado pelo autor, com o
mesmo conteúdo programático. A continuidade da fonte de capacitação é
registro relevante porque elimina, para os efeitos da comparação
metodológica que orienta este paper, a variável ``qualidade do
treinamento'' como possível explicação concorrente para os resultados. A
formação técnica oferecida foi a mesma; o que mudou foram o órgão, a
temática e a equipe.

A passagem das aulas para a operação real, no entanto, exigiu adaptações
específicas. A UCI não trabalha com processos disciplinares; trabalha
com fiscalização de execução financeira, análise de contratações e
auditoria de programas. O documento central da unidade, a nota técnica,
tem estrutura, complexidade e ritmo distintos da minuta de Decisão da
Controladoria Setorial da Saúde. As adaptações se distribuíram em três
frentes.

A primeira frente foi a curadoria das aplicações em que a IA seria
efetivamente usada. A estratégia de adoção priorizou casos de alto
impacto e baixo risco: atividades manuais e repetitivas que
historicamente consumiam tempo significativo da equipe técnica, mas que
não envolviam decisão crítica automatizada. O objetivo foi deixar a
equipe técnica livre para o que exige julgamento: análise crítica,
avaliação do caso concreto, tomada de decisão, com a IA cobrindo o
processamento de informações e a estruturação preliminar dos documentos.
As três aplicações que se consolidaram como rotina foram o auxílio na
elaboração de minutas de Notas Técnicas, a síntese de documentos
extensos, e a revisão gramatical e padronização de textos.

A partir de janeiro de 2025, a operação interna foi reforçada com a
construção de \textbf{IAs personalizadas próprias para a UCI},
inicialmente na plataforma Gemini (Gems) e posteriormente migradas para
o Claude (Projects). A estratégia foi modular: para cada matéria
recorrente nos processos analisados pela unidade, Termos de Fomento sob
a Lei nº 13.019/2014 (MROSC), Licitações sob a Lei nº 14.133/2021,
Tomadas de Contas Especiais (TCE), Despesas de Exercícios Anteriores
(DEA), entre outras, foi criada uma IA personalizada específica, com
base de conhecimento composta pela legislação federal, decretos
regulamentadores e normas de controle aplicáveis. Para a IA
personalizada de Licitações, por exemplo, a base reúne a Lei nº
14.133/2021, o Decreto nº 32.598/2010-DF, o Decreto nº 44.330/2023-DF, o
Decreto nº 45.933/2024-DF e a Portaria nº 29/2021-CGDF. O ferramental
permitia confecção de minutas de notas técnicas, conferência de
aderência à legislação aplicável e padronização da estrutura dos
pareceres, sempre dentro dos protocolos de anonimização formalizados no
Framework de Governança de IA da UCI.

A segunda frente foi a documentação formal do uso da IA na unidade. Foi
elaborado um \textbf{Framework de Governança de IA específico para a
UCI} (Doc. SEI 194251158), com orientações operacionais sobre o que pode
e o que não pode ser submetido a ferramentas de IA, em que momentos a
revisão humana é obrigatória, e quais protocolos de anonimização de
dados devem ser observados antes de qualquer interação. O Framework
traduz princípios gerais de governança em regras operacionais aderentes
à matéria de controle interno geral, à Lei de Proteção de Dados, e aos
princípios da Administração Pública.

A terceira frente foi a adoção incremental. A introdução da IA na rotina
não se deu por bloco único, e sim em ondas: começou pelas tarefas mais
simples e estáveis, avançou para atividades de maior complexidade
conforme a equipe ganhava confiança e domínio, e foi consolidando
padrões internos à medida que cada bom resultado virava referência para
a próxima aplicação.

Quatro aprendizados atravessaram 2025 e vale registrá-los, porque
separam um método maduro de uma adoção entusiasmada. O primeiro é que a
qualidade do resultado depende diretamente da qualidade do comando:
prompts bem estruturados, com contexto adequado e instruções claras,
acompanhados de notas técnicas-exemplo, produzem resultados
significativamente melhores. O segundo é que a IA acelera, mas não
substitui a necessidade de conhecimento técnico nem de avaliação humana
qualificada. A revisão crítica por profissional habilitado permaneceu
indispensável para garantir precisão e adequação das análises. O
terceiro é que documentar boas práticas cria multiplicadores: a criação
de IAs personalizadas e a padronização dos prompts mais eficazes
permitiram que toda a equipe se beneficiasse das descobertas
individuais. O quarto é que a resistência inicial, comum a qualquer
mudança de processo, se dissolveu mais rápido com resultados práticos do
que com argumentos teóricos.

Com o curso ministrado, o Framework de Governança consolidado e a rotina
operacional adaptada, a unidade operou em regime do método ao longo de
todo o ano de 2025. Os números estão na Seção X.

\hypertarget{x.-resultados-ucisedet-20242025}{%
\section{X. Resultados: UCI/SEDET
(2024→2025)}\label{x.-resultados-ucisedet-20242025}}

O ano de 2025 foi o primeiro ano da UCI/SEDET operando dentro do método.
Os números abaixo são integralmente extraídos das estatísticas oficiais
do SEI-GDF e consolidados no Relatório Nº 1/2026 da SEDET/GAB/UCI (Doc.
SEI 193384267, processo 04035-00000853/2026-24), assinado pelo Chefe da
Unidade, bem como de todos os Documentos e Notas Técnicas emitidos em
2025.

A comparação 2024 → 2025 traz ganhos expressivos e simultâneos em todos
os indicadores agregados. O tempo médio de tramitação caiu de 34 dias
para 17 dias (−50\%). A capacidade de processamento subiu de 256 para
336 processos tramitados (+31\%). A produção documental total cresceu de
251 para 419 documentos (+67\%). A produção da peça mais densa, a nota
técnica, subiu de 66 para 122 (+85\%), conforme o total consolidado no relatório gerencial assinado de 2025. Lida mês a mês (Figura 14), essa produção manteve-se num patamar baixo em 2023 e 2024 e deu um salto em 2025, primeiro ano sob o método.

\begin{figure}[H]
\centering
\includegraphics[width=0.80\linewidth,height=7cm,keepaspectratio]{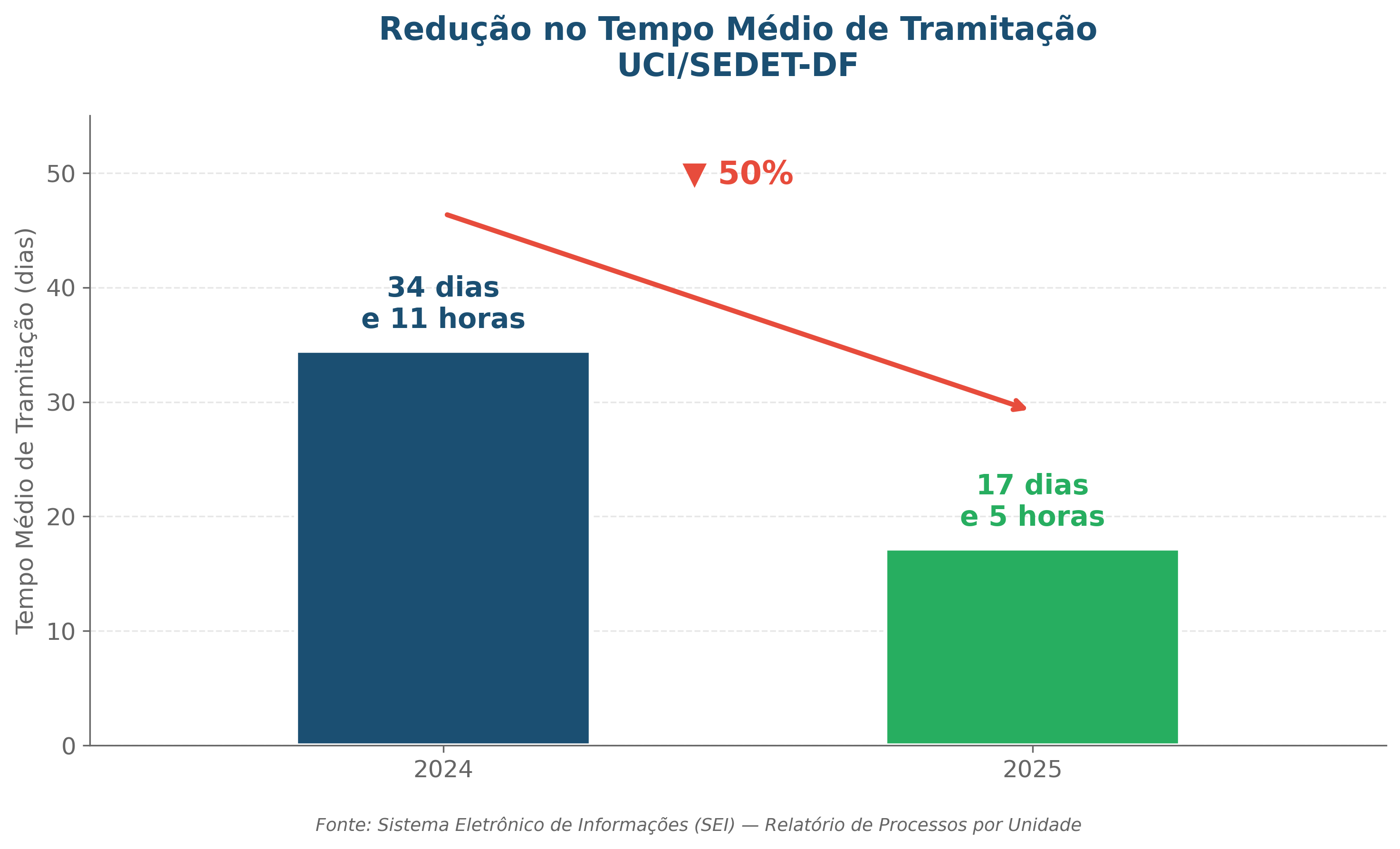}
\par\smallskip
\textit{Figura 10: Redução no tempo médio de tramitação na UCI/SEDET (2024 vs. 2025).}
\end{figure}

\begin{figure}[H]
\centering
\includegraphics[width=0.80\linewidth,height=7cm,keepaspectratio]{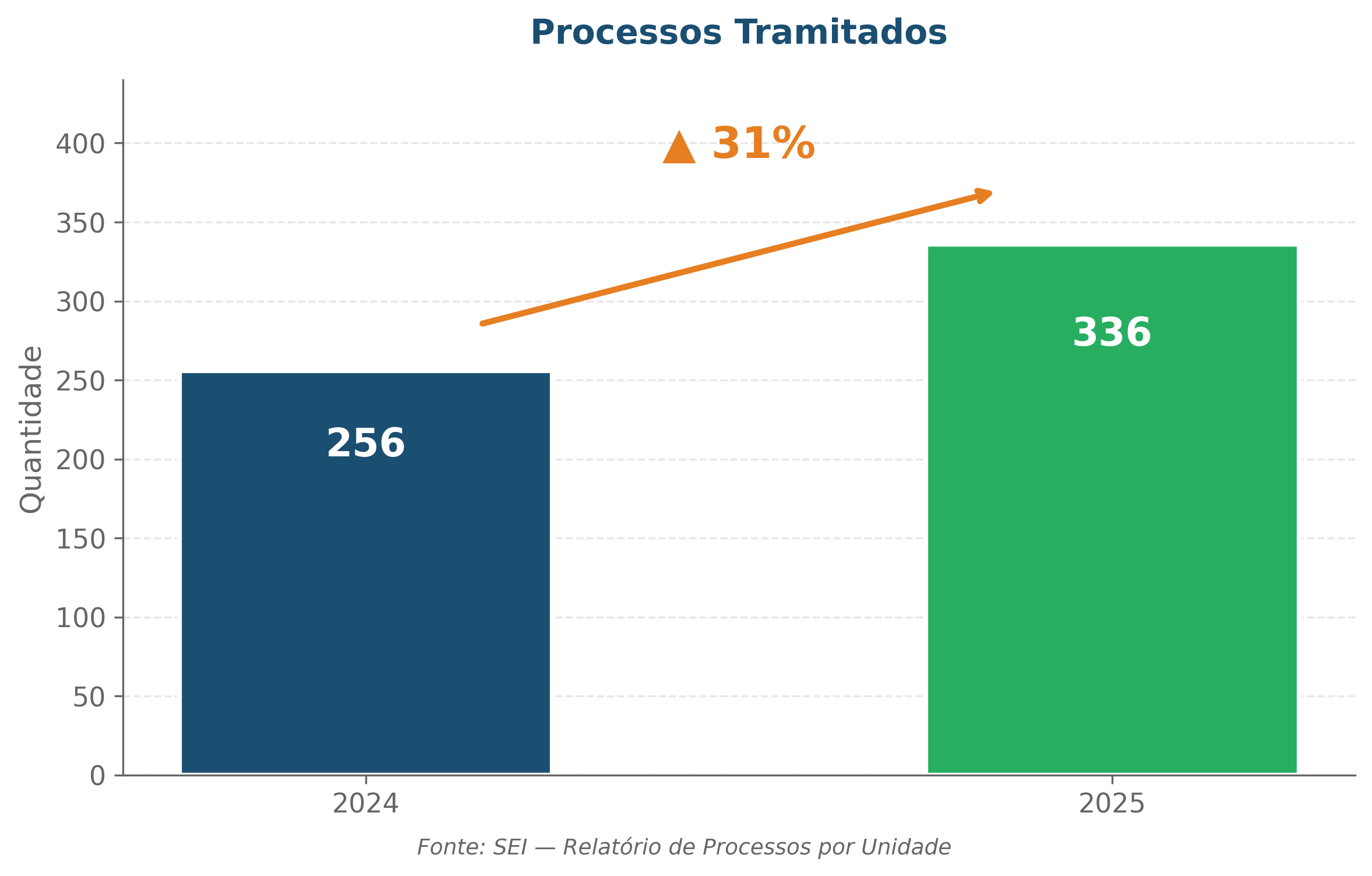}
\par\smallskip
\textit{Figura 11: Processos tramitados na UCI/SEDET (2024 vs. 2025).}
\end{figure}

\begin{figure}[H]
\centering
\includegraphics[width=0.80\linewidth,height=7cm,keepaspectratio]{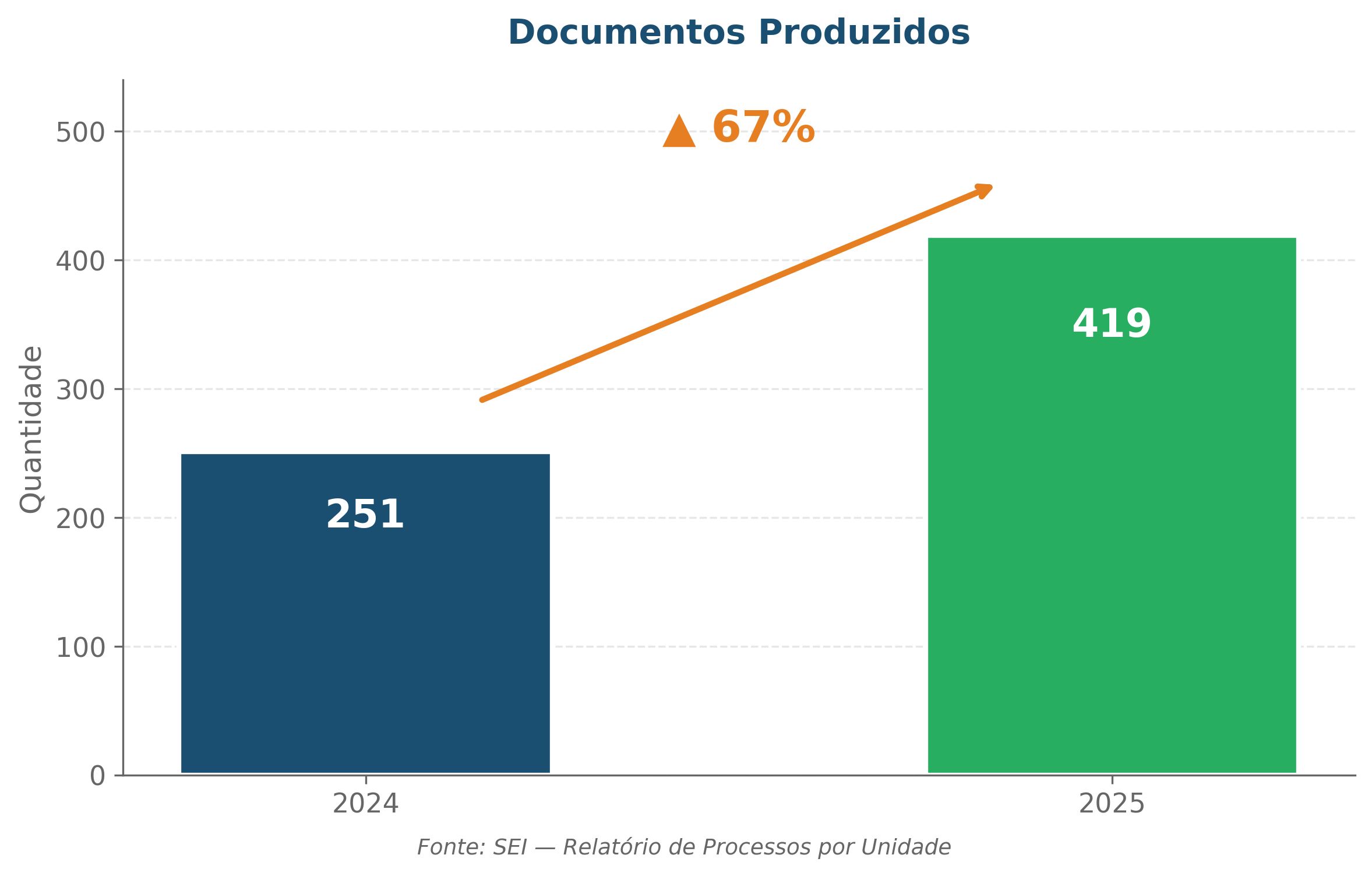}
\par\smallskip
\textit{Figura 12: Documentos produzidos na UCI/SEDET (2024 vs. 2025).}
\end{figure}

\begin{figure}[H]
\centering
\includegraphics[width=0.80\linewidth,height=7cm,keepaspectratio]{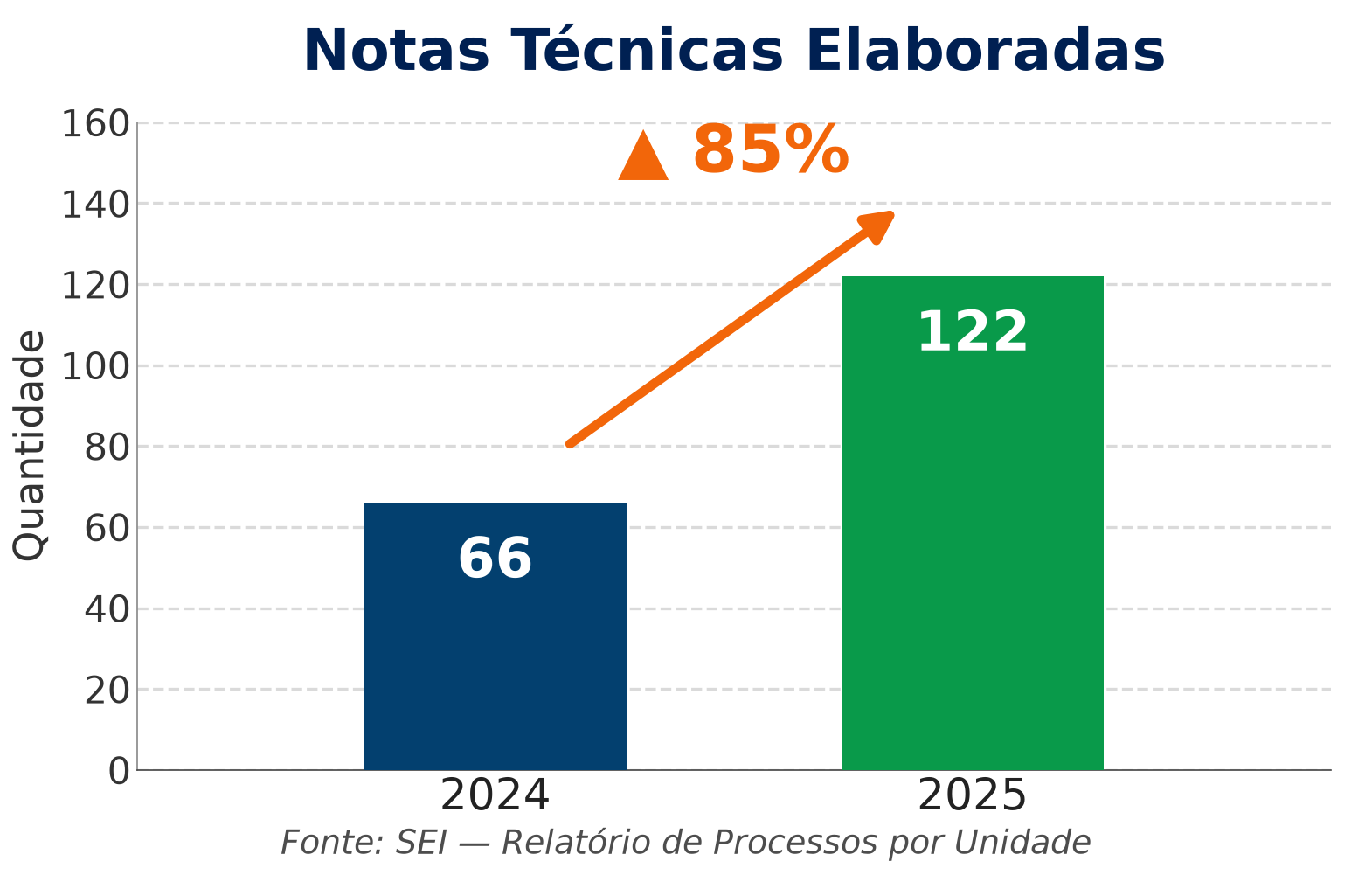}
\par\smallskip
\textit{Figura 13: Notas técnicas elaboradas na UCI/SEDET: 66 em 2024 e 122 em 2025, aumento de aproximadamente 85\%. Fonte: o valor de 2024 é a estatística bruta de documentos do SEI-GDF; o valor de 2025 é o total consolidado no relatório gerencial assinado da UCI/SEDET 2025 (ver a nota de conciliação no Apêndice A.0).}
\end{figure}

\begin{figure}[H]
\centering
\includegraphics[width=0.80\linewidth,height=7cm,keepaspectratio]{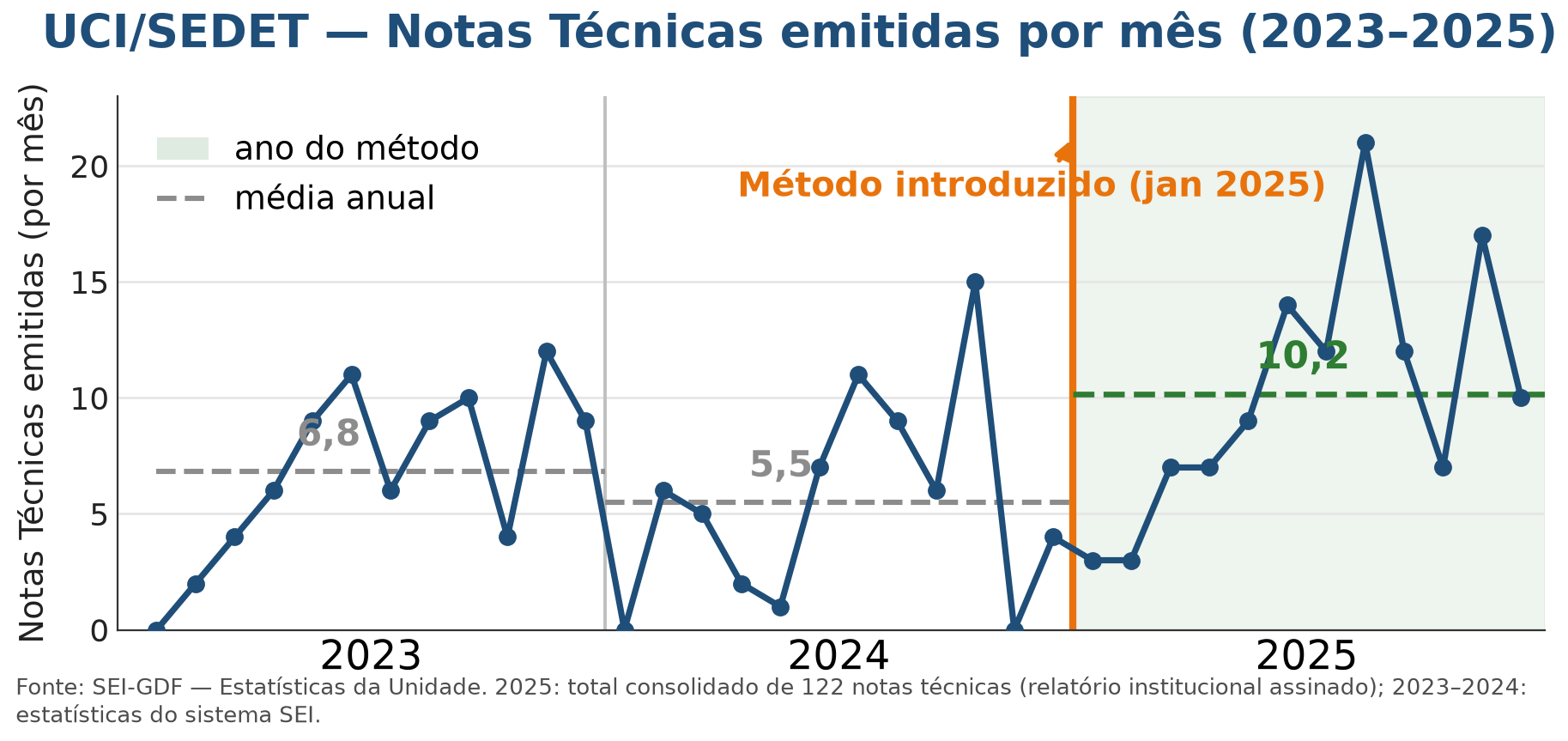}
\par\smallskip
\textit{Figura 14: Notas Técnicas emitidas por mês na UCI/SEDET (2023--2025). A média anual sobe de 6,8 (2023) e 5,5 (2024) para 10,2 em 2025, primeiro ano sob o método; o degrau concentra-se no ano do método, não nos dois anos de base. Fonte: os valores mensais de 2023--2024 são a estatística bruta de documentos do SEI-GDF; a série mensal de 2025 é o total consolidado no relatório gerencial assinado da UCI/SEDET 2025, somando 122 (ver a nota de conciliação no Apêndice A.0).}
\end{figure}

\begin{longtable}[]{@{}
  >{\raggedright\arraybackslash}p{(\linewidth - 6\tabcolsep) * \real{0.4468}}
  >{\raggedright\arraybackslash}p{(\linewidth - 6\tabcolsep) * \real{0.1383}}
  >{\raggedright\arraybackslash}p{(\linewidth - 6\tabcolsep) * \real{0.2191}}
  >{\raggedright\arraybackslash}p{(\linewidth - 6\tabcolsep) * \real{0.1958}}@{}}
\toprule
Indicador & 2024 & 2025 (com IA) & Variação\tabularnewline
\midrule
\endhead
Tempo médio de tramitação & 34 dias & 17 dias &
\textbf{−50\%}\tabularnewline
Processos tramitados & 256 & 336 & \textbf{+31\%}\tabularnewline
Documentos produzidos & 251 & 419 & \textbf{+67\%}\tabularnewline
Notas técnicas elaboradas & 66 & 122 & \textbf{+85\%}\tabularnewline
Despachos produzidos & 102 & 163 & +60\%\tabularnewline
\bottomrule
\end{longtable}

Na Controladoria, a redução de tempo veio com produção documental estável. Na UCI, a redução de tempo veio acompanhada de
aumento simultâneo do volume documental e da intensidade analítica. A
Seção XI retoma a diferença em chave comparativa.

As 122 notas técnicas consolidadas no relatório gerencial assinado de 2025 resultaram na expedição de 286
recomendações formais aos gestores da pasta. O dado é importante porque significa que
o aumento de produção não foi apenas quantitativo: a unidade conseguiu,
simultaneamente, produzir mais peças e aprofundar a análise dentro de
cada peça, identificando mais pontos de melhoria, mais riscos potenciais
e mais oportunidades de aprimoramento da gestão por processo examinado.

A revisão item a item das recomendações identificadas a partir do texto integral de cada uma das 122 notas
técnicas consolidadas no relatório gerencial assinado de 2025, classificadas pelos autores por natureza e materialidade (Apêndice A.4), produziu um valor em risco bruto modelado de R\$ 240.685.238,49 (R\$ 240,7 milhões). Aplicando matriz de probabilidade
calibrada com referência em literatura internacional de auditoria
pública, a estimativa de mitigação potencial agregada varia entre R\$ 6.094.150,54 (R\$ 6,1
milhões, cenário conservador) e R\$ 28.682.721,75 (R\$ 28,7 milhões, cenário otimista), com
estimativa central de R\$ 15.235.376,36 (R\$ 15,2 milhões): uma estimativa modelada, não economia realizada ou auditada. O detalhamento metodológico, os critérios de
classificação e a matriz de probabilidade estão descritos no
Apêndice A.4.

Em termos financeiros, as 122 notas técnicas consolidadas em
2025 dizem respeito a processos e matérias cujo valor total, segundo a
estatística oficial assinada no relatório gerencial de 2025 (Doc. SEI 193436555), somou
\textbf{R\$ 521.358.798,09} (R\$ 521,3 milhões): o valor das matérias submetidas à análise técnica. A distribuição
temporal do montante foi heterogênea, com meses de baixo volume e meses
de alta carga. A Tabela A.3.2 do apêndice organiza o valor analisado em cada mês.

O mês de agosto merece destaque. Foi o mês de maior volume tanto em número de processos analisados (21) quanto em volume financeiro do
ano: R\$ 105,1 milhões analisados em um único mês. Segundo o relatório institucional, a unidade absorveu esse pico sem gerar gargalos ou atrasos e sem comprometer a profundidade ou a qualidade técnica das análises realizadas.

O ganho de eficiência não foi restrito a um tipo de processo. A própria
UCI registra, no relatório oficial, que a melhoria operacional apareceu
em todas as frentes de trabalho da unidade, da análise rotineira aos
processos de maior complexidade técnica e financeira. Não houve,
portanto, troca de qualidade por velocidade em nenhum segmento do
trabalho.

O balanço de 2025 da UCI/SEDET pode ser sintetizado em quatro afirmações
verificáveis pelo SEI-GDF: a unidade dobrou a velocidade média de
tramitação; aumentou em 31\% o volume de processos tramitados; aumentou
em 85\% a produção de notas técnicas e em 60\% a de despachos;
e analisou R\$ 521,3 milhões em volume financeiro, com pico de R\$ 105,1
milhões absorvido sem gerar gargalos ou atrasos. A metodologia Human-in-the-Loop foi mantida como
condição inegociável, com vedação expressa à expedição de qualquer documento sem revisão técnica
humana integral e minuciosa, operacionalizada por meio de Tripla Revisão em três etapas (profissional redator, supervisor direto e autoridade signatária).

A Seção XI discute o que os dois casos, lidos juntos, ensinam sobre o
método.

\hypertarget{xi.-discussuxe3o-integrada}{%
\section{XI. Discussão integrada}\label{xi.-discussuxe3o-integrada}}

Lidos juntos, os Seções VI e X trazem um conjunto de evidências sobre os ganhos observados quando o método foi instalado corretamente em uma unidade pública. Os dois casos não são repetições idênticas. A
Controladoria Setorial da Saúde e a Unidade de Controle Interno da
SEDET-DF operam em Secretarias distintas, com mandato finalístico
distinto, com temática processual distinta e com equipes distintas. O
que se manteve constante entre as duas foi o método, o instrutor, o
curso pelo qual a equipe foi capacitada (o curso oficial da EGOV-DF), o
uso exclusivo de versões gratuitas das plataformas comerciais de IA, e o
princípio inegociável da supervisão humana integral. O método e a
estrutura institucional de capacitação foram constantes; o órgão, a
temática e a equipe variaram. É essa configuração que permite ler os
dois conjuntos de números como evidência convergente.

O quadro a seguir condensa, lado a lado, os principais indicadores dos
dois casos, para apoio à leitura comparativa nas seções seguintes.

\begin{longtable}[]{@{}
  >{\raggedright\arraybackslash}p{(\linewidth - 4\tabcolsep) * \real{0.3370}}
  >{\raggedright\arraybackslash}p{(\linewidth - 4\tabcolsep) * \real{0.3758}}
  >{\raggedright\arraybackslash}p{(\linewidth - 4\tabcolsep) * \real{0.2872}}@{}}
\toprule
\textbf{Dimensão} & \textbf{SES/CONT (2023 → 2024)} & \textbf{UCI/SEDET
(2024 → 2025)}\tabularnewline
\midrule
\endhead
\textbf{Tempo médio de tramitação} & 17,92 → 14,66 dias
(\textbf{−18,2\%}) & 34 → 17 dias (\textbf{−50,0\%})\tabularnewline
\textbf{Capacidade liberada estimada} & \textbf{+22,3\%} (≈352
processos adicionais) & \textbf{+31\%} (256 → 336
processos)\tabularnewline
\textbf{Estabilidade da produção mensal dos documentos mais importantes
da unidade} & CV reduzido em 3 de 4 tipos (Decisão, Julgamento e Ofício)
& CV reduzido em ambos os tipos (Despacho e Nota Técnica)\tabularnewline
\textbf{Peça analítica: produção mensal} & Julgamentos: 31,0/mês vs.
30,4/mês (\textbf{+2,1\%}, ex-mês atípico) & Notas Técnicas: 66 → 122
(\textbf{+84,8\%})\tabularnewline
\textbf{Função preventiva} & Comunicados: 28 → 214 (\textbf{+664\%}) &
286 recomendações formais expedidas\tabularnewline
\textbf{Articulação institucional} & Ofícios: 434 → 532
(\textbf{+22,6\%}) & n/a (perfil de controle interno
geral)\tabularnewline
\textbf{Tipos com maior melhoria de tempo} & Cumprimento de Decisão:
\textbf{−95,5\%}; Solicitação de Informação: \textbf{−93,9\%}; Ação
Judicial: \textbf{−88,3\%} & Melhoria transversal em todas as
atividades\tabularnewline
\textbf{Problemas/controles de IA documentados no relatório institucional} & Nenhum relatado & Não tratado como contagem de incidentes; revisão humana obrigatória antes da expedição (Human-in-the-Loop, Tripla Revisão)\tabularnewline
\bottomrule
\end{longtable}

A convergência mais visível, e a que mais diretamente sustenta o argumento central do paper, é
a magnitude do ganho de tempo. A Controladoria reduziu o tempo médio de
tramitação em 18,2\%; a UCI reduziu em 50\%. Os percentuais diferem, mas
a direção é a mesma, em dois órgãos com perfis operacionais opostos. Os
ganhos observados são temporalmente coincidentes com a introdução do
método e compatíveis com a hipótese de relação causal: em ambos os casos
a equipe permaneceu a mesma entre o ano-base e o ano com IA, a produção
documental por processo não diminuiu (estável na Controladoria,
crescente na UCI) e a base normativa das duas unidades não foi alterada.

Duas dessas alternativas merecem ser nomeadas explicitamente, porque ambas acompanharam o método e nenhuma pode ser inteiramente separada dele num desenho observacional. A primeira é a mudança de gestão: o autor assumiu a direção de cada unidade no início do respectivo ciclo, de modo que a nova gestão coincidiu com a introdução do método. A segunda é a reorganização dos fluxos de elaboração e revisão de documentos, ocorrida à medida que o método foi adotado. Não se afirma isolar o método desses fatores concomitantes.

O próprio desenho dos dois casos ajuda a afastar as explicações
alternativas mais imediatas. A repetição se deu em outra unidade, com
outra matéria e outro ponto de partida, o que é difícil atribuir à
simples acomodação de um ano atípico. A produção por processo não caiu
(ao contrário, manteve-se ou cresceu), o que afasta a hipótese de troca
de qualidade por velocidade. E o ganho não se espalhou de forma difusa:
concentrou-se nos tipos de documento em que o método efetivamente atua,
e não no conjunto, como seria de esperar se viesse apenas da evolução
das ferramentas. Nada disso prova causalidade; a evidência é
observacional. Ainda assim, no conjunto, esses pontos deixam a
capacitação como a explicação mais econômica para o que se observou nas
duas unidades.

A divergência mais visível entre os dois casos está no perfil de saída.
Na Controladoria, observou-se redução de tempo com produção documental
por processo praticamente estável (2,77 para 2,78 documentos D+D+J por
processo) após a adoção do método, e a capacidade liberada se converteu
em saneamento do passivo represado, estabilização da produção mensal e
expansão da função preventiva (Comunicados +664\%). Na UCI, a redução de
tempo veio acompanhada de aumento simultâneo do volume documental total
(+67\%) e da peça analítica mais densa (Notas Técnicas +85\%, com 286
recomendações formais expedidas). A diferença é institucional. A
Controladoria operava no limite da capacidade, com passivo represado, e
os indicadores pós-adoção são compatíveis com a liberação dessa
capacidade represada. A UCI operava com capacidade técnica
subaproveitada por falta de tempo para a peça analítica densa, e os
indicadores pós-adoção são compatíveis com a expressão dessa capacidade
em mais peças e mais profundas. O ganho observado, em cada caso, é
compatível com o gargalo do ponto de partida. Há, ainda, um ganho
qualitativo comum aos dois casos: ambas as unidades passaram a operar
simultaneamente em chave repressiva (controle posterior) e em chave
preventiva e corretiva, os Comunicados expedidos pela SES/CONT (+664\%)
e as 286 recomendações formais da UCI/SEDET são a manifestação
documentada dessa virada, em que o controle interno passa a sinalizar
aos gestores os fluxos e práticas que originaram as irregularidades,
para que possam ser saneados antes do próximo incidente.

O mês de agosto, em cada um dos dois ciclos, ilustra a mudança
estrutural de forma particularmente clara. Em agosto de 2023, a
Controladoria absorveu um pico extraordinário de 69 Julgamentos por meio
de uma força-tarefa emergencial: servidores trabalhando acima da
capacidade normal, parte da produção feita em dias não úteis, risco real
de perda de processos por prescrição. Em agosto de 2025, a UCI absorveu
o pico do ano tanto em número de processos analisados (21) quanto em volume financeiro (R\$ 105,1 milhões em um único mês), e o relatório institucional registra que isso ocorreu sem gerar gargalos ou atrasos e sem
comprometer a profundidade da análise técnica. Mesmo calendário; outra geometria operacional.

É importante registrar o que o método não faz, com base nos aprendizados
consolidados no Relatório UCI. O método não substitui o conhecimento
técnico: a IA acelera a produção, mas a revisão crítica por profissional
habilitado permaneceu indispensável em ambos os casos para garantir
precisão e adequação. O método não dispensa esforço de comando: prompts
bem estruturados, com contexto adequado, instruções claras e exemplos
pertinentes, produzem resultados significativamente melhores do que
prompts genéricos, e essa diferença não se resolve sozinha; exige
treinamento e prática. O método não é instantâneo: a adoção foi
incremental, da tarefa simples para a tarefa complexa, e a resistência
inicial cedeu mais rápido a resultados práticos do que a argumentos
teóricos. E o método não opera no vácuo: cada unidade precisou de uma
camada de proteção documentada formalmente (a IA personalizada com base
legal na Controladoria, o Framework de Governança de IA específico na
UCI), aderente à matéria que aquela unidade trata.

Há também o resultado que os dois casos produziram em conjunto: nenhum
incidente identificado. Nos dois ciclos anuais, os mecanismos internos
de controle das duas unidades não identificaram vazamento de dado
sensível, decisão administrativa anulada por falha decorrente do uso de
IA, ou questionamento formal por órgão de controle externo quanto à
conformidade do método com a Legislação de Proteção de Dados ou com os
princípios da Administração Pública. A leitura compatível com a
evidência é que esse resultado decorreu de a preocupação com a parte de
Governança ter sido construído desde o início, com a base construída
explicando os motivos de a IA errar, por exemplo, e de a supervisão
humana integral ter sido tratada como condição de funcionamento, não
como medida adicional.

A estabilidade da produção, mencionada qualitativamente nas Seções VI e
X, é quantificável pelo coeficiente de variação (CV) da produção mensal
por tipo de documento, conforme as tabelas A.2.4 e A.3.4 do apêndice. O
quadro abaixo resume a variação do CV entre o ano-base e o ano com
adoção do método nas duas unidades nos documentos mais importantes:

\begin{longtable}[]{@{}
  >{\raggedright\arraybackslash}p{(\linewidth - 12\tabcolsep) * \real{0.1580}}
  >{\raggedright\arraybackslash}p{(\linewidth - 12\tabcolsep) * \real{0.1886}}
  >{\raggedright\arraybackslash}p{(\linewidth - 12\tabcolsep) * \real{0.1404}}
  >{\raggedright\arraybackslash}p{(\linewidth - 12\tabcolsep) * \real{0.1601}}
  >{\raggedright\arraybackslash}p{(\linewidth - 12\tabcolsep) * \real{0.1095}}
  >{\raggedright\arraybackslash}p{(\linewidth - 12\tabcolsep) * \real{0.1030}}
  >{\raggedright\arraybackslash}p{(\linewidth - 12\tabcolsep) * \real{0.1404}}@{}}
\toprule
Unidade & Tipo & Ano-base & CV ano-base & Ano com IA & CV ano IA &
Variação\tabularnewline
\midrule
\endhead
SES/CONT & Decisão & 2023 & 39,8\% & 2024 & 17,3\% &
\textbf{−56,5\%}\tabularnewline
SES/CONT & Julgamento & 2023 & 52,4\% & 2024 & 26,5\% &
−49,4\%\tabularnewline
SES/CONT & Ofício & 2023 & 19,8\% & 2024 & 13,6\% &
−31,3\%\tabularnewline
UCI/SEDET & Despacho & 2024 & 73,3\% & 2025 & 49,3\% &
−32,7\%\tabularnewline
UCI/SEDET & Nota Técnica & 2024 & 83,3\% & 2025 (consolidado) & 53,0\% &
\textbf{−36,4\%}\tabularnewline
\bottomrule
\end{longtable}

\hypertarget{xii.-conclusuxf5es}{%
\section{XII. Conclusões}\label{xii.-conclusuxf5es}}

A síntese, portanto, tem quatro frentes verificáveis. A primeira é a replicabilidade: ganhos de eficiência acompanharam o método em dois órgãos com
perfis institucionais distintos, com equipes distintas e com temáticas
processuais distintas, sem reformulação estrutural entre os dois casos.
A segunda é que o método operou, no escopo do estudo, sem que os
mecanismos internos de controle identificassem incidentes, com
governança documentada formalmente em ambas as unidades. A terceira é
que o método é acessível: os resultados foram obtidos com versões
gratuitas das plataformas comerciais, o que abre a possibilidade de
replicação para outros entes públicos com restrições orçamentárias. A
quarta concerne o efeito financeiro potencial do método, que modelamos em vez de medir diretamente: sob uma matriz de probabilidade construída sobre premissas explícitas, ancoradas na literatura, a revisão item a item das recomendações identificadas nas 122 notas técnicas consolidadas no relatório gerencial assinado de 2025, classificadas pelos autores para este exercício (Apêndice A.4), situa a mitigação potencial numa faixa de R\$ 6,1 milhões (conservador) a R\$ 28,7 milhões (otimista), com estimativa central de R\$ 15,2 milhões. Trata-se de uma faixa modelada, não de economia realizada ou auditada, e é distinta dos R\$ 521,3 milhões em volume financeiro que a estatística oficial assinada da unidade registra como submetido à análise técnica.

\hypertarget{apuxeandice}{%
\section{APÊNDICE}\label{apuxeandice}}

\hypertarget{apuxeandice-a-referuxeancias-documentais-e-tabelas-de-dados-oficiais}{%
\section{Apêndice A: Referências Documentais e Tabelas de Dados
Oficiais}\label{apuxeandice-a-referuxeancias-documentais-e-tabelas-de-dados-oficiais}}

Os dados e análises apresentados neste paper estão integralmente
ancorados em documentos institucionais do Sistema Eletrônico de
Informações do Governo do Distrito Federal (SEI-GDF), todos
identificados por número de documento único e processo correspondente, e
portanto verificáveis por terceiros. Este apêndice consolida as
referências documentais utilizadas e replica, em formato auditável, as
tabelas de dados oficiais que sustentam o texto principal.

\hypertarget{a.0-nota-metodoluxf3gica-de-extrauxe7uxe3o-dos-dados}{%
\subsection{A.0 Nota metodológica de extração dos
dados}\label{a.0-nota-metodoluxf3gica-de-extrauxe7uxe3o-dos-dados}}

Os dados quantitativos apresentados neste paper foram extraídos do
\textbf{Sistema Eletrônico de Informações do Governo do Distrito Federal
(SEI-GDF)}, sistema oficial de tramitação processual da Administração
Pública distrital.

A extração foi realizada pela funcionalidade nativa
\textbf{``Estatísticas da Unidade''}, com filtragem por unidade emissora
(SES/CONT ou UCI/SEDET) e por ano-calendário inteiro. A funcionalidade
gera, em formato PDF, três blocos de informação por ano: (i) processos
gerados no período, por tipo; (ii) documentos gerados no período, por
tipo (Decisão, Despacho, Julgamento, Ofício, Comunicado, Nota Técnica e
demais tipos disponíveis ao perfil da unidade); (iii) tempos médios de
tramitação por tipo de processo. O \textbf{indicador de tempo médio}
corresponde ao tempo bruto, em horas corridas, entre o recebimento do
processo na unidade e o despacho que encerra a etapa de tramitação
naquela unidade, conforme metodologia interna do SEI-GDF.

O cotejo entre os dois ciclos comparativos (SES/CONT 2023×2024;
UCI/SEDET 2024×2025) foi feito sobre extrações realizadas com o mesmo
procedimento, sem alterações conhecidas na metodologia de cálculo do
sistema entre os anos comparados. A unidade emissora dos relatórios é a
mesma em cada par de anos, o que preserva a base de comparação. Não
foram aplicados filtros adicionais sobre os PDFs extraídos.

O \textbf{pós-processamento estatístico} das tabelas de produção mensal
(médias, medianas, desvios-padrão, quartis, IQR e coeficientes de
variação reportados nas tabelas A.2.4 e A.3.4) foi realizado por
pipeline determinístico mantido pelo autor, com auditoria cruzada
confirmando, para todas as séries reportadas, a identidade entre a soma
dos 12 valores mensais extraídos e o total declarado nos PDFs do SEI. O
pipeline encontra-se disponível mediante solicitação ao autor.

\textbf{Conciliação da contagem de notas técnicas da UCI/SEDET 2025 (122 versus 127).} A planilha oficial da página 48 do relatório gerencial assinado da UCI/SEDET 2025 (``Estatística Notas Técnicas Unidade de Controle Interno, UCI/SEDET, 2025'', Doc. SEI 193436555) consolida notas técnicas que analisam processos de contratação, pagamento em contrato ou parceria, tipificados como CM, PM, IN, CL ou PL, e registra um total de 122 dessas notas para 2025, com valor combinado de R\$ 521.358.798,09. Uma consulta bruta e separada à funcionalidade ``Estatísticas da Unidade'' do SEI-GDF (p.~59 do mesmo relatório) identificou 127 documentos rotulados ``Nota Técnica'' emitidos pela unidade em 2025. O relatório gerencial de 2025 assinado consolidou 122 notas técnicas, ante 66 em 2024, aumento aproximado de 85\%. Os dois universos foram conciliados item a item, e o total consolidado de 122 do relatório é adotado como medida oficial de desempenho. A diferença de cinco documentos reflete notas técnicas que não eram análises de processos de contratação, pagamento em contrato ou parceria: uma nota de acompanhamento da implementação das recomendações de 2024, uma análise de minuta de instrumento normativo, uma resposta a ofício correicional externo, e dois processos de tomada de contas especial, conforme identificado na conciliação item a item. O mesmo relatório registrou 286 recomendações formais (p.~5). Por essa razão, este paper utiliza 122 notas técnicas e 286 recomendações em todo o texto, preservando a contagem bruta do SEI de 127 apenas para documentar o procedimento de conciliação.

\hypertarget{a.1-documentos-sei-gdf-citados-no-paper}{%
\subsection{A.1 Documentos SEI-GDF citados no
paper}\label{a.1-documentos-sei-gdf-citados-no-paper}}

\hypertarget{a.1.1-controladoria-setorial-da-sauxfade-sescont}{%
\subsubsection{A.1.1 Controladoria Setorial da Saúde
(SES/CONT)}\label{a.1.1-controladoria-setorial-da-sauxfade-sescont}}

\textbf{Doc. SEI 146621014: Memorando Nº 4/2024, SES/CONT/ASJULG}

\begin{longtable}[]{@{}
  >{\raggedright\arraybackslash}p{(\linewidth - 2\tabcolsep) * \real{0.1783}}
  >{\raggedright\arraybackslash}p{(\linewidth - 2\tabcolsep) * \real{0.8217}}@{}}
\toprule
Campo & Conteúdo\tabularnewline
\midrule
\endhead
Processo SEI & 00060-00314546/2024-15\tabularnewline
Data & 25 de junho de 2024\tabularnewline
Assunto & Proposta de criação do curso \emph{Inteligência Artificial
para Servidores Públicos}\tabularnewline
\bottomrule
\end{longtable}

\textbf{Doc. SEI 144321884: Plano de Aula do curso}

\begin{longtable}[]{@{}
  >{\raggedright\arraybackslash}p{(\linewidth - 2\tabcolsep) * \real{0.1373}}
  >{\raggedright\arraybackslash}p{(\linewidth - 2\tabcolsep) * \real{0.8627}}@{}}
\toprule
Campo & Conteúdo\tabularnewline
\midrule
\endhead
Processo SEI & 00060-00314546/2024-15\tabularnewline
Conteúdo & Estrutura programática completa das cinco aulas presenciais
do curso \emph{Inteligência Artificial no setor público: técnicas,
riscos e aplicações} (20 horas)\tabularnewline
\bottomrule
\end{longtable}

\textbf{Doc. SEI 197403428: Relatório Gerencial Executivo: Impacto da
Inteligência Artificial na Eficiência Operacional}

\begin{longtable}[]{@{}
  >{\raggedright\arraybackslash}p{(\linewidth - 2\tabcolsep) * \real{0.1788}}
  >{\raggedright\arraybackslash}p{(\linewidth - 2\tabcolsep) * \real{0.8212}}@{}}
\toprule
Campo & Conteúdo\tabularnewline
\midrule
\endhead
Processo SEI & 00060-00582291/2024-11\tabularnewline
Data & 2 de janeiro de 2025\tabularnewline
Unidade emissora & Controladoria Setorial da Saúde
(SES/CONT)\tabularnewline
Conteúdo & Análise comparativa 2023 vs.~2024 dos indicadores
operacionais da unidade após implementação do método\tabularnewline
\bottomrule
\end{longtable}

\textbf{Relatório Nº 12/2024: SES/CONT/ASJULG}

\begin{longtable}[]{@{}
  >{\raggedright\arraybackslash}p{(\linewidth - 2\tabcolsep) * \real{0.1290}}
  >{\raggedright\arraybackslash}p{(\linewidth - 2\tabcolsep) * \real{0.8710}}@{}}
\toprule
Campo & Conteúdo\tabularnewline
\midrule
\endhead
Data & 14 de dezembro de 2024\tabularnewline
Conteúdo & Registro institucional de que a unidade não enfrentava
acúmulo de processos, com prazo máximo de 10 dias para elaboração de
minutas de decisão\tabularnewline
\bottomrule
\end{longtable}

\hypertarget{a.1.2-unidade-de-controle-interno-ucisedet-df}{%
\subsubsection{A.1.2 Unidade de Controle Interno
(UCI/SEDET-DF)}\label{a.1.2-unidade-de-controle-interno-ucisedet-df}}

\textbf{Doc. SEI 194251158: Framework de Governança de IA, UCI}

\begin{longtable}[]{@{}
  >{\raggedright\arraybackslash}p{(\linewidth - 2\tabcolsep) * \real{0.1979}}
  >{\raggedright\arraybackslash}p{(\linewidth - 2\tabcolsep) * \real{0.8021}}@{}}
\toprule
Campo & Conteúdo\tabularnewline
\midrule
\endhead
Unidade emissora & Unidade de Controle Interno, SEDET-DF\tabularnewline
Conteúdo & Documento institucional com orientações operacionais para
utilização de ferramentas de IA na UCI\tabularnewline
\bottomrule
\end{longtable}

\textbf{Doc. SEI 193384267, Relatório Nº 1/2026: Relatório Gerencial
de Resultados 2025, UCI/GAB/SEDET}

\begin{longtable}[]{@{}
  >{\raggedright\arraybackslash}p{(\linewidth - 2\tabcolsep) * \real{0.1788}}
  >{\raggedright\arraybackslash}p{(\linewidth - 2\tabcolsep) * \real{0.8212}}@{}}
\toprule
Campo & Conteúdo\tabularnewline
\midrule
\endhead
Processo SEI & 04035-00000853/2026-24\tabularnewline
Data & 28 de janeiro de 2026\tabularnewline
Unidade emissora & Unidade de Controle Interno, SEDET-DF\tabularnewline
Conteúdo & Análise comparativa 2024 vs.~2025 dos indicadores
operacionais da unidade após implementação do método\tabularnewline
\bottomrule
\end{longtable}

\hypertarget{a.1.3-escola-de-governo-do-distrito-federal-egov-df}{%
\subsubsection{A.1.3 Escola de Governo do Distrito Federal
(EGOV-DF)}\label{a.1.3-escola-de-governo-do-distrito-federal-egov-df}}

\textbf{Doc. SEI 146526272: Memorando Nº 166/2024,
SEEC/SEGEA/EGOV}

\begin{longtable}[]{@{}
  >{\raggedright\arraybackslash}p{(\linewidth - 2\tabcolsep) * \real{0.1131}}
  >{\raggedright\arraybackslash}p{(\linewidth - 2\tabcolsep) * \real{0.8869}}@{}}
\toprule
Campo & Conteúdo\tabularnewline
\midrule
\endhead
Processo SEI & 04044-00021535/2024-26\tabularnewline
Data & 22 de julho de 2024\tabularnewline
Assunto & Aprovação institucional da realização do curso
\emph{Inteligência Artificial no setor público: técnicas, riscos e
aplicações}, com ônus, por meio de instrutoria interna, na modalidade de
ensino presencial, destinado aos agentes públicos dos órgãos e entidades
da Administração Direta e Indireta do Distrito Federal\tabularnewline
\bottomrule
\end{longtable}

\hypertarget{a.1.4-base-normativa-da-ia-personalizada-sescont}{%
\subsubsection{A.1.4 Base normativa da IA personalizada
SES/CONT}\label{a.1.4-base-normativa-da-ia-personalizada-sescont}}

A IA personalizada da Controladoria Setorial da Saúde (descrita na Seção
IV) operava sobre uma base de conhecimento composta por dez normativos
jurídicos diretamente aplicáveis à matéria correcional da unidade. Esses
dez normativos não estão isolados na Lista de Referências por não serem
invocados argumentativamente no corpo do paper; figuram nesta subseção
exclusivamente como rastreabilidade documental do material operacional
carregado na IA personalizada:

• Lei Orgânica do Distrito Federal. Câmara Legislativa do DF, 1993.

• Decreto nº 37.297, de 29 de abril de 2016: Código de Ética dos
Servidores e Empregados Públicos Civis do Poder Executivo do DF.

• Controladoria-Geral do DF. Instrução Normativa nº 2, de 25 de julho de
2016: procedimentos de mediação no âmbito da administração pública
distrital.

• Controladoria-Geral do DF. Instrução Normativa nº 1, de 12 de março de
2021: Termo de Ajustamento de Conduta (TAC).

• Controladoria-Geral do DF. Instrução Normativa nº 2, de 19 de outubro
de 2021: Juízo de Admissibilidade e Procedimento Investigativo
Preliminar (PIP).

• Controladoria-Geral do DF. Manual Teórico de Procedimentos
Disciplinares. Brasília, DF.

• Controladoria-Geral do DF. Manual Prático de Procedimentos
Disciplinares. Brasília, DF.

• Lei Complementar nº 840, de 23 de dezembro de 2011: Regime jurídico
dos servidores públicos civis do Distrito Federal, das autarquias e das
fundações públicas distritais.

• Lei nº 4.938, de 19 de setembro de 2012: Institui o Sistema de
Correição do Distrito Federal (SICOR).

• Decreto nº 39.701, de 7 de março de 2019: Normas de procedimentos
disciplinares no âmbito do Distrito Federal.

\hypertarget{a.1.5-base-normativa-da-ia-personalizada-ucisedet-licitauxe7uxf5es}{%
\subsubsection{A.1.5 Base normativa da IA personalizada UCI/SEDET
(Licitações)}\label{a.1.5-base-normativa-da-ia-personalizada-ucisedet-licitauxe7uxf5es}}

A UCI/SEDET-DF operou, a partir de janeiro de 2025, com IAs
personalizadas modulares por matéria (descritas na Seção IX). Para a IA
personalizada de Licitações, a base de conhecimento reuniu a legislação
federal e distrital aplicável. Esses normativos figuram nesta subseção
como rastreabilidade documental do material operacional carregado, em
paralelo à A.1.4 (base normativa da IA personalizada SES/CONT):

• Lei nº 13.019, de 31 de julho de 2014: Marco Regulatório das
Organizações da Sociedade Civil (MROSC), Termos de Fomento.

• Lei nº 14.133, de 1º de abril de 2021: Lei de Licitações e
Contratos Administrativos

• Decreto nº 32.598, de 15 de dezembro de 2010: normas de
planejamento, orçamento, finanças, patrimônio e contabilidade do
Distrito Federal.

• Decreto nº 44.330, de 7 de abril de 2023: regulamenta a Lei nº
14.133/2021 no âmbito do Distrito Federal.

• Decreto nº 45.933, de 18 de junho de 2024: atualização do
regulamento de licitações e contratos do DF.

• Controladoria-Geral do Distrito Federal. Portaria nº 29, de 2021:
diretrizes aplicáveis ao controle interno das licitações distritais.

\hypertarget{a.2-tabelas-de-dados-oficiais-controladoria-setorial-da-sauxfade-sescont}{%
\subsection{A.2 Tabelas de dados oficiais: Controladoria Setorial da
Saúde
(SES/CONT)}\label{a.2-tabelas-de-dados-oficiais-controladoria-setorial-da-sauxfade-sescont}}

As tabelas desta seção foram extraídas do Relatório Gerencial Executivo
(Doc. SEI 197403428).

\hypertarget{a.2.1-indicadores-gerais-2023-vs.-2024}{%
\subsubsection{A.2.1 Indicadores gerais 2023
vs.~2024}\label{a.2.1-indicadores-gerais-2023-vs.-2024}}

\begin{longtable}[]{@{}
  >{\raggedright\arraybackslash}p{(\linewidth - 6\tabcolsep) * \real{0.4189}}
  >{\raggedright\arraybackslash}p{(\linewidth - 6\tabcolsep) * \real{0.1786}}
  >{\raggedright\arraybackslash}p{(\linewidth - 6\tabcolsep) * \real{0.2156}}
  >{\raggedright\arraybackslash}p{(\linewidth - 6\tabcolsep) * \real{0.1869}}@{}}
\toprule
Indicador & 2023 & 2024 (com IA) & Variação\tabularnewline
\midrule
\endhead
Tempo médio de tramitação & 17,92 dias & 14,66 dias &
\textbf{−18,2\%}\tabularnewline
Processos tramitados & 1.752 & 1.581 & −9,8\%\tabularnewline
Total D+D+J produzidos & 4.846 & 4.393 & −9,3\%\tabularnewline
D+D+J por processo & 2,77 & 2,78 & +0,4\%\tabularnewline
Despachos por processo & 1,57 & 1,65 & +5,1\%\tabularnewline
\bottomrule
\end{longtable}

\hypertarget{a.2.2-tipos-de-processo-com-maiores-reduuxe7uxf5es-de-tempo-alta-complexidade-analuxedtica}{%
\subsubsection{A.2.2 Tipos de processo com maiores reduções de tempo
(alta complexidade
analítica)}\label{a.2.2-tipos-de-processo-com-maiores-reduuxe7uxf5es-de-tempo-alta-complexidade-analuxedtica}}

\begin{longtable}[]{@{}
  >{\raggedright\arraybackslash}p{(\linewidth - 6\tabcolsep) * \real{0.5629}}
  >{\raggedright\arraybackslash}p{(\linewidth - 6\tabcolsep) * \real{0.1554}}
  >{\raggedright\arraybackslash}p{(\linewidth - 6\tabcolsep) * \real{0.1344}}
  >{\raggedright\arraybackslash}p{(\linewidth - 6\tabcolsep) * \real{0.1473}}@{}}
\toprule
Tipo de Processo & 2023 & 2024 & Redução\tabularnewline
\midrule
\endhead
Cumprimento de Decisão & 17,82 dias & 0,81 dias &
\textbf{−95,5\%}\tabularnewline
Solicitação de Informação (Controle Interno) & 22,81 dias & 1,38 dias &
\textbf{−93,9\%}\tabularnewline
Ação Judicial: Cumprimento & 24,96 dias & 2,92 dias &
\textbf{−88,3\%}\tabularnewline
Consultas Órgãos Jurídicos & 9,27 dias & 2,82 dias &
−69,6\%\tabularnewline
Orientações e Diretrizes & 15,74 dias & 5,92 dias &
−62,4\%\tabularnewline
Apuração Preliminar & 7,39 dias & 3,16 dias & −57,2\%\tabularnewline
Consultas Órgãos de Controle & 12,96 dias & 7,00 dias &
−46,0\%\tabularnewline
\bottomrule
\end{longtable}

\hypertarget{a.2.3-produuxe7uxe3o-de-julgamentos-anuxe1lise-normalizada}{%
\subsubsection{A.2.3 Produção de Julgamentos: análise
normalizada}\label{a.2.3-produuxe7uxe3o-de-julgamentos-anuxe1lise-normalizada}}

\begin{longtable}[]{@{}
  >{\raggedright\arraybackslash}p{(\linewidth - 6\tabcolsep) * \real{0.5490}}
  >{\raggedright\arraybackslash}p{(\linewidth - 6\tabcolsep) * \real{0.1175}}
  >{\raggedright\arraybackslash}p{(\linewidth - 6\tabcolsep) * \real{0.1175}}
  >{\raggedright\arraybackslash}p{(\linewidth - 6\tabcolsep) * \real{0.2160}}@{}}
\toprule
Cenário & Total & Meses & Média mensal\tabularnewline
\midrule
\endhead
2023 (incluído agosto, mês atípico) & 403 & 12 & 33,58\tabularnewline
2023 (excluído agosto, mês atípico)¹ & 334 & 11 & 30,36\tabularnewline
2024 (todos os meses, com IA) & 372 & 12 & 31,00\tabularnewline
\bottomrule
\end{longtable}

¹ \emph{Agosto de 2023 desconsiderado por se tratar de volume atípico
decorrente de força-tarefa emergencial, não representativo da operação
regular. Ver detalhamento na Seção II.}

\hypertarget{a.2.4-estatuxedstica-descritiva-da-produuxe7uxe3o-mensal-por-tipo-de-documento-sescont}{%
\subsubsection{A.2.4 Estatística descritiva da produção mensal por tipo
de documento:
SES/CONT}\label{a.2.4-estatuxedstica-descritiva-da-produuxe7uxe3o-mensal-por-tipo-de-documento-sescont}}

Base: 12 valores mensais por tipo/ano, extraídos das estatísticas
oficiais do SEI-GDF da unidade. DP = desvio-padrão amostral; CV =
coeficiente de variação (DP/média × 100); IQR = intervalo interquartil
(Q3 − Q1).

\begin{longtable}[]{@{}
  >{\raggedright\arraybackslash}p{(\linewidth - 22\tabcolsep) * \real{0.1485}}
  >{\raggedright\arraybackslash}p{(\linewidth - 22\tabcolsep) * \real{0.0941}}
  >{\raggedright\arraybackslash}p{(\linewidth - 22\tabcolsep) * \real{0.0743}}
  >{\raggedright\arraybackslash}p{(\linewidth - 22\tabcolsep) * \real{0.0891}}
  >{\raggedright\arraybackslash}p{(\linewidth - 22\tabcolsep) * \real{0.1040}}
  >{\raggedright\arraybackslash}p{(\linewidth - 22\tabcolsep) * \real{0.0743}}
  >{\raggedright\arraybackslash}p{(\linewidth - 22\tabcolsep) * \real{0.0446}}
  >{\raggedright\arraybackslash}p{(\linewidth - 22\tabcolsep) * \real{0.0891}}
  >{\raggedright\arraybackslash}p{(\linewidth - 22\tabcolsep) * \real{0.0891}}
  >{\raggedright\arraybackslash}p{(\linewidth - 22\tabcolsep) * \real{0.0446}}
  >{\raggedright\arraybackslash}p{(\linewidth - 22\tabcolsep) * \real{0.0891}}
  >{\raggedright\arraybackslash}p{(\linewidth - 22\tabcolsep) * \real{0.0594}}@{}}
\toprule
Tipo & Ano & Total & Média & Mediana & DP & Mín & Q1 & Q3 & Máx & IQR &
CV\%\tabularnewline
\midrule
\endhead
\textbf{Decisão} & 2022 & 1.885 & 157,08 & 122,0 & 72,68 & 91 & 102,25 &
222,50 & 291 & 120,25 & 46,3\tabularnewline
& 2023 & 1.684 & 140,33 & 120,0 & 55,92 & 91 & 104,00 & 151,25 & 266 &
47,25 & 39,8\tabularnewline
& \textbf{2024 (com IA)} & 1.408 & 117,33 & 124,0 & \textbf{20,34} & 87
& 100,50 & 135,25 & 142 & 34,75 & \textbf{17,3}\tabularnewline
\textbf{Despacho} & 2022 & 2.856 & 238,00 & 238,5 & 45,42 & 164 & 206,50
& 267,25 & 315 & 60,75 & 19,1\tabularnewline
& 2023 & 2.759 & 229,92 & 231,5 & 20,37 & 193 & 217,25 & 244,75 & 259 &
27,50 & 8,9\tabularnewline
& \textbf{2024 (com IA)} & 2.613 & 217,75 & 226,5 & 33,19 & 152 & 202,00
& 238,25 & 258 & 36,25 & 15,2\tabularnewline
\textbf{Julgamento} & 2022 & 180 & 15,00 & 10,0 & 14,82 & 1 & 7,75 &
17,75 & 56 & 10,00 & 98,8\tabularnewline
& 2023 & 403 & 33,58 & 31,5 & 17,58 & 13 & 18,50 & 47,25 & 69 & 28,75 &
52,4\tabularnewline
& \textbf{2024 (com IA)} & 372 & 31,00 & 30,5 & \textbf{8,21} & 20 &
23,00 & 36,75 & 45 & 13,75 & \textbf{26,5}\tabularnewline
\textbf{Ofício} & 2022 & 357 & 29,75 & 27,0 & 8,81 & 18 & 24,00 & 36,00
& 45 & 12,00 & 29,6\tabularnewline
& 2023 & 434 & 36,17 & 35,5 & 7,16 & 22 & 32,75 & 38,50 & 50 & 5,75 &
19,8\tabularnewline
& \textbf{2024 (com IA)} & 532 & 44,33 & 43,5 & \textbf{6,04} & 30 &
41,75 & 49,00 & 52 & 7,25 & \textbf{13,6}\tabularnewline
\bottomrule
\end{longtable}

\textbf{Nota sobre Comunicados (SES/CONT):} o tipo Comunicado teve
volume residual em 2022 (n=8) e 2023 (n=28), e cresceu para 214 em 2024,
refletindo a virada da função preventiva discutida na Seção VI. Para
esse tipo, o coeficiente de variação não é métrica significativa para
comparação entre os anos da série, e foi omitido desta tabela.

\hypertarget{a.2.5-evoluuxe7uxe3o-da-produuxe7uxe3o-de-ofuxedcios}{%
\subsubsection{A.2.5 Evolução da produção de
Ofícios}\label{a.2.5-evoluuxe7uxe3o-da-produuxe7uxe3o-de-ofuxedcios}}

\begin{longtable}[]{@{}
  >{\raggedright\arraybackslash}p{(\linewidth - 6\tabcolsep) * \real{0.5655}}
  >{\raggedright\arraybackslash}p{(\linewidth - 6\tabcolsep) * \real{0.1207}}
  >{\raggedright\arraybackslash}p{(\linewidth - 6\tabcolsep) * \real{0.1207}}
  >{\raggedright\arraybackslash}p{(\linewidth - 6\tabcolsep) * \real{0.1931}}@{}}
\toprule
Indicador & 2023 & 2024 & Variação\tabularnewline
\midrule
\endhead
Ofícios produzidos & 434 & 532 & \textbf{+22,6\%}\tabularnewline
Processos tramitados (referência) & 1.752 & 1.581 &
−9,8\%\tabularnewline
Ofícios por processo tramitado & 0,25 & 0,34 &
\textbf{+36,0\%}\tabularnewline
\bottomrule
\end{longtable}

\hypertarget{a.2.6-suxedntese-das-eviduxeancias-sescont}{%
\subsubsection{A.2.6 Síntese das evidências
SES/CONT}\label{a.2.6-suxedntese-das-eviduxeancias-sescont}}

\begin{longtable}[]{@{}
  >{\raggedright\arraybackslash}p{(\linewidth - 4\tabcolsep) * \real{0.3365}}
  >{\raggedright\arraybackslash}p{(\linewidth - 4\tabcolsep) * \real{0.1828}}
  >{\raggedright\arraybackslash}p{(\linewidth - 4\tabcolsep) * \real{0.4807}}@{}}
\toprule
Evidência & Resultado & Significado\tabularnewline
\midrule
\endhead
Tempo médio geral & −18,2\% & De 17,92 para 14,66 dias por
processo\tabularnewline
Capacidade liberada estimada & +22,3\% & Equivalente a 352 processos
adicionais (5.155 dias)\tabularnewline
Passivo processual & Zerado & Caixa sem processos com mais de 10 dias ao
final de 2024\tabularnewline
Processos de alta complexidade & Até −95,5\% & 70\% dos tipos analisados
melhoraram (7 de 10)\tabularnewline
Julgamentos (análise normalizada) & +2,1\% & Média de 31,00/mês superou
os 30,36/mês de 2023 sem força-tarefa\tabularnewline
Julgamentos (estabilidade) & −55,4\% amplitude & Produção mais regular,
sem picos extraordinários\tabularnewline
Produtividade por processo & +0,4\% & Mesma entrega documental em 18,2\%
menos tempo\tabularnewline
Ofícios & +22,6\% & Maior capacidade de articulação
institucional\tabularnewline
Comunicados & +664\% & Virada da postura punitiva para
preventiva\tabularnewline
Problemas relacionados a IA relatados & Nenhum & Relatório institucional\tabularnewline
\bottomrule
\end{longtable}

\hypertarget{a.3-tabelas-de-dados-oficiais-unidade-de-controle-interno-ucisedet-df}{%
\subsection{A.3 Tabelas de dados oficiais: Unidade de Controle
Interno
(UCI/SEDET-DF)}\label{a.3-tabelas-de-dados-oficiais-unidade-de-controle-interno-ucisedet-df}}

Os dados de 2024--2025 das tabelas desta seção foram extraídos do Relatório Nº 1/2026 (Doc. SEI
193384267); a série de 2023 da Tabela A.3.4 provém das estatísticas oficiais do SEI-GDF da unidade (ver a nota daquela tabela).

\hypertarget{a.3.1-indicadores-gerais-2024-vs.-2025}{%
\subsubsection{A.3.1 Indicadores gerais 2024
vs.~2025}\label{a.3.1-indicadores-gerais-2024-vs.-2025}}

\begin{longtable}[]{@{}
  >{\raggedright\arraybackslash}p{(\linewidth - 6\tabcolsep) * \real{0.4468}}
  >{\raggedright\arraybackslash}p{(\linewidth - 6\tabcolsep) * \real{0.1383}}
  >{\raggedright\arraybackslash}p{(\linewidth - 6\tabcolsep) * \real{0.2191}}
  >{\raggedright\arraybackslash}p{(\linewidth - 6\tabcolsep) * \real{0.1958}}@{}}
\toprule
Indicador & 2024 & 2025 (com IA) & Variação\tabularnewline
\midrule
\endhead
Tempo médio de tramitação & 34 dias & 17 dias &
\textbf{−50\%}\tabularnewline
Processos tramitados & 256 & 336 & \textbf{+31\%}\tabularnewline
Documentos produzidos & 251 & 419 & \textbf{+67\%}\tabularnewline
Notas técnicas elaboradas & 66 & 122 & \textbf{+85\%}\tabularnewline
Despachos produzidos & 102 & 163 & +60\%\tabularnewline
\bottomrule
\end{longtable}

\hypertarget{a.3.2-distribuiuxe7uxe3o-mensal-de-processos-e-volume-financeiro-analisado-em-2025}{%
\subsubsection{A.3.2 Distribuição mensal do volume financeiro
analisado em 2025}\label{a.3.2-distribuiuxe7uxe3o-mensal-de-processos-e-volume-financeiro-analisado-em-2025}}

\begin{longtable}[]{@{}
  >{\raggedright\arraybackslash}p{(\linewidth - 2\tabcolsep) * \real{0.5000}}
  >{\raggedright\arraybackslash}p{(\linewidth - 2\tabcolsep) * \real{0.5000}}@{}}
\toprule
Mês & Valor analisado\tabularnewline
\midrule
\endhead
Janeiro & R\$ 5,0 mi\tabularnewline
Fevereiro & R\$ 72,1 mi\tabularnewline
Março & R\$ 37,8 mi\tabularnewline
Abril & R\$ 20,1 mi\tabularnewline
Maio & R\$ 44,1 mi\tabularnewline
Junho & R\$ 64,6 mi\tabularnewline
Julho & R\$ 23,6 mi\tabularnewline
\textbf{Agosto} & \textbf{R\$ 105,1 mi}\tabularnewline
Setembro & R\$ 23,2 mi\tabularnewline
Outubro & R\$ 64,4 mi\tabularnewline
Novembro & R\$ 26,8 mi\tabularnewline
Dezembro & R\$ 34,7 mi\tabularnewline
\textbf{Total 2025} & \textbf{R\$ 521,3 mi}\tabularnewline
\bottomrule
\end{longtable}

Nota: os valores mensais aparecem arredondados para R\$ 0,1 milhão; a
soma das parcelas pode, portanto, não fechar exatamente com o total
anual, que é R\$ 521.358.798,09 (dado-fonte sem arredondamento). Fonte: planilha oficial ``Estatística Notas Técnicas Unidade de Controle Interno, UCI/SEDET, 2025'', p.~48 do relatório gerencial assinado da UCI/SEDET 2025 (Doc. SEI 193436555).

\hypertarget{a.3.3-intensidade-analuxedtica-e-governanuxe7a}{%
\subsubsection{A.3.3 Intensidade analítica e
governança}\label{a.3.3-intensidade-analuxedtica-e-governanuxe7a}}

\begin{longtable}[]{@{}
  >{\raggedright\arraybackslash}p{(\linewidth - 2\tabcolsep) * \real{0.7056}}
  >{\raggedright\arraybackslash}p{(\linewidth - 2\tabcolsep) * \real{0.2944}}@{}}
\toprule
Indicador & 2025\tabularnewline
\midrule
\endhead
Notas técnicas elaboradas & 122\tabularnewline
Recomendações formais expedidas aos gestores & 286\tabularnewline
Volume financeiro analisado & R\$ 521,3 milhões\tabularnewline
Procedimento de revisão humana antes da expedição & Obrigatório (política Human-in-the-Loop; Tripla Revisão em três etapas)\tabularnewline
\bottomrule
\end{longtable}

\hypertarget{a.3.4-estatuxedstica-descritiva-da-produuxe7uxe3o-mensal-por-tipo-de-documento-ucisedet}{%
\subsubsection{A.3.4 Estatística descritiva da produção mensal por tipo
de documento:
UCI/SEDET}\label{a.3.4-estatuxedstica-descritiva-da-produuxe7uxe3o-mensal-por-tipo-de-documento-ucisedet}}

Base: 12 valores mensais por tipo/ano. Salvo indicação em contrário, as séries são a estatística bruta do SEI-GDF da unidade (extração nativa ``Estatísticas da Unidade'' descrita no Apêndice A.0). A série de Notas Técnicas de 2025 é a exceção: é o total consolidado no relatório gerencial assinado (122; ver Apêndice A.0), não a contagem bruta do SEI-GDF.

\begin{longtable}[]{@{}
  >{\raggedright\arraybackslash}p{(\linewidth - 22\tabcolsep) * \real{0.1519}}
  >{\raggedright\arraybackslash}p{(\linewidth - 22\tabcolsep) * \real{0.1133}}
  >{\raggedright\arraybackslash}p{(\linewidth - 22\tabcolsep) * \real{0.0816}}
  >{\raggedright\arraybackslash}p{(\linewidth - 22\tabcolsep) * \real{0.0816}}
  >{\raggedright\arraybackslash}p{(\linewidth - 22\tabcolsep) * \real{0.1143}}
  >{\raggedright\arraybackslash}p{(\linewidth - 22\tabcolsep) * \real{0.0653}}
  >{\raggedright\arraybackslash}p{(\linewidth - 22\tabcolsep) * \real{0.0490}}
  >{\raggedright\arraybackslash}p{(\linewidth - 22\tabcolsep) * \real{0.0653}}
  >{\raggedright\arraybackslash}p{(\linewidth - 22\tabcolsep) * \real{0.0816}}
  >{\raggedright\arraybackslash}p{(\linewidth - 22\tabcolsep) * \real{0.0490}}
  >{\raggedright\arraybackslash}p{(\linewidth - 22\tabcolsep) * \real{0.0816}}
  >{\raggedright\arraybackslash}p{(\linewidth - 22\tabcolsep) * \real{0.0653}}@{}}
\toprule
Tipo & Ano & Total & Média & Mediana & DP & Mín & Q1 & Q3 & Máx & IQR &
CV\%\tabularnewline
\midrule
\endhead
\textbf{Despacho} & 2023 & 85 & 7,08 & 6,5 & 4,56 & 1 & 4,25 & 10,50 &
14 & 6,25 & 64,4\tabularnewline
& 2024 & 102 & 8,50 & 9,0 & 6,23 & 0 & 3,25 & 13,25 & 18 & 10,00 &
73,3\tabularnewline
& \textbf{2025 (com IA)} & 163 & 13,58 & 12,5 & 6,69 & 5 & 8,00 & 19,25
& 25 & 11,25 & \textbf{49,3}\tabularnewline
\textbf{Nota Técnica} & 2023 & 82 & 6,83 & 7,5 & 3,76 & 0 & 4,00 & 9,25
& 12 & 5,25 & 55,1\tabularnewline
& 2024 & 66 & 5,50 & 5,5 & 4,58 & 0 & 1,75 & 7,50 & 15 & 5,75 &
83,3\tabularnewline
& \textbf{2025 (com IA, consolidado)} & 122 & 10,17 & 9,5 & 5,39 & 3 & 7,00 & 12,50
& 21 & 5,50 & \textbf{53,0}\tabularnewline
\bottomrule
\end{longtable}

\hypertarget{a.4-apurauxe7uxe3o-da-economia-hipotuxe9tica-das-nts-ucisedet-2025}{%
\subsection{A.4 Apuração da economia hipotética das NTs UCI/SEDET
2025}\label{a.4-apurauxe7uxe3o-da-economia-hipotuxe9tica-das-nts-ucisedet-2025}}

\hypertarget{a.4.1-universo-e-princuxedpio-de-classificauxe7uxe3o}{%
\subsubsection{A.4.1 Universo e princípio de
classificação}\label{a.4.1-universo-e-princuxedpio-de-classificauxe7uxe3o}}

A apuração abrange as 122 Notas Técnicas consolidadas no relatório gerencial assinado da UCI/SEDET 2025 (Doc. SEI 193384267), o mesmo universo oficial adotado em todo este paper (ver a nota de conciliação no Apêndice A.0). Cada uma dessas 122 notas tem valor oficial registrado na própria planilha do relatório (Doc. SEI 193436555, p.~48), cuja soma é R\$ 521.358.798,09 (R\$ 521,3 milhões): o valor das matérias submetidas à análise técnica.

Para os fins do modelo financeiro deste apêndice, os valores em risco foram atribuídos recomendação a recomendação, a partir do texto integral de cada uma das 122 notas técnicas, seguindo o critério moderado descrito em A.4.3, e auditados contra os documentos originais no SEI.

A classificação foi feita pelo conteúdo de cada recomendação, não pelo
veredito formal da NT. Em sistema não-vinculante de controle interno, é
comum que notas com conclusão favorável contenham recomendações
materiais cuja adoção evitaria perda futura; classificar pelo veredito
formal subestimaria sistematicamente a efetividade do controle.

A granularidade da análise é a recomendação individual, não a NT como
unidade. As recomendações identificadas e classificadas pelos autores a partir do texto integral de cada uma das 122 notas técnicas alimentam este modelo financeiro (ver A.4.7). Essa classificação é própria dos autores, construída estritamente para fins de modelagem, e não substitui o total institucional de 286 recomendações formais informado no relatório gerencial assinado de 2025 (p.~5), que rege todas as demais afirmações deste paper (ver Apêndice A.0). A diferença reflete a forma como os trechos textuais foram segmentados em recomendações discretas e passíveis de avaliação individual para fins de atribuição de valor em risco, e não uma divergência quanto à contagem oficial.

\hypertarget{a.4.2-tipologia-aplicada}{%
\subsubsection{A.4.2 Tipologia
aplicada}\label{a.4.2-tipologia-aplicada}}

Cada recomendação foi classificada em duas dimensões. A primeira é a
natureza:

\begin{longtable}[]{@{}
  >{\raggedright\arraybackslash}p{(\linewidth - 2\tabcolsep) * \real{0.3150}}
  >{\raggedright\arraybackslash}p{(\linewidth - 2\tabcolsep) * \real{0.6850}}@{}}
\toprule
Natureza & Definição\tabularnewline
\midrule
\endhead
I: Procedimental/documental & Recomendações sobre documentação,
descrição, fluxo, padronização, controles internos formais ou melhorias
prospectivas. Sem implicação financeira direta sobre o ato sob
análise.\tabularnewline
II: Conformidade legal & Descumprimento de exigência legal específica
que pode causar nulidade ou questionamento por controle externo
(certidão expirada, ausência de empenho, vício de cláusula obrigatória,
divergência documental sem valor quantificado).\tabularnewline
III: Anomalia quantitativa & Divergência de valor, sobrepreço,
pagamento indevido, duplicidade, glosa específica. Valor monetário
mensurável em risco.\tabularnewline
IV: Irregularidade material & Indícios de fraude, conluio,
favorecimento, ou recomendação explícita de PAD/sindicância/comunicação
ao MP/TCDF/MPC. Casos raros em controle preventivo; nenhuma das 122 notas técnicas
do exercício se enquadrou nesta categoria.\tabularnewline
\bottomrule
\end{longtable}

A segunda dimensão é a materialidade:

\begin{longtable}[]{@{}
  >{\raggedright\arraybackslash}p{(\linewidth - 2\tabcolsep) * \real{0.2021}}
  >{\raggedright\arraybackslash}p{(\linewidth - 2\tabcolsep) * \real{0.7979}}@{}}
\toprule
Materialidade & Definição\tabularnewline
\midrule
\endhead
Alta & Natureza IV; natureza III com valor explícito; natureza II com
vício de validade afetando pagamento/contrato de valor
significativo.\tabularnewline
Média & Natureza III sem valor direto explícito; natureza II com
cláusula que pode gerar nulidade contratual.\tabularnewline
Baixa & Natureza II isolada (descumprimento legal sanável
menor).\tabularnewline
Nula & Apenas natureza I (procedimental pura).\tabularnewline
\bottomrule
\end{longtable}

\hypertarget{a.4.3-crituxe9rio-moderado-para-atribuiuxe7uxe3o-de-valor-em-risco}{%
\subsubsection{A.4.3 Critério moderado para atribuição de valor em
risco}\label{a.4.3-crituxe9rio-moderado-para-atribuiuxe7uxe3o-de-valor-em-risco}}

O valor em risco de cada recomendação foi atribuído segundo seis casos
exaustivos:

\begin{enumerate}
\def\labelenumi{\arabic{enumi}.}
\tightlist
\item
  \textbf{Valor explícito} mencionado na recomendação (glosa específica,
  devolução, ressarcimento quantificado): usa esse valor.
\item
  \textbf{Vício que afeta a validade do ato} (certidão expirada,
  ausência de empenho, divergência atesto×NF, ausência de designação
  formal de fiscal por ato publicado, cláusula contratual em desacordo
  com edital, ausência de cláusula obrigatória, descumprimento de
  exigência legal de habilitação): usa o valor principal do
  pagamento/contrato em análise.
\item
  \textbf{Divergência quantitativa sem valor especificado:} usa o valor
  principal como teto máximo de exposição, com justificativa explícita
  de que o risco real pode ser inferior.
\item
  \textbf{Recomendação procedimental pura} (padronização, detalhamento,
  melhoria de relatório, designação preventiva para futuras
  contratações): valor zero.
\item
  \textbf{Recomendação de melhoria futura, vigilância ou manutenção} de
  boas práticas: valor zero.
\item
  \textbf{Observância genérica de legalidade} sem apontamento de falha
  específica: valor zero.
\end{enumerate}

Adicionalmente, quando múltiplas recomendações de uma mesma NT apontam
vícios diferentes, mas afetam o mesmo pagamento/contrato, o valor é
atribuído à recomendação mais grave; as demais recebem valor zero com
justificativa ``vício acessório à recomendação X.Y, evitando dupla
contagem''. Esta regra é necessária porque a perda efetiva sobre um
mesmo objeto financeiro só pode ocorrer uma vez, independentemente de
quantos vícios formais a UCI identifique simultaneamente.

\hypertarget{a.4.4-matriz-de-probabilidade}{%
\subsubsection{A.4.4 Matriz de
probabilidade}\label{a.4.4-matriz-de-probabilidade}}

A cada recomendação é aplicada uma probabilidade de materialização do
risco, modulada por natureza e materialidade:

\begin{longtable}[]{@{}
  >{\raggedright\arraybackslash}p{(\linewidth - 6\tabcolsep) * \real{0.4540}}
  >{\raggedright\arraybackslash}p{(\linewidth - 6\tabcolsep) * \real{0.2310}}
  >{\raggedright\arraybackslash}p{(\linewidth - 6\tabcolsep) * \real{0.1470}}
  >{\raggedright\arraybackslash}p{(\linewidth - 6\tabcolsep) * \real{0.1680}}@{}}
\toprule
Natureza & Conservador & Central & Otimista\tabularnewline
\midrule
\endhead
I: Procedimental & 0\% & 0\% & 0\%\tabularnewline
II: Conformidade legal & 2\% & 5\% & 10\%\tabularnewline
III: Anomalia quantitativa & 10\% & 25\% & 40\%\tabularnewline
IV: Irregularidade material & 30\% & 50\% & 70\%\tabularnewline
\bottomrule
\end{longtable}

O modificador de materialidade ajusta cada probabilidade: alta = 1,00;
média = 0,75; baixa = 0,50; nula = 0,00.

A estimativa de mitigação potencial por recomendação é calculada como:
\textbf{valor em risco × probabilidade (natureza, cenário) × modificador
(materialidade)}.

A estimativa agregada é o somatório sobre as recomendações
classificadas nesta apuração (Apêndice A.4.7), para cada um dos três cenários.

\hypertarget{a.4.5-calibragem-das-probabilidades-fontes-utilizadas}{%
\subsubsection{A.4.5 Calibragem das probabilidades: fontes
utilizadas}\label{a.4.5-calibragem-das-probabilidades-fontes-utilizadas}}

A matriz de probabilidades (Tabela A.4.4) foi calibrada com base
prudencial: parte do princípio de que o controle prévio orientativo,
característico da UCI, não substitui a decisão do gestor, o que limita a
fração de recomendações que efetivamente impede a perda. As
probabilidades foram fixadas em três cenários, conservador, central e
otimista, ancorados na seguinte literatura de referência:

\textbf{(i) Auditoria pública federal americana.} O U.S. Government
Accountability Office (GAO), órgão federal de auditoria externa
equivalente funcional do TCU, publica anualmente estimativas de
\emph{improper payments} (US\$ 162 bilhões em FY 2024, US\$ 186 bilhões
em FY 2025; aproximadamente US\$ 3 trilhões acumulados desde 2003), e a
literatura associada de \emph{payment integrity} documenta taxas típicas
de recuperação efetiva consideravelmente inferiores ao volume
identificado.

\textbf{(ii) Estudos de fraude ocupacional.} O \emph{Report to the
Nations} da ACFE (1.921 casos em 138 países, edição 2024) documenta que
organizações com controles internos estruturados, particularmente
\emph{surprise audits}, \emph{financial statement audits}, hotlines de
denúncia e análise proativa de dados, apresentam reduções substanciais
(até 50\%) em perdas e duração de fraude comparadas a organizações sem
tais controles.

A combinação dessas referências sustenta uma matriz
\textbf{intencionalmente conservadora}: o cenário central da natureza II
(conformidade legal) opera com 5\% de probabilidade de materialização, o
da natureza III (anomalia quantitativa) com 25\%, e o da natureza IV
(irregularidade material) com 50\%. Esses valores não são estimados
diretamente da literatura; são suposições prudenciais, baseadas em
julgamento, consistentes com as faixas sugeridas pelos estudos
agregados. O cenário otimista representa o limite superior da literatura
agregada; o conservador, o piso prudencial.

\hypertarget{a.4.6-processo-de-validauxe7uxe3o}{%
\subsubsection{A.4.6 Processo de
validação}\label{a.4.6-processo-de-validauxe7uxe3o}}

O resultado passou por quatro fases sequenciais de validação. A primeira
foi a leitura integral nota a nota. A segunda foi a reconciliação
focada, NTs com texto extenso e baixa contagem inicial de recomendações
foram re-examinadas para incorporar recomendações adicionais
identificadas. A terceira foi a auditoria cruzada, uma amostra
estratificada cobrindo todos os blocos de processamento e todas as
naturezas foi revisada, com correção das divergências identificadas em
natureza, materialidade e valor em risco. A quarta foi a última
validação, amostra dos principais contribuintes ao agregado financeiro
foi revisada pelo autor, com aplicação das observações.

\hypertarget{a.4.7-resultados-consolidados}{%
\subsubsection{A.4.7 Resultados
consolidados}\label{a.4.7-resultados-consolidados}}

\textbf{Tier 1: Valor oficial do universo analisado:}

R\$ 521.358.798,09 (R\$ 521,3 milhões), soma do valor oficial de cada uma das 122 notas técnicas registrado na própria planilha do relatório gerencial assinado (Doc. SEI 193436555, p.~48; ver Apêndice A.0).

\textbf{Tier 2: Recomendações por natureza (classificação própria dos autores para fins de modelagem; ver Apêndice A.0 para o total institucional de 286 recomendações):}

\begin{longtable}[]{@{}
  >{\raggedright\arraybackslash}p{(\linewidth - 2\tabcolsep) * \real{0.5000}}
  >{\raggedright\arraybackslash}p{(\linewidth - 2\tabcolsep) * \real{0.5000}}@{}}
\toprule
Natureza & Valor em risco bruto\tabularnewline
\midrule
\endhead
IV: Irregularidade material & N/D\tabularnewline
III: Anomalia quantitativa & R\$ 18.601.185,78 (R\$ 18,6 milhões)\tabularnewline
II: Conformidade legal & R\$ 222.084.052,71 (R\$ 222,1 milhões)\tabularnewline
I: Procedimental & N/D\tabularnewline
\textbf{Total} & \textbf{R\$ 240.685.238,49 (R\$ 240,7
milhões)}\tabularnewline
\bottomrule
\end{longtable}

\textbf{Tier 3: Estimativa de mitigação potencial:}

\begin{longtable}[]{@{}
  >{\raggedright\arraybackslash}p{(\linewidth - 2\tabcolsep) * \real{0.4477}}
  >{\raggedright\arraybackslash}p{(\linewidth - 2\tabcolsep) * \real{0.5523}}@{}}
\toprule
Cenário & Total\tabularnewline
\midrule
\endhead
Conservador & \textbf{R\$ 6.094.150,54 (R\$ 6,1 milhões)}\tabularnewline
Central & \textbf{R\$ 15.235.376,36 (R\$ 15,2 milhões)}\tabularnewline
Otimista & \textbf{R\$ 28.682.721,75 (R\$ 28,7 milhões)}\tabularnewline
\bottomrule
\end{longtable}

Essas estimativas não são economia realizada ou auditada: trata-se de uma faixa modelada de mitigação potencial, condicionada à matriz de probabilidade de A.4.4 e à efetiva adoção das recomendações pelos gestores, o que este estudo não acompanha.

\hypertarget{a.4.8-limitauxe7uxf5es-da-estimativa}{%
\subsubsection{A.4.8 Limitações da
estimativa}\label{a.4.8-limitauxe7uxf5es-da-estimativa}}

\begin{enumerate}
\def\labelenumi{\arabic{enumi}.}
\item
  \textbf{Caráter não-vinculante das recomendações UCI.} A UCI emite
  controle prévio orientativo; não há registro sistemático de quais
  recomendações foram efetivamente acatadas pelos gestores. A estimativa
  quantifica potencial de mitigação, não realização documentada.
\item
  \textbf{Counterfactual não observável.} Não é possível estabelecer com
  certeza o cenário alternativo sem controle UCI. As probabilidades
  aplicadas são derivadas de literatura internacional, não calibradas
  com dados específicos do GDF.
\item
  \textbf{Subestimação por classificação conservadora.} Recomendações
  classificadas como Natureza I (procedimental) recebem probabilidade
  zero, mesmo quando podem prevenir perdas indiretas, opção deliberada
  por rigor metodológico.
\item
  \textbf{Benefícios não-monetários não capturados.} Ganhos em
  transparência, conformidade processual, padronização documental e
  capacidade técnica da equipe não são quantificados pela estimativa
  financeira.
\item
  \textbf{Critério moderado é interpretativo.} A atribuição do valor
  inteiro do pagamento/contrato a vícios de validade é defensável, mas
  representa uma escolha metodológica que poderia ser apresentada de
  forma mais ou menos generosa em interpretações alternativas.
\end{enumerate}

\hypertarget{referuxeancias}{%
\section{Referências}\label{referuxeancias}}

Documentos Institucionais SEI-GDF

{[}1{]} DISTRITO FEDERAL. Secretaria de Estado de Saúde. Controladoria
Setorial da Saúde (SES/CONT). \textbf{Relatório Gerencial Executivo:
Impacto da Inteligência Artificial na Eficiência Operacional.} Doc. SEI
nº 197403428, Processo SEI 00060-00582291/2024-11. Brasília, DF, 2 de
janeiro de 2025.

{[}2{]} DISTRITO FEDERAL. Secretaria de Estado de Desenvolvimento
Econômico, Trabalho e Renda. Unidade de Controle Interno (UCI/SEDET-DF).
\textbf{Relatório Nº 1/2026: Relatório Gerencial de Resultados 2025,
UCI/GAB/SEDET.} Doc. SEI nº 193384267, Processo SEI
04035-00000853/2026-24. Brasília, DF, 28 de janeiro de 2026.

{[}3{]} DISTRITO FEDERAL. Secretaria de Estado de Desenvolvimento
Econômico, Trabalho e Renda. Unidade de Controle Interno (UCI/SEDET-DF).
\textbf{Framework de Governança de IA, UCI.} Doc. SEI nº 194251158.
Brasília, DF.

{[}4{]} DISTRITO FEDERAL. Secretaria de Estado de Saúde. Controladoria
Setorial da Saúde, Assessoria de Apoio aos Julgamentos de Processos
Administrativos (SES/CONT/ASJULG). \textbf{Memorando Nº 4/2024:
Proposta de criação do curso "Inteligência Artificial no setor público:
técnicas, riscos e aplicações".} Doc. SEI nº 146621014, Processo SEI
00060-00314546/2024-15. Signatário: Vinicius Santana Gomes. Brasília,
DF, 25 de junho de 2024.

{[}5{]} DISTRITO FEDERAL. Secretaria Executiva da Casa Civil,
Subsecretaria de Gestão Administrativa, Escola de Governo
(SEEC/SEGEA/EGOV). \textbf{Memorando Nº 166/2024: Aprovação
institucional do curso "Inteligência Artificial no setor público:
técnicas, riscos e aplicações".} Doc. SEI nº 146526272, Processo SEI
04044-00021535/2024-26. Signatária: Raquel Aben Athar de Sousa (Diretora
Executiva substituta). Brasília, DF, 22 de julho de 2024.

{[}6{]} DISTRITO FEDERAL. Secretaria de Estado de Saúde. Controladoria
Setorial da Saúde, Assessoria de Apoio aos Julgamentos de Processos
Administrativos (SES/CONT/ASJULG). \textbf{Relatório Nº 12/2024:
Registro institucional de zeração do passivo de processos da unidade.}
Brasília, DF, 14 de dezembro de 2024.

Normativas Brasileiras

{[}7{]} BRASIL. \textbf{Lei nº 13.709, de 14 de agosto de 2018.} Lei
Geral de Proteção de Dados Pessoais (LGPD). Diário Oficial da União,
Brasília, DF.

Auditoria Pública

{[}8{]} U.S. GOVERNMENT ACCOUNTABILITY OFFICE (GAO). \textbf{Improper
Payments: Information on Agencies' Fiscal Year 2024 Estimates.}
GAO-25-107753. Washington, DC: GAO, 2025. Disponível em:
https://www.gao.gov/products/gao-25-107753.

{[}9{]} U.S. GOVERNMENT ACCOUNTABILITY OFFICE (GAO). \textbf{Payment
Integrity: Agencies' Estimated Improper Payments Increased to \$186
Billion in Fiscal Year 2025.} GAO-26-108694. Washington, DC: GAO, 2026.
Disponível em: https://www.gao.gov/products/gao-26-108694.

{[}10{]} ASSOCIATION OF CERTIFIED FRAUD EXAMINERS (ACFE).
\textbf{Occupational Fraud 2024: A Report to the Nations.} Austin, TX:
ACFE, 2024. Disponível em:
https://www.acfe.com/-/media/files/acfe/pdfs/rttn/2024/2024-report-to-the-nations.pdf.

Frameworks Internacionais de Governança de Inteligência Artificial

{[}11{]} NATIONAL INSTITUTE OF STANDARDS AND TECHNOLOGY (NIST).
\textbf{Artificial Intelligence Risk Management Framework (AI RMF 1.0).}
NIST AI 100-1. Gaithersburg, MD: U.S. Department of Commerce, 2023.
Disponível em: https://www.nist.gov/itl/ai-risk-management-framework.

{[}12{]} U.S. GOVERNMENT ACCOUNTABILITY OFFICE (GAO). \textbf{Artificial
Intelligence: An Accountability Framework for Federal Agencies and Other
Entities.} GAO-21-519SP. Washington, DC: GAO, 2021. Disponível em:
https://www.gao.gov/products/gao-21-519sp.

{[}13{]} EUROPEAN PARLIAMENT AND COUNCIL OF THE EUROPEAN UNION.
\textbf{Regulation (EU) 2024/1689 of 13 June 2024 laying down harmonised
rules on artificial intelligence (Artificial Intelligence Act).}
Official Journal of the European Union, Brussels: European Union, 2024.
Disponível em: https://eur-lex.europa.eu/eli/reg/2024/1689/oj.

{[}14{]} ORGANISATION FOR ECONOMIC CO-OPERATION AND DEVELOPMENT (OECD).
\textbf{Recommendation of the Council on Artificial Intelligence.}
OECD/LEGAL/0449. Paris: OECD, adopted 22 May 2019, amended 8 November
2023 and 3 May 2024. Disponível em: https://oecd.ai/en/ai-principles.

Referências Metodológicas

{[}15{]} WEI, Jason; WANG, Xuezhi; SCHUURMANS, Dale; BOSMA, Maarten;
ICHTER, Brian; XIA, Fei; CHI, Ed H.; LE, Quoc V.; ZHOU, Denny.
\textbf{Chain-of-Thought Prompting Elicits Reasoning in Large Language
Models.} Advances in Neural Information Processing Systems (NeurIPS),
2022. Disponível em: https://arxiv.org/abs/2201.11903 (DOI:
https://doi.org/10.48550/arXiv.2201.11903).

{[}16{]} BROWN, Tom B.; MANN, Benjamin; RYDER, Nick et
al.~\textbf{Language Models are Few-Shot Learners.} Advances in Neural
Information Processing Systems (NeurIPS), 2020. Disponível em:
https://arxiv.org/abs/2005.14165 (DOI:
https://doi.org/10.48550/arXiv.2005.14165).

Adoção Tecnológica e Capacitação em IA

{[}17{]} COHEN, Wesley M.; LEVINTHAL, Daniel A. \textbf{Absorptive
Capacity: A New Perspective on Learning and Innovation.} Administrative
Science Quarterly, v. 35, n.~1, p.~128--152, 1990. Disponível em:
https://doi.org/10.2307/2393553.

{[}18{]} DUNLEAVY, Patrick; MARGETTS, Helen; BASTOW, Simon; TINKLER,
Jane. \textbf{New Public Management Is Dead: Long Live Digital-Era
Governance.} Journal of Public Administration Research and Theory, v.
16, n.~3, p.~467--494, 2006. Disponível em:
https://doi.org/10.1093/jopart/mui057.

{[}19{]} HEEKS, Richard. \textbf{Most eGovernment-for-development
projects fail: How can risks be reduced?} Manchester: Institute for
Development Policy and Management, University of Manchester, 2003.
Disponível em: https://ssrn.com/abstract=3540052.

{[}20{]} LONG, Duri; MAGERKO, Brian. \textbf{What is AI literacy?
Competencies and design considerations.} Proceedings of the 2020 CHI
Conference on Human Factors in Computing Systems, p. 1--16, 2020.
Disponível em: https://doi.org/10.1145/3313831.3376727.

\end{document}